\documentclass[nojss]{jss}

\usepackage{thumbpdf,lmodern}
\usepackage{amsmath}
\usepackage{amssymb}
\usepackage{color}
\usepackage{framed}
\usepackage{subfigure}
\usepackage{multirow}

\author{Mar\'ia Xos\'e Rodr\'iguez-\'Alvarez\\BCAM - Basque Center for Applied Mathematics \\and IKERBASQUE, Basque Foundation for Science \And Vanda In\'acio\\University of Edinburgh}
\Plainauthor{Mar\'ia Xos\'e Rodr\'iguez-\'Alvarez, Vanda In\'acio}

\title{\pkg{ROCnReg}: An \proglang{R} Package for Receiver Operating Characteristic Curve Inference with and
without Covariate Information}
\Plaintitle{ROCnReg: An R Package for Receiver Operating Characteristic Curve Inference with and without Covariate Information}
\Shorttitle{\pkg{ROCnReg}: An \proglang{R} Package for ROC Inference}

\Abstract{
The receiver operating characteristic (ROC) curve is the most popular tool used to evaluate the discriminatory capability of diagnostic tests/biomarkers measured on a continuous scale  when distinguishing between two alternative disease states (e.g, diseased and nondiseased). In some circumstances, the test's performance and its discriminatory ability may vary according to subject-specific characteristics or different test settings. In such cases, information-specific accuracy measures, such as the covariate-specific and the covariate-adjusted ROC curve are needed, as ignoring covariate information may lead to biased or erroneous results. This paper introduces the \proglang{R} package \pkg{ROCnReg} that allows estimating the pooled (unadjusted) ROC curve, the covariate-specific ROC curve, and the covariate-adjusted ROC curve by different methods, both from (semi) parametric and nonparametric perspectives and within Bayesian and frequentist paradigms. From the estimated ROC curve (pooled, covariate-specific or covariate-adjusted), several summary measures of discriminatory accuracy, such as the (partial) area under the ROC curve and the Youden index, can be obtained. The package also provides functions to obtain ROC-based optimal threshold values using several criteria, namely, the Youden index criterion and the criterion that sets a target value for the false positive fraction. For the Bayesian methods, we provide tools for assessing model fit via posterior predictive checks, while model choice can be carried out via several information criteria. Numerical and graphical outputs are provided for all methods. The package is illustrated through the analyses of data from an endocrine study where the aim is to assess the capability of the body mass index to detect the presence or absence of cardiovascular disease risk factors. The package is available from the Comprehensive \proglang{R} Archive Network (CRAN) at \url{https://CRAN.R-project.org/package=ROCnReg}.
}

\Plainkeywords{accuracy measures, Bayesian, diagnostic tests, ROC curve, covariate-specific ROC curve, covariate-adjusted ROC curve, optimal thresholds, R}

\Address{ Mar\'ia Xos\'e Rodr\'iguez-\'Alvarez\\
  BCAM - Basque Center for Applied Mathematics\\
  \emph{and}\\
  IKERBASQUE, Basque Foundation for Science\\
  Alameda de Mazarredo, 14\\
  E-48009 Bilbao, Basque Country, Spain\\
  E-mail: \email{mxrodriguez@bcamath.org}\\
  
  Vanda In\'acio\\
  School of Mathematics\\
  University of Edinburgh\\
  James Clerk Maxwell Building, The King's Buildings, Peter Guthrie Tait Road\\
  EH9 3FD Edinburgh, Scotland, United Kingdom\\
  E-mail: \email{vanda.inacio@ed.ac.uk}
}

\begin{document}

%% -- Introduction -------------------------------------------------------------

%% - In principle "as usual".
%% - But should typically have some discussion of both _software_ and _methods_.
%% - Use \proglang{}, \pkg{}, and \code{} markup throughout the manuscript.
%% - If such markup is in (sub)section titles, a plain text version has to be
%%   added as well.
%% - All software mentioned should be properly \cite-d.
%% - All abbreviations should be introduced.
%% - Unless the expansions of abbreviations are proper names (like "Journal
%%   of Statistical Software" above) they should be in sentence case (like
%%   "generalized linear models" below).

\section{Introduction} \label{sec:intro}
%The problem of binary classification is a recurrent topic in widely diverse fields, such as medicine, biology, and finance. In general, this problem may be described as follows: a set of individuals or objects can be classified, with no uncertainty, into two different categories or classes, and the aim is to determine the category to which individuals belong to, based on some observed-and more easily obtainable-information on each of them. The importance of or need for making such a classification will depend on the particular study situation in question but, as a general rule, it will afford relevant information that can be subsequently used. For instance, in the medical field, to be able to classify given individuals as diseased (or nondiseased) on the basis of some known clinical information about them would be the point of departure for their subsequent treatment.

%Associated with any classification rule is the need to evaluate its performance. For instance, classification of an individual's healthy/diseased status based on the result of a diagnostic test is not usually error-free; there is always the possibility of individuals being misclassified. Accordingly, before the routine application of any classifier in practice, any classification errors must be rigorously quantified. In other words, what is known as the accuracy of the classifier or its ability to discriminate between alternative classes/categories, must be assessed. 
Before a diagnostic test is approved for being routinely used in practice, its ability to distinguish say, diseased from nondiseased individuals, must be narrowly evaluated. Throughout we assume that the true disease status of the individuals is known and the task is, compared to the truth, to quantify how accurate the test being investigated is. Before proceeding, it is worth noting that although our focus is on medical diagnosis, the problem of binary classification is a much wider one, finding applications in fields as diverse as biology and finance, to name only two.

The receiver operating characteristic (ROC) curve \citep{Metz78} is, unarguably, the most popular tool used for evaluating the discriminatory ability of continuous-outcome diagnostic tests. The ROC curve displays the false positive fraction (FPF) against the true positive fraction (TPF) for all possible threshold values used to dichotomise the test result. The ROC curve thus provides a global description of the trade-off between the FPF and the TPF of the test as the threshold changes. Plenty of parametric and semi/nonparametric methods are available for estimating ROC curves, either from frequentist or Bayesian viewpoints and we refer the interested reader to \citet[][Chapter 5]{Pepe98}, \citet[][Chapter 4]{Zhou2011}, \cite{Goncalves2014} and \cite{Inacio2020}, and references therein.

However, it is known that in many situations, a test's outcome and, possibly, its discriminatory capacity, can be affected by additional information (covariates); \citet[pp 48--49]{Pepe03} provides several examples of covariates that can affect the result of a diagnostic test. For instance, patient characteristics, such as age and gender, are important covariates to be considered as diagnostic accuracy is likely to vary according to them. In these cases pooling the test outcomes regardless of their covariate values may lead to erroneous or, at least, oversimplified conclusions and decisions. Interest should therefore be focused on assessing the accuracy of the test, but taking into account covariate information. Two different ROC-based measures that incorporate covariate information have been proposed: the covariate-specific or conditional ROC curve \citep[see, e.g.,][Chapter 6]{Pepe03} and the covariate-adjusted ROC curve \citep{Janes09a}. The formal definition of both curves is given in Section \ref{sec:notation}. In brief, a covariate-specific ROC curve is an ROC curve that conditions on a specific covariate value, thus describing the accuracy of the test in the `subpopulation' defined by that covariate value. On the other hand, the covariate-adjusted ROC curve is a weighted average of covariate-specific ROC curves. Regarding estimation, since the seminal paper of \cite{Pepe98}, a plethora of methods have been proposed in the literature for the estimation of the covariate-specific ROC curve and associated summary measures. Without being exhaustive, we mention the work of \cite{Faraggi03}, \cite{MX11a,MX11b}, \cite{Inacio13}, and \cite{Inacio17}. A detailed review can be found in \cite{MX11c}, \cite{Pardo14} and \cite{Inacio2020}. With respect to the covariate-adjusted ROC curve, estimation has been discussed in \cite{Janes09a}, \cite{MX11a}, \cite{Zhong12}, and \cite{Inacio18}. %Pooling the test outcomes regardless their covariate values may lead to erroneous or, at least, oversimplified conclusions and decisions.

In a slightly different context, in the machine learning community, the topic of covariate-dependent classification has received much attention recently, being related to the concept of \textit{fairness} \citep[see, e.g.,][]{Hutchinson:2019}. As an example, \cite{Buolamwini17} reports that the evaluation of four gender classifiers revealed that a significant gap exists when comparing gender classification accuracies of females versus males. 

There are a few \proglang{R} \citep{R20} packages for ROC curve analysis available on the Comprehensive R Archive Network (CRAN) and, as far as we are aware, all of them implementing frequentist approaches. The package \pkg{sROC} \citep{package_sROC} contains functions to perform nonparametric, kernel-based, estimation of ROC curves, while \pkg{pROC} \citep{package_pROC} offers a set of tools to visualise, smooth, and compare ROC curves, and \pkg{nsROC} \citep{perez2018} also allows estimating ROC curves, building confidence bands, as well as comparing several curves both for dependent and independent data (i.e., data arising from paired and unpaired designs, respectively). However, covariate information cannot be explicitly taken into account in any of these packages. Packages \pkg{ROCRegression} (available at \url{https://bitbucket.org/mxrodriguez/rocregression}) and \pkg{npROCRegression} \citep{package_npROCRegression} provide routines to estimate semiparametrically and nonparametrically, under a frequentist framework, the covariate-specific ROC curve. We also mention  \pkg{OptimalCutpoints} \citep{Lopez2014} and  \pkg{ThresholdROC} \citep{Perez2017} that provide a collection of functions for point and interval estimation of optimal thresholds for continuous diagnostic tests. To the best of our knowledge, there is no statistical software package implementing Bayesian inference for ROC curves and associated summary indices and optimal thresholds. 

To close this gap, in this paper we introduce the \pkg{ROCnReg} package that allows conducting Bayesian inference for the (pooled or marginal) ROC curve, the covariate-specific ROC curve, and the covariate-adjusted ROC curve. For the sake of generality, frequentist approaches are also implemented. Specifically, in what concerns estimation of the pooled ROC curve, \pkg{ROCnReg} implements the frequentist empirical estimator described in \cite{Hsieh1996}, the kernel-based approach proposed of \cite{Zou97}, the Bayesian Bootstrap method of \cite{Gu2008}, and the Bayesian nonparametric method based on a Dirichlet process mixture of normal distributions model proposed by \cite{Erkanli2006}. Regarding the covariate-specific ROC curve, \pkg{ROCnReg} implements the frequentist normal method of \cite{Faraggi03} and its semiparametric counterpart as described in \cite{Pepe98}, the kernel-based approach of \cite{MX11a}, and the Bayesian nonparametric model, based on a single-weights dependent Dirichlet process mixture of normal distributions, proposed by \cite{Inacio13}. As for the covariate-adjusted ROC curve, the \pkg{ROCnReg} package allows estimation using the frequentist semiparametric approach of \cite{Janes09a}, the frequentist nonparametric method discussed in \cite{MX11a}, and the recently proposed Bayesian nonparametric estimator of \cite{Inacio18}. Table~\ref{tab:overview} shows a summary of all methods implemented in the package. In addition, \pkg{ROCnReg} also provides functions to obtain ROC-based optimal thresholds to perform the classification/diagnosis using two different criteria, namely, the Youden index and the criterion that sets a target value for the false positive fraction. These are implemented for both the ROC curve, the covariate-specific and the covariate-adjusted ROC curve. A detailed description of the methods is presented in Section \ref{sec:methods}.

\begin{table}[t!]
\centering
\begin{tabular}{p{3.5cm}p{10.5cm}}
\hline
Method & Description \\ \hline
\multicolumn{2}{l}{Pooled ROC curve} \\ \hline
emp & (Frequentist) empirical estimator \citep{Hsieh1996}.\\
kernel & (Frequentist) kernel-based approach \citep{Zou97}.\\
BB &  Bayesian bootstrap method \citep{Gu2008}. \\
dpm &  Nonparametric Bayesian approach based on a Dirichlet process mixture of normal distributions \citep{Erkanli2006}.\\ \hline
\multicolumn{2}{l}{Covariate-specific ROC curve} \\ \hline
sp & (Frequentist) parametric and semiparametric induced ROC regression approach \citep{Pepe98,Faraggi03}\\
kernel & Nonparametric (kernel-based) induced ROC regression approach \citep{MX11a}.\\
bnp & Nonparametric Bayesian model based on a single-weights dependent Dirichlet process mixture of normal distributions \citep{Inacio13}.\\ \hline
\multicolumn{2}{l}{Covariate-adjusted ROC curve} \\ \hline
sp & (Frequentist) semiparametric method \citep{Janes09a}.\\
kernel &  Nonparametric (kernel-based) induced ROC regression approach \citep{MX11a}.\\
bnp & Nonparametric Bayesian model based on a single-weights dependent Dirichlet process mixture of normal distributions and the Bayesian bootstrap \citep{Inacio18}. \\ \hline
\end{tabular}
\caption{\label{tab:overview} Overview of ROC estimation methods included in the \pkg{ROCnReg} package.}
\end{table}

The remainder of the paper is organised as follows. In Section \ref{sec:notation} we formally introduce the (pooled or marginal) ROC curve, the covariate-specific ROC curve, and the covariate-adjusted ROC curve. The description of the estimation methods implemented in the \pkg{ROCnReg} package is given in Section \ref{sec:methods}. In Section \ref{sec:illustration} the usage of the main functions and capabilities of \pkg{ROCnReg} are described and illustrated using a real example. The paper concludes with a discussion in Section \ref{sec:summary}.

\section{Notation and definitions} \label{sec:notation}
This section sets out the formal definition of the pooled or marginal ROC curve, the covariate-specific ROC curve, and the covariate-adjusted ROC curve. Also, it describes the most commonly used summary measures of discriminatory accuracy, namely, the area under the ROC curve, the partial area under the ROC curve, and the Youden index. For conciseness, we intentionally avoid giving too many details and refer the interested reader to \cite{Pepe03} (and references therein) for an extensive account of many aspects of ROC curves with and without covariates.

In what follows, we denote as $Y$ the outcome of the diagnostic test and as $D$ the binary variable indicating the presence ($D = 1$) or absence ($D = 0$) of disease. We also assume that along with $Y$ and the true disease status $D$, a covariate vector $\mathbf{X}$ is also available, and that it may encompass both continuous and categorical covariates. For ease of notation, the covariate vector $\mathbf{X}$ is assumed to be the same in both the diseased ($D = 1$) and nondiseased ($D = 0$) populations, although this is not always necessarily the case in practice (e.g., disease stage is, obviously, a disease-specific covariate). By a slight abuse of notation, we use the subscripts $D$ and $\bar{D}$ to denote (random) quantities conditional on, respectively, $D = 1$ and $D = 0$. For example, $Y_{D}$ and $Y_{\bar{D}}$ denote the test outcomes in the diseased and nondiseased populations.

\subsection{Pooled ROC curve}\label{sec:pooledroc}
\begin{figure}[ht!]
	\begin{center}
		\subfigure[]{\includegraphics[width=5.5cm]{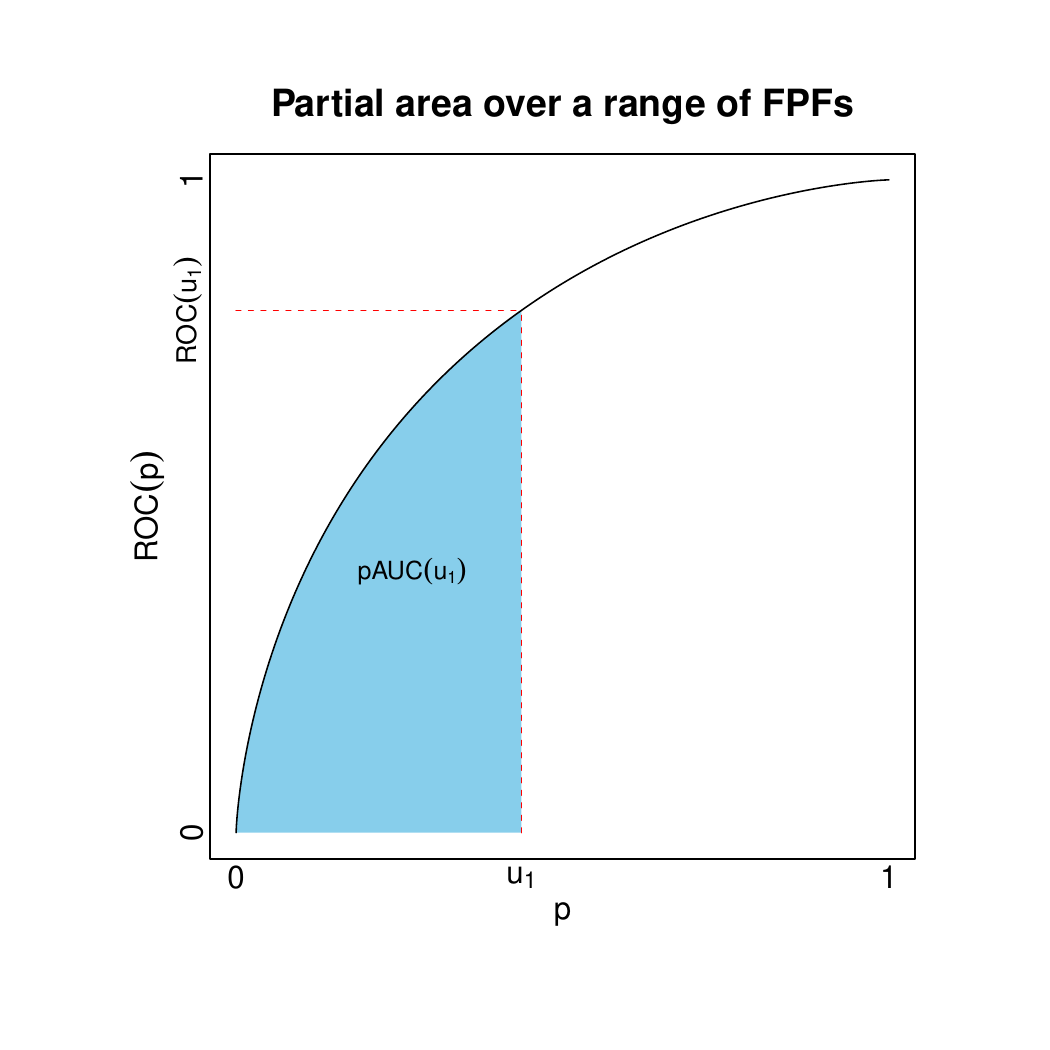}}\hspace{-0.6cm}
		\subfigure[]{\includegraphics[width=5.5cm]{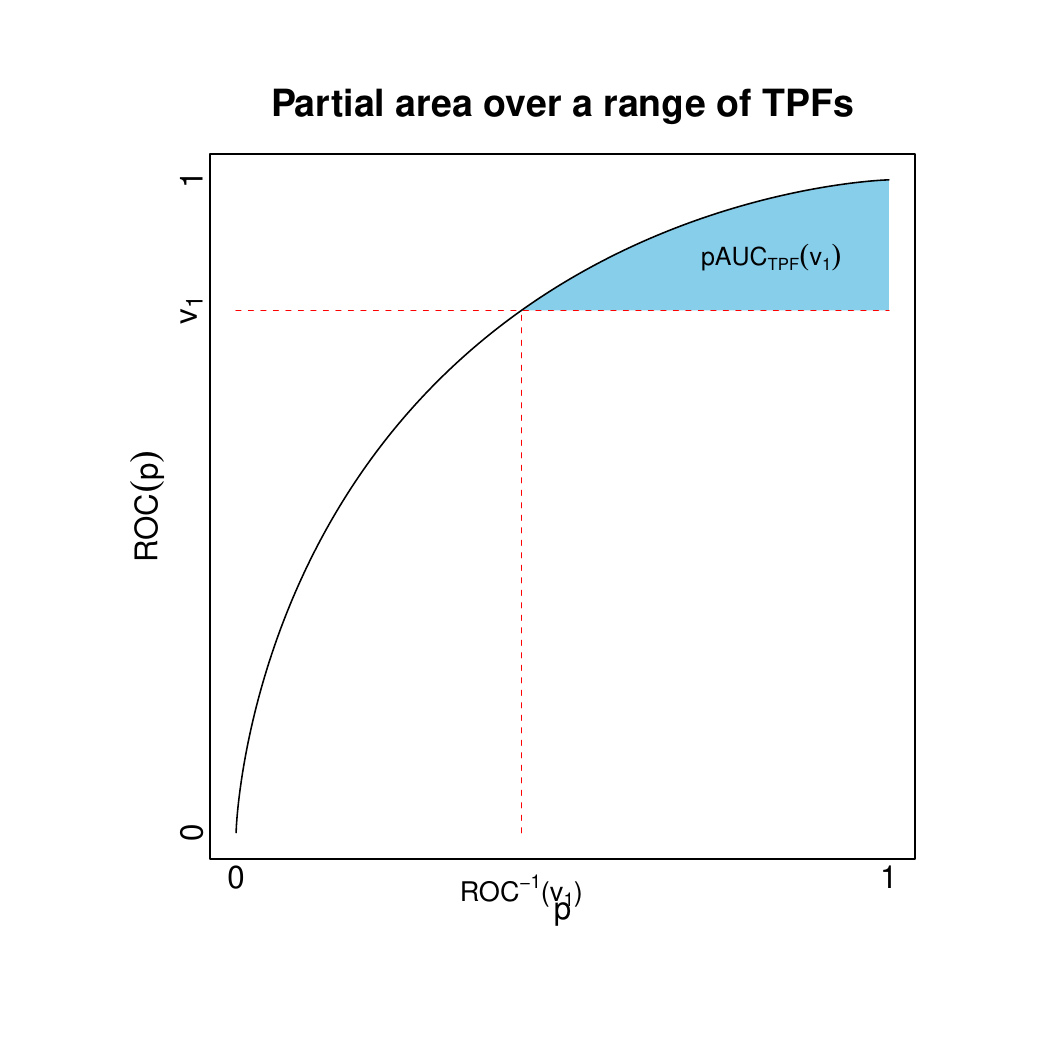}}\hspace{-0.6cm}
		\subfigure[]{\includegraphics[width=5.5cm]{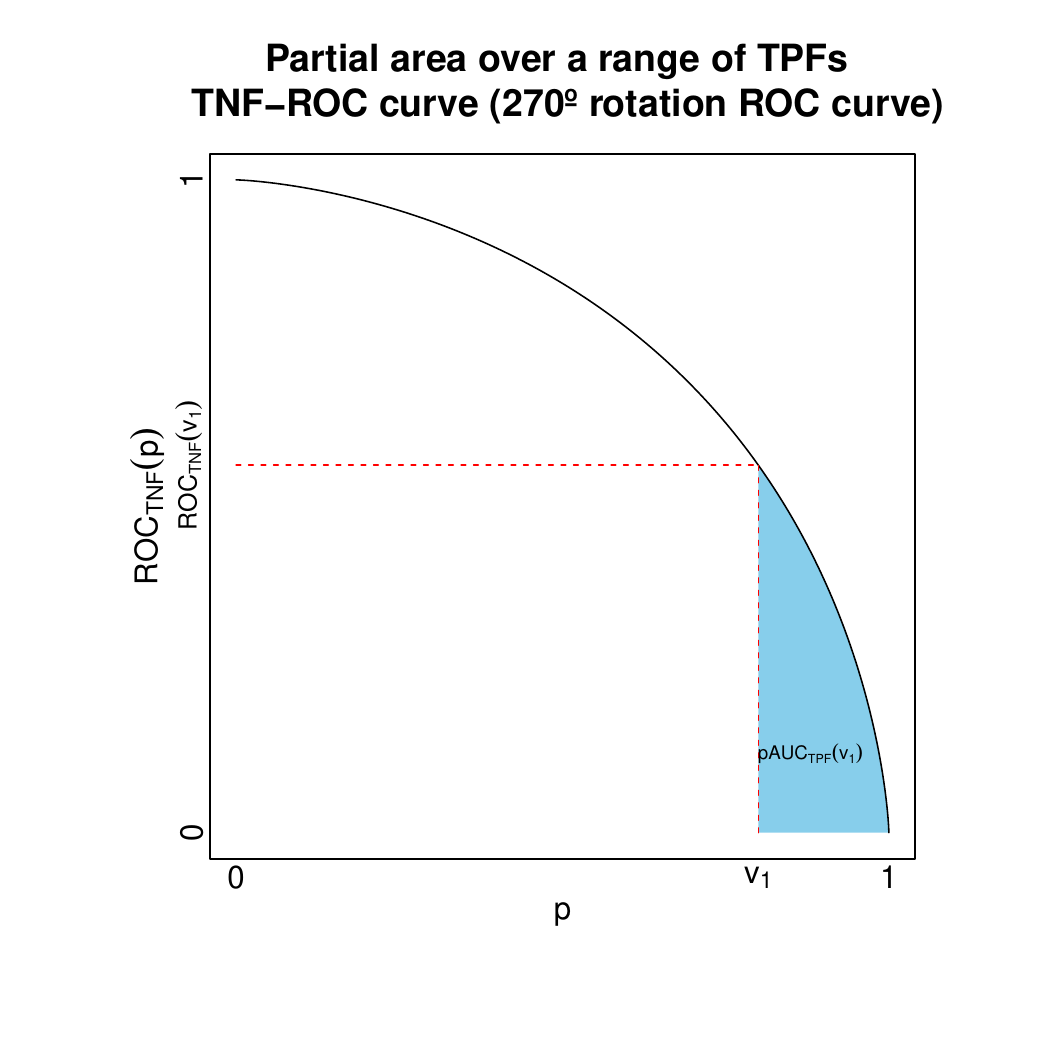}}
	\end{center}
	\caption{(a) Shaded area in blue represents the partial area under the ROC curve over the interval $(0,u_1)$ of FPFs. (b) Shaded area in blue represents the partial area under the ROC curve over the interval $(v_1,1)$ of TPFs. (c) The same as in (b) but now represented as an area over the true negative fraction ROC curve.}
	\label{pAUCs}
\end{figure}

In the case of a continuous-outcome diagnostic test, the classification is usually made by comparing the test result $Y$ against a threshold $c$. If the outcome is equal or above the threshold, $Y \geq c$, the subject will be considered as diseased. On the other hand, if the test result is below the threshold, $Y < c$, he/she will be classified as nondiseased. The ROC curve is then defined as the set of all possible false positive fractions, $\text{FPF}\left(c\right) = \Prob(Y \geq c \mid D = 0) = \Prob(Y_{\bar{D}} \geq c)$, and true positive fractions, $\text{TPF}\left(c\right) = \Prob(Y \geq c \mid D = 1) = \Prob(Y_D \geq c)$, that can be obtained by varying the threshold value $c$, i.e.,
\[
\left\{\left(\text{FPF}\left(c\right), \text{TPF}\left(c\right)\right): c \in \mathbb{R} \right\}.
\]
It is common to represent the ROC curve as $\left\{\left(p, \text{ROC}(p)\right): p \in [0,1] \right\}$, where
\begin{equation} 
p =\text{FPF}(c) = 1 -F_{\bar{D}}(c), \quad \text{ROC}(p) = 1 - F_D\left\{F_{\bar{D}}^{-1}(1-p)\right\},
\label{ROC}
\end{equation}
with $F_{D}\left(y\right) = \Prob(Y_{D} \leq y)$ and $F_{\bar{D}}\left(y\right) = \Prob(Y_{\bar{D}} \leq y)$ denoting the cumulative distribution function (CDF) of $Y$ in the nondiseased and diseased groups, respectively. Several indices can be used as global summary measures of the accuracy of a test. The most widely used is the area under the ROC curve (AUC), defined as
\begin{equation}
\text{AUC} = \int_{0}^{1}\text{ROC}\left(p\right)\text{d}p.
\label{AUC1}
\end{equation}
In addition to its geometric definition, the AUC has also a probabilistic interpretation \citep[see, e.g.,][p. 78]{Pepe03}
\begin{equation}
\text{AUC} = \Prob\left(Y_{D} \geq Y_{\bar{D}}\right),
\label{AUC2}
\end{equation}
that is, the AUC is the probability that a randomly selected diseased subject has a higher test outcome than that of a randomly selected nondiseased subject. The AUC takes values between $0.5$, in the case of an uninformative test that classifies individuals no better than chance, and $1.0$ for a perfect test. We note that an AUC below $0.5$ simply means that the classification rule should be reversed. As it is clear from its definition, the AUC integrates the ROC curve over the whole range of FPFs. Depending on the clinical circumstances, however, interest might lie only on a relevant interval of FPFs or TPFs, which leads to the notion of partial area under the ROC curve (pAUC). The pAUC over a range of FPFs $\left(0, u_1\right)$, where $u_1$ is typically low and represents the largest acceptable FPF, is defined as
\begin{equation}
\text{pAUC}\left(u_1\right) = \int_{0}^{u_1}\text{ROC}\left(p\right)\text{d}p.
\label{pAUC}
\end{equation}
%If, alternatively, one desires to calculate the partial area over the range $\left(u_1,u_2\right)$ of FPFs, we can simply do it as
%\begin{equation*}
%\text{pAUC}\left(u_1, u_2\right) = \text{pAUC}(u_2) - \text{pAUC}(u_1).
%\end{equation*}
On the other hand, the pAUC over a range of TPFs $(v_1,1)$, where $v_1$ is typically large and represents the lowest acceptable TPF, is defined as
\begin{equation}
\text{pAUC}_{\text{TPF}}\left(v_1\right) = \int_{v_1}^{1}\text{ROC}_{\text{TNF}}\left(p\right)\text{d}p,
\label{pAUC_TPF}
\end{equation}
where $\text{ROC}_{\text{TNF}}$ is a $270^\circ$ rotation of the ROC curve, which can be expressed as 
\begin{equation}
\text{ROC}_{\text{TNF}}(p) = F_{\bar{D}}\{F_{D}^{-1}(1-p)\}.
\label{ROC_TNF}
\end{equation}
The curve (\ref{ROC_TNF}) is referred to as the true negative fraction (TNF) ROC curve, since TNF ( $= 1 - \text{FPF}$) is plotted on the $y$-axis. We shall highlight that the argument $p$ in the ROC curve stands for a false positive fraction, whereas in the $\text{ROC}_{\text{TNF}}$ curve it stands for a true positive fraction. In Figure \ref{pAUCs} we graphically illustrate the two partial areas.

%Again, $\text{pAUC}_{\text{TPF}}\left(v_1, v_2\right) = \text{pAUC}_{\text{TPF}}(v_1) - \text{pAUC}_{\text{TPF}}(v_2)$. 
We shall highlight that the argument $p$ in the ROC curve stands for a false positive fraction, whereas in the $\text{ROC}_{\text{TNF}}$ curve it stands for a true positive fraction. 

Another summary index of diagnostic accuracy is the Youden index \citep{Shapiro99,Youden50}
\begin{eqnarray}
\text{YI} & = & \max_{c}\left\{\text{TPF}(c)-\text{FPF}(c)\right\} \label{YI1}\\
& = & \max_{c}\left\{F_{\bar{D}}\left(c\right) - F_{D}\left(c\right)\right\} \label{YI2}\\
& = & \max_{p}\left\{\text{ROC}(p)-p\right\} \label{YI3}.
\end{eqnarray}
The YI ranges from $0$ to $1$, taking the value of $0$ in the case of an uninformative test and $1$ for a perfect test. As for the AUC, a YI below $0$ means that the classification rule should be reversed. The value $c^{*}$ which maximises Equation \eqref{YI1} (or, equivalently, Equation \eqref{YI2}) is frequently used in practice to classify subjects as diseased or nondiseased. It should be noted that this index is equivalent to the Kolmogorov--Smirnov measure of distance between the distributions of $Y_{D}$ and $Y_{\bar{D}}$ \citep[p. 80]{Pepe03}.
\subsection{Covariate-specific ROC curve}\label{sec:croc}
The conditional or covariate-specific ROC curve, given a covariate value $\mathbf{x}$, is defined as
\begin{eqnarray}
\text{ROC}(p\mid\mathbf{x}) = 1-F_{D}\{F_{\bar{D}}^{-1}(1-p\mid\mathbf{x})\mid\mathbf{x}\},
\label{ROCConditional}
\end{eqnarray}
where $F_{\bar{D}}(y\mid\mathbf{x}) = \Pr(Y_{\bar{D}}\leq y\mid\mathbf{X}_{\bar{D}}=\mathbf{x})$ and $F_{D}(y\mid\mathbf{x}) = \Pr(Y_{D}\leq y\mid \mathbf{X}_{D}= \mathbf{x})$ are the conditional CDFs of the test in the nondiseased and diseased groups, respectively. In this case, a number of possibly different ROC curves (and therefore accuracies) may be obtained for different values of $\mathbf{x}$. Thus, the covariate-specific ROC curve is an important tool that helps to understand and determine the optimal and suboptimal populations where to apply the tests on (i.e., it allows determining the populations, defined by or homogeneous with respect to $\mathbf{x}$, where the diagnostic test has a `good' or `poor' discriminatory capacity). Similarly to the unconditional case, the covariate-specific TNF-ROC curve is given by %\textcolor{red}{the optimal and suboptimal populations - defined by or homogeneous with respect to $\mathbf{x}$ - where to apply the tests on}. Similarly to the unconditional case, the covariate-specific TNF-ROC curve is given by %
\begin{eqnarray}
\text{ROC}_{\text{TNF}}(p\mid\mathbf{x}) = F_{\bar{D}}\{F_{D}^{-1}(1-p\mid\mathbf{x})\mid\mathbf{x}\},
\label{ROCConditional_TNF}
\end{eqnarray}
and the covariate-specific AUC, pAUC, and Youden index are
\begin{eqnarray}
\text{AUC}(\mathbf{x}) & = & \int_{0}^{1}\text{ROC}(p\mid\mathbf{x})\text{d}p, \label{AUCConditional}\\
\text{pAUC}(u_1 \mid \mathbf{x}) & = & \int_{0}^{u_1}\text{ROC}(p\mid\mathbf{x})\text{d}p, \label{pAUCConditional}\\
\text{pAUC}_{\text{TPF}}(v_1 \mid \mathbf{x}) & = & \int_{v_1}^{1}\text{ROC}_{\text{TNF}}(p\mid\mathbf{x})\text{d}p, \label{pAUCConditional2}\\
\text{YI}(\mathbf{x}) & = & \max_{c} \lvert \text{TPF}(c\mid\mathbf{x}) - \text{FPF}(c \mid \mathbf{x}) \rvert \label{YIConditional1}\\
& = &  \max_{c}\lvert F_{\bar{D}}(c\mid\mathbf{x}) - F_{D}(c\mid\mathbf{x}) \rvert \label{YIConditional2}\\
& = &  \max_{p}\lvert \text{ROC}(p\mid\mathbf{x})- p \rvert \label{YIConditional3}.
\end{eqnarray}
The value $c^{*}_{\mathbf{x}}$ that achieves the maximum in (\ref{YIConditional1}) (or (\ref{YIConditional2})) is called the optimal covariate-specific YI threshold and can be used to classify a subject, with covariate value $\mathbf{x}$, as diseased or nondiseased.
\subsection{Covariate-adjusted ROC curve}\label{sec:aroc}
The covariate-specific ROC curve and associated AUC, pAUCs, and YI described in Section \ref{sec:croc} depict the accuracy of the test for specific covariate values. However, it would be undoubtedly useful to have a global summary measure that also takes covariate information into account. Such summary measure was developed by \cite{Janes09a}, who proposed the covariate-adjusted ROC (AROC) curve, defined as
\begin{equation}
\text{AROC}(p)=\int \text{ROC}(p\mid\mathbf{x})\text{d}H_{D}(\mathbf{x}),
\label{aroc1}
\end{equation}
where $H_{D}(\mathbf{x}) = \Prob(\mathbf{X}_{D}\leq \mathbf{x})$ is the CDF of $\mathbf{X}_{D}$. That is, the AROC curve is a weighted average of covariate-specific ROC curves, weighted according to the distribution of the covariates in the diseased group. Equivalently, as shown by \cite{Janes09a}, the AROC curve can also be expressed as
\begin{align}\label{aroc2}
\text{AROC}(p) & = \Prob\{Y_{D}>F_{\bar{D}}^{-1}(1-p\mid \mathbf{X}_{D})\} \nonumber \\
& = \Prob\{1-F_{\bar{D}}(Y_D\mid\mathbf{X}_{D})\leq p\}.
\end{align}
As will be seen in Section \ref{sec:methods}, Expression \eqref{aroc2} is very convenient when it comes to estimating the AROC curve. Also, it emphasises that the AROC curve at a FPF of $p$ is the overall TPF when the thresholds used for defining a positive test result are covariate-specific and chosen to ensure that the FPF is $p$ in each subpopulation defined by the covariate values.

In contrast to the pooled ROC curve (see Expressions \eqref{ROC} and \eqref{ROC_TNF}) and the covariate-specific ROC curve (see Expressions \eqref{ROCConditional} and \eqref{ROCConditional_TNF}), the AROC curve (and its $270^\circ$ rotation) cannot be expressed in terms of the (conditional) CDFs of the test in each group. This does not, however, preclude the possibility of defining AROC-based summary accuracy measures, yet more care is needed. Thus, for the AROC curve, the area under the AROC, as well as the partial areas and YI are expressed as follows
\begin{eqnarray}
\text{AAUC} & = & \int_{0}^{1}\text{AROC}(p)\text{d}p, \label{AAUCConditional}\\
\text{pAAUC}(u_1) & = & \int_{0}^{u_1}\text{AROC}(p)\text{d}p, \label{pAAUCConditional}\\
\text{pAAUC}_{\text{TPF}}(v_1) & = & \int_{\text{AROC}^{-1}(v_1)}^{1}\text{AROC}(p)\text{d}p - \{1-\text{AROC}^{-1}(v_1)\}v_1, \label{pAUCConditional3}\\
\text{YI}_{\text{AROC}} & = &  \max_{p}\left\{\text{AROC}(p)-p\right\} \label{YIAROC}.
\end{eqnarray}
Note, in particular, that the expressions for both the partial area under the AROC curve over a range of TPFs (see also Figure \ref{pAUCs}) and for the YI are defined in terms of the AROC curve. For the YI, once the value that achieves the maximum in \eqref{YIAROC} is calculated, say $p^{*}$, covariate-specific threshold values can be obtained as follows
\[
c^{*}_{\mathbf{x}} = F_{\bar{D}}^{-1}(1-p^{*}\mid \mathbf{X}_{D} = \mathbf{x}).
\]
Note that, by construction, these threshold values will ensure that the FPF is $p^{*}$ in each subpopulation defined by the covariate values; however, the TPF may vary with the covariate values, i.e.,
\[
\text{TPF}\left(c^{*}_{\mathbf{x}}\right) = 1 - F_{D}\left(c^{*}_{\mathbf{x}} \mid \mathbf{X}_{D} = \mathbf{x} \right).
\]
To finish this part, we mention that when the accuracy of a test is not affected by covariates, this does not necessarily means that the covariate-specific ROC curve (which in this case is the same for all covariate values) coincides with the pooled ROC curve. It does coincide, however, with the AROC curve \citep[see][for more details]{Janes09a, Pardo14, Inacio18}. As such, in all cases where covariates affect the results of the test, even though they might not affect its discriminatory capacity, inferences based on the pooled ROC curve might be misleading. In such cases, the AROC curve should be used instead. This also applies to the selection of (optimal) threshold values, which might be covariate-specific (i.e., possibly different for different covariate values). 

%The covariate-adjusted AUC (AAUC) is thus expressed as
%\begin{equation*}
%\text{AAUC}=\int_{0}^{1}\text{AROC}(t)\text{d}t=\int_{0}^{1}\int \text{ROC}(t\mid\mathbf{x})\text{d}H_{D}(\mathbf{x}) \text{d}t=\int \text{AUC}(\mathbf{x})\text{d}H_{D}(\mathbf{x}).
%\end{equation*}
%Note that the AAUC is also a weighted average of covariate-specific AUCs. \cite{Janes09a} shown that the AROC can be equivalently represented as

%% -- Manuscript ---------------------------------------------------------------

%% - In principle "as usual" again.
%% - When using equations (e.g., {equation}, {eqnarray}, {align}, etc.
%%   avoid empty lines before and after the equation (which would signal a new
%%   paragraph.
%% - When describing longer chunks of code that are _not_ meant for execution
%%   (e.g., a function synopsis or list of arguments), the environment {Code}
%%   is recommended. Alternatively, a plain {verbatim} can also be used.
%%   (For executed code see the next section.)
\section{Methods} \label{sec:methods}
In this section we describe the different methods for ROC curve inference (with and without covariate information) implemented in the \pkg{ROCnReg} package.
\subsection{Pooled ROC curve}
Let $\{y_{\bar{D}i}\}_{i=1}^{n_{\bar{D}}}$ and $\{y_{Dj}\}_{j=1}^{n_D}$ be two independent random samples of test outcomes from the nondiseased and diseased groups of size $n_{\bar{D}}$ and $n_D$, respectively.

\subsubsection*{Empirical estimator}
The function \code{pooledROC.emp} estimates the pooled ROC curve using the empirical estimator proposed by \cite{Hsieh1996}, which consists in estimating the CDFs of the test in each group by its empirical counterpart, that is,
\begin{equation*}
\widehat{F}_{\bar{D}}(y) = \frac{1}{n_{\bar{D}}}\sum_{i=1}^{n_{\bar{D}}}I(y_{\bar{D}i}\leq y),\qquad \widehat{F}_{D}(y) = \frac{1}{n_D}\sum_{j=1}^{n_D}I(y_{Dj}\leq y).
\end{equation*}
These empirical estimates are then plugged into Equations \eqref{ROC} and \eqref{ROC_TNF} to obtain, respectively, an estimate of the ROC and $\text{ROC}_{\text{TNF}}$ curves.

In what concerns estimation of the AUC (Expression \eqref{AUC2}), pAUC, and  $\text{pAUC}_{\text{TNF}}$ (Expressions \eqref{pAUC} and \eqref{pAUC_TPF}, respectively) these are computed empirically by means of the Mann--Whitney U statistic. With respect to the Youden index (and associated threshold value), it is obtained by maximising, over a grid of possible threshold values, the expression in \eqref{YI2}, with $F_D$ and $F_{\bar{D}}$ being replaced by their empirical estimators. 
% In this case the AUC is the Mann-Whitney U statistic,
%\begin{equation*}
%\widehat{\text{AUC}}_{\text{emp}} = \frac{1}{n_{\bar{D}}n_{D}}\sum_{i=1}^{n_{\bar{D}}}\sum_{j=1}^{n_D}\left\{I\left(y_{Dj} >y_{\bar{D}i}\right)+\frac{1}{2}I\left(y_{Dj} = y_{\bar{D}i}\right)\right\}.
%\end{equation*}
%\textcolor{red}{will add partial areas for all methods}
\subsubsection*{Kernel estimator}
The function \code{pooledROC.kernel} estimates the pooled ROC curve using the kernel-based estimator proposed by \cite{Zou97} and \cite{Zou98}, which is based on estimating the CDFs of the test as follows
\begin{equation*}
\widehat{F}_{\bar{D}}(y)=\frac{1}{n_{\bar{D}}}\sum_{i=1}^{n_{\bar{D}}}\Phi\left(\frac{y-y_{\bar{D}i}}{h_{\bar{D}}}\right),\qquad \widehat{F}_{D}(y)=\frac{1}{n_D}\sum_{j=1}^{n_D}\Phi\left(\frac{y-y_{Dj}}{h_D}\right),
\end{equation*}
where $\Phi(y)$ stands for the standard normal distribution evaluated at $y$. For the bandwidths, $h_{\bar{D}}$ and $h_D$, which control the amount of smoothing, two options are popular. Silverman's rule of thumb \citep[][p.~48]{Silverman1986}, which sets the bandwidth as
\begin{equation*}
h_d=0.9\min\{\text{SD}(\textbf{y}_d), \text{IQR}(\textbf{y}_d)/1.34\}n_d^{-0.2},\quad d\in\{\bar{D},D\},
\end{equation*}
where $\text{SD}(\textbf{y}_d)$ and $\text{IQR}(\textbf{y}_d)$ are the standard deviation and interquantile range, respectively, of $\textbf{y}_d=(y_{d1},\ldots,y_{dn_{d}})$. Another alternative criterion is to select the bandwidth by using least squares cross-validation \citep[][Chapter 3]{Wand1994}.

Here, both the AUC, pAUC, and  $\text{pAUC}_{\text{TNF}}$ (Expressions \eqref{AUC1}, \eqref{pAUC}, and \eqref{pAUC_TPF}) are computed numerically using Simpson's rule. Regarding the Youden index (and associated threshold value), it is obtained by maximising, over a grid of possible threshold values, Expression \eqref{YI2}, with $F_D$ and $F_{\bar{D}}$ being replaced by their kernel estimators. 

Uncertainty estimation for both the empirical and kernel estimators is conducted through bootstrap resampling.

\subsubsection*{Bayesian bootstrap estimator}
The function \code{pooledROC.bb} implements the Bayesian bootstrap (BB) approach proposed by \cite{Gu2008}. Their estimator relies on the notion of placement value \citep[][Chapter 5]{Pepe03}, which is simply a standardisation of the test outcomes with respect to a reference group. Specifically, $U_D= 1-F_{\bar{D}}(Y_D)$ is to be interpreted as a standardisation of a diseased test outcome with respect to the distribution of test results in the nondiseased population. The ROC curve can be regarded as the CDF of $U_D$
\begin{equation}\label{rocbb}
\Prob(U_D\leq p) = \Prob\{1-F_{\bar{D}}(Y_D)\leq p\}=1-F_{D}\{F_{\bar{D}}^{-1}(1-p)\}=\text{ROC}(p),\quad 0\leq p \leq 1.
\end{equation}
This representation in \eqref{rocbb} of the ROC curve provided the rationale for the two-step algorithm of \cite{Gu2008}, which can be described as follows. Let $S$ be the number of iterations.
\begin{description}
\item [\textbf{Step 1:}] \textbf{Computation of the placement value  based on the BB.}\\
For $s=1,\ldots,S$, let
\begin{equation*}
U_{Dj}^{(s)}=\sum_{i=1}^{n_{\bar{D}}}q_{1i}^{(s)}I\left(y_{\bar{D}i}\geq y_{Dj}\right), \quad j=1,\ldots,n_{D},
\end{equation*}
where $\left(q_{11}^{(s)},\ldots,q_{1n_{\bar{D}}}^{(s)}\right)\sim\text{Dirichlet}(n_{\bar{D}};1,\ldots,1)$.
\item [\textbf{Step 2:}] \textbf{Generate a realisation of the ROC curve.}
Based on \eqref{rocbb}, generate a realisation of $\text{ROC}^{(s)}(p)$, the cumulative distribution function of $(U_{D1}^{(s)},\ldots,U_{Dn_D}^{(s)})$, where
\begin{equation*}
\text{ROC}^{(s)}(p)=\sum_{j=1}^{n_D}q_{2j}^{(s)}I\left(U_{Dj}^{(s)}\leq p\right),\quad \left(q_{21}^{(s)},\ldots,q_{2n_{D}}^{(s)}\right)\sim\text{Dirichlet}(n_D;1,\ldots,1).
\end{equation*}
%and where $p$ is a span grid over $[0,1]$.
\end{description}
%The BB estimate of the ROC curve, denoted as $\widehat{\text{ROC}}^{\text{BB}}(p)$, is obtained by averaging over the ensemble of ROC curves $\{\text{ROC}^{(1)}(p),\ldots, \text{ROC}^{(S)}(p)\}$, that is,
%A point estimate for $\text{ROC}(p)$ can be obtained by computing the mean (or the median) of the ensemble $\{\text{ROC}^{(1)}(p),\ldots,\text{ROC}^{(S)}(p)\}$, e.g.,
%The $S$ posterior samples give rise to an ensemble of ROC curves $\{\text{ROC}^{(1)}(p),\ldots,\text{ROC}^{(S)}(p)\}$ from which the posterior mean (or median) can be computed, e.g.,
The BB estimate of the ROC curve is obtained by averaging over the ensemble of ROC curves $\{\text{ROC}^{(1)}(p),\ldots,\text{ROC}^{(S)}(p)\}$, that is,
\begin{equation*}
\widehat{\text{ROC}}^{\text{BB}}(p)=\frac{1}{S}\sum_{s=1}^{S}\text{ROC}^{(s)}(p),
\end{equation*}
and a $(1-\alpha)\times 100\%$ pointwise credible band can be obtained from the $\alpha/2\times 100\%$ and $(1- \alpha/2)\times 100\%$ percentiles of the same ensemble ($\alpha \in (0,1)$). Note that these pointwise credible bands for the ROC curve are to be interpreted as credible intervals for the TPFs.

The Bayesian bootstrap estimator leads to closed-form expressions for the AUC and pAUC, which are, respectively, given by
%with pointwise credible bands derived from the percentiles of the ensemble. In addition, the Bayesian bootstrap estimator leads to closed-form expressions for the AUC and pAUC, which are, respectively, given by
\begin{align*}
\text{AUC}^{(s)} &= \int_0^1\text{ROC}^{(s)}(p)\text{d}p = 1-\sum_{j=1}^{n_D}q_{2j}^{(s)}U_{Dj}^{(s)},\\
\text{pAUC}^{(s)}(u_1)&=\int_{0}^{u_1}\text{ROC}^{(s)}(p)\text{d}p = u_1-\sum_{j=1}^{n_D}q_{2j}^{(s)}\min\left\{u_1,U_{Dj}^{(s)}\right\}.
\end{align*}
It is easy to show that
\[
\text{pAUC}^{(s)}_{\text{TPF}}(v_1) = \int_{v_1}^{1}\text{ROC}^{(s)}_{\text{TNF}}\left(p\right)\text{d}p = \sum_{i=1}^{n_{\bar{D}}}q_{1i}^{(s)}\max\left\{v_1,U_{\bar{D}i}^{(s)}\right\} - v_1,
\]
where
\begin{equation*}
U_{\bar{D}i}^{(s)}=\sum_{j=1}^{n_D}q_{2j}^{(s)}I\left(y_{Dj}\geq y_{\bar{D}i}\right), \quad i=1,\ldots,n_{\bar{D}},
\end{equation*}
and it is also easy to demonstrate that the $\text{ROC}_{\text{TNF}}$ curve is the survival function of the placement value $U_{\bar{D}}=1-F_{D}(Y_{\bar{D}})$. With respect to the Youden index, it is obtained by maximising, over a grid of possible threshold values, the following expression
\[
\text{YI}^{(s)} = \max_{c}\left\{F^{(s)}_{\bar{D}}\left(c\right) - F^{(s)}_{D}\left(c\right)\right\},
\]
where
\[
F^{(s)}_{\bar{D}}\left(c\right) = \sum_{i=1}^{n_{\bar{D}}}q_{1i}^{(s)}I\left(y_{\bar{D}i} \leq c\right) \;\;\; \mbox{and} \;\;\; F^{(s)}_{D}\left(c\right) = \sum_{j=1}^{n_{D}}q_{2j}^{(s)}I\left(y_{Dj} \leq c\right).
\]
As for the ROC curve, point estimates for the AUC, pAUC, $\text{pAUC}_{\text{TPF}}$, YI, and $c^{*}$ can be obtained by averaging over the respective ensembles of $S$ realisations, with credible bands derived from the percentiles of such ensembles.
\subsubsection*{Dirichlet process mixture of normal distributions estimator}
The Bayesian nonparametric approach, based on a Dirichlet process mixture (DPM) of normal distributions, for estimating the pooled ROC curve \citep{Erkanli2006} is implemented in the \code{pooledROC.dpm} function. In this case, as implicit by the name, the CDFs of the test outcomes in each group are estimated via  a Dirichlet process mixture of normal distributions, that is, it is assumed that the CDF, say in the diseased group (the one in the nondiseased group, $\bar{D}$, follows analogously), is of the form
\begin{equation}\label{cdfdpm1}
F_{D}(y)=\int \Phi(y\mid \mu,\sigma^2)\text{d}G_{D}(\mu,\sigma^2), \qquad G_D\sim\text{DP}(\alpha_D,G_D^{*}(\mu,\sigma^2)),
\end{equation}
where $\Phi(y\mid \mu,\sigma^2)$ denotes the CDF of the normal distribution with mean $\mu$ and variance $\sigma^2$ evaluated at $y$. Here $G_D\sim\text{DP}(\alpha_D,G_D^{*})$ is used to denote that the mixing distribution $G_D$ follows a Dirichlet process (DP) \citep{Ferguson73} with centring distribution $G_D^{*}$, for which $E(G_D)=G_D^{*}$, and precision parameter $\alpha_D$.  Usually, due to conjugacy reasons, $G_D^{*}(\mu,\sigma^2)\equiv \text{N}(\mu\mid m_{D0},S_{D0})\Gamma(\sigma^{-2}\mid a_D,b_D)$ and this is the centring distribution used by the \code{pooledROC.dpm} function. Note that here $S_{D0}$ denotes the variance of the normal distribution and $a_D$ and $b_D$ are, respectively, the shape and rate parameters of the gamma distribution. All hyperparameter values are fixed.

For ease of posterior simulation and because it provides a highly accurate approximation, we make use of the truncated stick-breaking representation of the DP \citep{Ishwaran2001}, according to which $G_D$ can be written as
\begin{equation*}
G_{D}(\cdot)=\sum_{l=1}^{L_D}\omega_{Dl}\delta_{(\mu_{Dl},\sigma^2_{Dl})}(\cdot),
\end{equation*}
where $(\mu_{Dl},\sigma^2_{Dl})\overset{\text{iid}}\sim G_D^{*}(\mu,\sigma^2)$, for $l=1,\ldots,L_D$, and the weights follow the so-called (truncated) stick-breaking construction: $\omega_{D1}=v_{D1}$, $\omega_{Dl}=v_{Dl}\prod_{r<l}(1-v_{Dr})$, $l=2,\ldots,L_D$, and $v_{D1},\ldots,v_{D,L_{D}-1}\overset{\text{iid}}\sim\text{Beta}(1,\alpha_D)$. Further, one must set $v_{DL_{D}}=1$ in order to ensure that the weights add up to one. Further, one must set $v_{DL_{D}}=1$ in order to ensure that the weights add up to one. With regard to the parameter $\alpha_D$, in \pkg{ROCnReg} a prior distribution is placed on it. In particular, and due to conjugacy reasons, a gamma distribution is considered, i.e., $\alpha_D\sim\Gamma(a_{\alpha_{D}},b_{\alpha_{D}})$. The CDF in \eqref{cdfdpm1} can therefore be written as %With regard to the parameter $\alpha_D$, it can be either fixed or a prior distribution placed on it. In the latter case, due to conjugacy reasons, a gamma distribution is the most popular choice, $\alpha_D\sim\Gamma(a_{\alpha_{D}},b_{\alpha_{D}})$. The CDF in \eqref{cdfdpm1} can therefore be written as
\begin{equation*}
F_{D}(y)=\sum_{l=1}^{L_D}\omega_{Dl}\Phi(y\mid\mu_{Dl},\sigma_{Dl}^{2}),
\end{equation*}
where we shall note that $L_D$ is not the exact number of components expected to be observed, but rather an upper bound on it, as some of the components may be unoccupied. Some comments are in order regarding the specification of the hyperparameters' values. In what concerns the centring distribution, $m_{D0}$ represents the prior belief about the components' means and $S_{D0}$ represents the confidence in such prior belief. Similarly, the values of $a_D$ and $b_D$ can be chosen to represent prior belief about the components' variance. Of course, when setting these parameters, it is crucial to consider the measurement scale of the data. By default, test outcomes are standardised (so that the resulting mean is zero and the variance is one) in the \code{pooledROC.dpm} function and the default values are as follows
\begin{equation*}
m_{D0} = 0,\quad S_{D0} = 10, \quad a_D = 2, \quad b_D = 0.5.
\end{equation*}
Because test outcomes are standardised we expect the means of the components to be near zero and hence  why $m_{D0}=0$. The parameter $S_{D0}$ then controls where the drawn $\mu_{Dl}$ can lie and the value of $10$ implies that approximately $95\%$ of the values roughly lie within $-6$ and $6$. Further, note that $a_D = 2$ leads to a prior with infinite variance that is centred around a finite mean ($b_D = 0.5$) and therefore favours variances less than one. Considering that the standardised data have a variance of one, it is reasonable to expect the within component variance to be smaller than the overall variance. The option of not standardising the test outcomes is also available in \code{pooledROC.dpm} and in such a case the  defaults for the centring distribution hyperparameters' values are as following
\begin{equation*}
 m_{D0}=\bar{y}_D,\quad S_{D0}=100 s^2_D/n_D, \quad a_D=2, \quad b_D=s^2_D/2, 
\end{equation*}
with $\bar{y}_D=\frac{1}{n_D}\sum_{j=1}^{n_D}y_{Dj}$ and $s_D^2=\frac{1}{n_D-1}\sum_{j=1}^{n_D}(y_{Dj}-\bar{y}_D)^2$. Regarding the precision parameter of the DP, $\alpha_D$, it has a direct relationship with the number of occupied mixture components. One possible strategy for specifying $\alpha_D$ is to fix it to a small value to favour a small number of occupied components relative to the sample size. In the \code{pooledROC.dpm} function we set $\alpha_D=1$, a commonly used default value \citep[][p.~553]{Gelman2013}. Lastly, by default $L_D=10$. Before proceeding we shall emphasise that these two configurations of hyperparameters values (for standardised and not standardised test outcomes) have proved to work well for a different range of test outcomes distributions but it is certainly not our goal to encourage users to use it blindly and thought should be dedicated to this important task. Nevertheless, output from the function \code{pooledROC.dpm} may be post processed and (informal) model fit diagnostics obtained; see more in Section \nameref{sec:illustration} and in the Appendices.

Because the full conditional distributions for all model parameters are available in closed-form, posterior simulation can be easily conducted through Gibbs sampler (see the details, for instance, in \citealt{Ishwaran2002}). At iteration $s$ of the Gibbs sampler procedure, the ROC curve is computed as
\begin{equation*} 
\text{ROC}^{(s)}(p)=1-F_D^{(s)}\left\{F_{\bar{D}}^{-1(s)}(1-p)\right\},  \quad s=1,\ldots, S,
\end{equation*}
with
\begin{equation} \label{cdfdpm_2} 
F_D^{(s)}(y)=\sum_{l=1}^{L_D}\omega_{Dl}^{(s)}\Phi\left(y\mid\mu_{Dl}^{(s)},\sigma_{Dl}^{2(s)}\right),\quad
F_{\bar{D}}^{(s)}(y)=\sum_{k=1}^{L_{\bar{D}}}\omega_{\bar{D}k}^{(s)}\Phi\left(y\mid\mu_{\bar{D}k}^{(s)},\sigma_{\bar{D}k}^{2(s)}\right),
\end{equation}
and where the inversion is performed numerically. There is a closed-form expression for the AUC \citep{Erkanli2006} given by
\begin{equation*}
\text{AUC}^{(s)}=\sum_{k=1}^{L_{\bar{D}}}\sum_{l=1}^{L_D}\omega_{\bar{D}k}^{(s)}\omega_{Dl}^{(s)}\Phi\left(\frac{b_{kl}^{(s)}}{\sqrt{1+a_{kl}^{2(s)}}}\right),\quad b_{kl}^{(s)}=\frac{\mu_{Dl}^{(s)}-\mu_{\bar{D}k}^{(s)}}{\sigma_{Dl}^{(s)}},\quad a_{kl}^{(s)}=\frac{\sigma_{\bar{D}k}^{(s)}}{\sigma_{Dl}^{(s)}}.
\end{equation*}
Also, when $L_D = L_{\bar{D}}$ = 1, there are closed-form expressions for the pAUC and $\text{pAUC}_{\text{TPF}}$ which are used in the package \citep[see][]{Hillis12}. For the pAUC/$\text{pAUC}_{\text{TPF}}$, when $L_D > 1$ or $L_{\bar{D}} > 1$, the integrals are approximated numerically using Simpson's rule. The Youden index/optimal threshold is computed as in the Bayesian bootstrap method, with the obvious difference that here the CDFs are expressed as in \eqref{cdfdpm_2}. At the end of the sampling procedure, we have an ensemble of $S$ ROC curves and AUCs/pAUCs/$\text{pAUC}_{\text{TPF}}\text{s}$/YIs/optimal thresholds which, as before, allows obtaining point and interval estimates.

\subsection{Covariate-specific ROC curve}\label{sec:croc_methods}
We now let $\{(\mathbf{x}_{\bar{D}i},y_{\bar{D}i})\}_{i=1}^{n_{\bar{D}}}$ and $\{(\mathbf{x}_{Dj},y_{Dj})\}_{j=1}^{n_D}$ be two independent random samples of test outcomes and covariates from the nondiseased and diseased groups of size $n_{\bar{D}}$ and $n_D$, respectively. Further, for all $i = 1,\ldots,n_{\bar{D}}$ and $j = 1,\ldots,n_D$, let $\mathbf{x}_{\bar{D}i}=(x_{\bar{D}i,1},\ldots, x_{\bar{D}i,q})^{\top}$ and $\mathbf{x}_{Dj}=(x_{Dj,1},\ldots, x_{Dj,q})^{\top}$ be $q$-dimensional vectors of covariates, which can be either continuous or categorical.

\subsubsection*{Induced semiparametric linear model}
The function \code{cROC.sp} implements the induced ROC approaches proposed by \cite{Faraggi03} and \cite{Pepe98}. Both authors assume a location-scale regression model of the following form for the test outcomes in each group
\begin{equation}
Y_{\bar{D}}=\tilde{\mathbf{X}}_{\bar{D}}^{\top}\boldsymbol{\beta}_{\bar{D}}+\sigma_{\bar{D}}\varepsilon_{\bar{D}},\qquad Y_D=\tilde{\mathbf{X}}_D^{\top}\boldsymbol{\beta}_D+\sigma_D\varepsilon_D,
\label{loc-scale-sp}
\end{equation}
where $\tilde{\mathbf{X}}_{\bar{D}}^{\top}=(1,\mathbf{X}_{\bar{D}}^{\top})$ and $\boldsymbol{\beta}_{\bar{D}}=(\beta_{\bar{D}0},\ldots,\boldsymbol{\beta}_{\bar{D}q})^{\top}$ is a $(q+1)$-dimensional vector of (unknown) regression coefficients; $\tilde{\mathbf{X}}_{D}$ and $\boldsymbol{\beta}_{D}$ are analogously defined. The error terms $\varepsilon_{\bar{D}}$ and $\varepsilon_{D}$ have mean zero, variance one, are independent of each other and of the covariate, and have distribution functions given by $F_{\varepsilon_{\bar{D}}}$ and $F_{\varepsilon_{D}}$, respectively. Under these assumptions, we have
\begin{equation}
F_{\bar{D}}(y\mid\mathbf{x}) = F_{\varepsilon_{\bar{D}}}\left(\frac{y - \tilde{\mathbf{x}}^{\top}\boldsymbol{\beta}_{\bar{D}}}{\sigma_{\bar{D}}}\right)\;\;\;\mbox{and}\;\;\;F_{D}(y\mid\mathbf{x}) = F_{\varepsilon_{D}}\left(\frac{y - \tilde{\mathbf{x}}^{\top}\boldsymbol{\beta}_{D}}{\sigma_D}\right),
\label{cROC_YI_sp}
\end{equation}
with $\tilde{\mathbf{x}}^{\top}=(1,\mathbf{x}^{\top})$.

The approaches of \cite{Faraggi03} and \cite{Pepe98} differ in the assumptions made about the error terms. More concretely, \cite{Faraggi03}'s method assumes that the error term in both groups follows a standard normal distribution, i.e., $F_{\varepsilon_{\bar{D}}}(y)=F_{\varepsilon_{D}}(y)=\Phi(y)$, and can be summarised by the following three steps:
\begin{enumerate}
\item Estimate the regression coefficients $\boldsymbol{\beta}_{\bar{D}}$ and $\boldsymbol{\beta}_D$ by ordinary least squares, on the basis of the samples $\{(\mathbf{x}_{\bar{D}i},y_{\bar{D}i})\}_{i=1}^{n_{\bar{D}}}$ and $\{(\mathbf{x}_{Dj},y_{Dj})\}_{j=1}^{n_D}$, respectively. 
\item Estimate $\widehat{\sigma}_{D}^{2}$ as
\begin{equation*}
\widehat{\sigma}_{D}^{2}=\frac{\sum_{j=1}^{n_D}\left(y_{Dj}-\tilde{\mathbf{x}}_{Dj}^{\top}\widehat{\boldsymbol{\beta}}_D\right)^{2}}{n_D-q-1},
\end{equation*}
with $\widehat{\sigma}_{\bar{D}}^{2}$ similarly estimated.
\item For a given covariate vector $\mathbf{x}$, compute the covariate-specific ROC curve as follows
\begin{equation}
\widehat{\text{ROC}}(p\mid\mathbf{x})=1-\Phi\left\{a(\mathbf{x})+b\Phi^{-1}(1-p)\right\},
\label{cROC-a-b-sp}
\end{equation}
where
\begin{equation}
a(\mathbf{x}) =\tilde{\mathbf{x}}^{\top}\frac{(\widehat{\boldsymbol{\beta}}_{\bar{D}}-\widehat{\boldsymbol{\beta}}_D)}{\widehat{\sigma}_D},\quad \text{and}\quad b =\frac{\widehat{\sigma}_{\bar{D}}}{\widehat{\sigma}_D}.
\label{a-b-sp}
\end{equation}
\end{enumerate}
Regarding the covariate-specific AUC, pAUC, and $\text{pAUC}_{\text{TPF}}$ (Expressions \eqref{AUCConditional}, \eqref{pAUCConditional}, and \eqref{pAUCConditional2}) they admit closed-form expressions \cite[see][]{Hillis12}.

As an alternative, \cite{Pepe98} suggests to estimate the CDF of the errors in each group by the corresponding empirical CDF of the estimated standardised residuals. Therefore, the first two steps of the estimation procedure remain the same, but now we have the following extra step
\begin{equation*}
\widehat{F}_{\varepsilon_{D}}(y)=\frac{1}{n_D}\sum_{j=1}^{n_D}I(\widehat{\varepsilon}_{Dj}\leq y), \qquad \widehat{\varepsilon}_{Dj}=\frac{y_{Dj}-\tilde{\mathbf{x}}_{Dj}^{\top}\widehat{\boldsymbol{\beta}}_D}{\widehat{\sigma}_D}. 
\end{equation*}
The empirical CDF of the standardised residuals in the nondiseased group is estimated in a similar fashion. The covariate-specific ROC curve is finally computed in an analogous way as for the method of \cite{Faraggi03} as
\begin{equation*}
\widehat{\text{ROC}}(p\mid\mathbf{x})=1-\widehat{F}_{\varepsilon_{D}}\left\{a(\mathbf{x})+b\widehat{F}_{\varepsilon_{\bar{D}}}^{-1}(1-p)\right\}.
\end{equation*}
Here, the covariate-specific AUC and pAUC (Expressions \eqref{AUCConditional} and \eqref{pAUCConditional}) also admit closed forms. However, especially for large datasets, their calculation can be very time-consuming. As a consequence, in \pkg{ROCnReg} they are computed numerically using Simpson's rule; in our experience Simpson's rule provides almost identical results to the ones obtained using the closed-form expressions. In what concerns the covariate-specific $\text{pAUC}_{\text{TNF}}$, it is interesting to note that
\[
\widehat{\text{ROC}}_{\text{TNF}}(p\mid\mathbf{x}) = 1-\widehat{F}_{\varepsilon_{D}}\left\{a^{*}(\mathbf{x})+b^{*}\widehat{F}_{\varepsilon_{\bar{D}}}^{-1}(1-p)\right\},
\]
with
\[
a^{*}(\mathbf{x}) =\tilde{\mathbf{x}}^{\top}\frac{(\widehat{\boldsymbol{\beta}}_{D}-\widehat{\boldsymbol{\beta}}_{\bar{D}})}{\widehat{\sigma}_{\bar{D}}}\;\;\;\mbox{and}\;\;\; b^{*} =\frac{\widehat{\sigma}_D}{\widehat{\sigma}_{\bar{D}}}.
\]
The covariate-specific $\text{pAUC}_{\text{TNF}}$ (Expression \eqref{pAUCConditional2}) is then computed numerically using Simpson's rule based on the previous expressions. 

Finally, in pretty much the same way as for the pooled ROC curve, the covariate-specific Youden index (and associated threshold value) is obtained by maximising, over a grid of possible threshold values, the expression in \eqref{YIConditional2}, making use of result \eqref{cROC_YI_sp}.
\subsubsection*{Induced kernel-based approach}
The kernel-based approach of \cite{Gonzalez2011} and \cite{MX11a} is available in the \code{cROC.kernel}. Differently to all the other estimating approaches for the covariate-specific ROC curve presented in this section, it can only deal with one continuous covariate. Similarly to the approaches of \cite{Pepe98} and \cite{Faraggi03}, it also assumes a location-scale regression model for the test outcomes in each group, but the effect of the covariate is not assumed to be linear and the variance is allowed to depend on the covariate. Specifically, the models postulated in each group are as follows
\begin{equation}\label{locscaleker}
Y_{\bar{D}}=\mu_{\bar{D}}(X_{\bar{D}})+\sigma_{\bar{D}}(X_{\bar{D}})\varepsilon_{\bar{D}},\qquad Y_D=\mu_{D}(X_{D})+\sigma_D(X_{D})\varepsilon_D,
\end{equation}
where $\mu_{D}(x)=\E(Y_D\mid X_D=x)$ and $\sigma_{D}^2(x) = \VAR(Y_D\mid X_D=x)$ are the regression and variance functions, respectively, with $\mu_{\bar{D}}(x)$ and $\sigma_{\bar{D}}^2(x)$ being analogously defined. The error terms $\varepsilon_{\bar{D}}$ and $\varepsilon_{D}$ have mean zero, variance one, are independent of each other and of the covariate, and have distribution functions given by $F_{\varepsilon_{\bar{D}}}$ and $F_{\varepsilon_{D}}$, respectively. 

Both the regression and variance functions are estimated using local polynomial kernel smoothers \citep{Fan1996}. In particular, local constant (Nadaraya--Watson) or local linear estimators are employed for the regression function, whereas for the variance function only local constant estimators are used. Estimation in \pkg{ROCnReg} makes use of the \proglang{R} package \pkg{np} by \cite{Hayfield08}. We note that estimation proceeds in a sequential manner: 1) the regression function, say in the diseased group and denoted by $\widehat{\mu}_{D}$, is estimated first on the basis of $\{(x_{Dj},y_{Dj})\}_{j=1}^{n_{D}}$, and 2) the variance function is estimated next on the basis of the sample $\{(x_{Dj},[y_{Dj}-\widehat{\mu}_{D}(x_{Dj})]^{2})\}$. Both steps involve the selection of a bandwidth parameter which is done via cross-validation. As in the model of \cite{Pepe98}, the CDFs $F_{\varepsilon_{D}}$ and $F_{\varepsilon_{\bar{D}}}$ are estimated via the empirical CDF of the standardised residuals, that is,
\begin{equation*}
\widehat{F}_{\varepsilon_{D}}(y)=\frac{1}{n_D}\sum_{j=1}^{n_D}I(\widehat{\varepsilon}_{Dj}\leq y), \qquad \widehat{\varepsilon}_{Dj}=\frac{y_{Dj}-\widehat{\mu}_D(x_{Dj})}{\widehat{\sigma}_D(x_{Dj})}. 
\end{equation*}
with the empirical CDF of the standardised residuals in the nondiseased group estimated analogously. Finally, the covariate-specific ROC curve is computed in an analogous way as before as
\begin{equation*}
\widehat{\text{ROC}}(p\mid x)=1-\widehat{F}_{\varepsilon_{D}}\left\{a(x)+b(x)\widehat{F}_{\varepsilon_{\bar{D}}}^{-1}(1-p)\right\},
\end{equation*}
where
\[
a(x) = \frac{\widehat{\mu}_{\bar{D}}(x) - \widehat{\mu}_{D}(x)}{\widehat{\sigma}_{D}(x)}\;\;\;\mbox{and}\;\;\; b(x) = \frac{\widehat{\sigma}_{\bar{D}}(x)}{\widehat{\sigma}_{D}(x)}.
\]
Estimation of the covariate-specific AUC, pAAUC, $\text{pAUC}_{\text{TNF}}$, and YI follows a similar reasoning as the one described previously for the induced semiparametric linear model when no assumptions are made regarding the distribution of the error terms.

For both the induced semiparametric linear model and the induced kernel approach, uncertainty quantification is done through a bootstrap of the residuals. For further details see, for instance, \cite{MX11a}.

\subsubsection*{Bayesian nonparametric approach based on a dependent Dirichlet process mixture of normal distributions}
The Bayesian nonparametric approach for conducting inference about the covariate-specific ROC curve of \cite{Inacio13}, which is based on a single-weights dependent Dirichlet process mixture of normal distributions \cite{Iorio2009}, is implemented in the function \code{cROC.bnp}. By opposition to the previously described approaches to ROC regression, this method rests on directly modelling the CDF of test outcomes separately in the diseased and nondiseased groups. In a single-weights dependent Dirichlet process mixture of normals model \citep{Iorio2009}, the conditional CDF in the diseased group is modelled as follows
\begin{equation*}
F_{D}(y_{Dj}\mid\mathbf{X}_D=\mathbf{x}_{Dj})=\int \Phi(y_{Dj}\mid \mu_{D}(\mathbf{x}_{Dj},\boldsymbol{\beta}),\sigma^2)
\text{d}G_D(\boldsymbol{\beta},\sigma^2),\qquad G_D\sim\text{DP}(\alpha_D,G_{D}^{*}(\boldsymbol{\beta},\sigma^2)),
\end{equation*}
with the conditional CDF in the nondiseased group $\bar{D}$ following in an analogous manner. As in the no-covariate case, by making use of Sethuraman's truncated representation of the DP, we can write the conditional CDF as
\begin{align*}
&F_{D}(y_{Dj}\mid\mathbf{x}_{Dj}) = \sum_{l=1}^{L_D}\omega_{Dl}\Phi(y_{Dj}\mid \mu_{D}(\mathbf{x}_{Dj},\boldsymbol{\beta}_{Dl}),\sigma_{Dl}^2),\\
&\omega_{D1}=v_{D1},\quad \omega_{Dl}=v_{Dl}\prod_{r<l}(1-v_{Dr}),\quad l=2,\ldots, L_{D},\\
&v_{Dl}\overset{\text{iid}}\sim\text{Beta}(1,\alpha_D),\quad l = 1,\ldots,L_{D}-1,\quad v_{DL_{D}}=1.
\end{align*}
\begin{equation*}
\text{var}(y_{Dj}\mid\mathbf{x}_{Dj})=\sum_{l=1}^{L_D}\omega_{Dl}\sigma_{Dl}^{2}+\sum_{l=1}^{L_D}\omega_{Dl}\left\{\mu_{D}(\mathbf{x}_{Dj},\boldsymbol{\beta}_{Dl})-\left(\sum_{l=1}^{L_D}\omega_{Dl}\mu_{D}(\mathbf{x}_{Dj},\boldsymbol{\beta}_{Dl})\right)^2\right\}.
\end{equation*}
Note that by assuming that the weights, $w_{Dl}$, do not vary with covariates, the model might has limited flexibility in practice \citep{Maceachern2000}. This issue can, however, be largely mitigated by using a flexible formulation for $\mu_{D}(\mathbf{x}_{Dj},\boldsymbol{\beta}_{Dl})$, which is needed not only for the model to be able to recover nonlinear trends, but also to recover flexible shapes that might arise due to a dependence of the weights on the covariates. As such, the function \code{cROC.bnp} in \pkg{ROCnReg} allows to model the mean function of each component using an additive smooth structure 
\begin{equation}\label{additive}
\mu_{D}(\mathbf{x}_{Dj},\boldsymbol{\beta}_{Dl})=\beta_{Dl0}+f_{Dl1}(x_{Dj,1})+\ldots+f_{Dlq}(x_{Dj,q}),\qquad l=1,\ldots, L_D,
\end{equation}
where the smooth functions, $f_{Dlm}$ $(m = 1, \ldots, q)$, are approximated using a linear combination of B-splines basis functions. To avoid notational burden we have assumed that all $q$ covariates are continuous and modelled in a flexible way. However, the function \code{cROC.bnp} cal also deal with categorical covariates, linear effects of continuous covariates, as well as interactions. For the reasons mentioned before, we recommend that all continuous covariates are modelled as in \eqref{additive}. Nonetheless, posterior predictive checks, as illustrated in Section \nameref{sec:illustration}, can also be used to informally validate the fitted model. We write 
\begin{equation}\label{mu_ddp}
\mu_{D}(\mathbf{x}_{Dj},\boldsymbol{\beta}_{Dl})=\mathbf{z}_{Dj}^{\top}\boldsymbol{\beta}_{Dl}, \quad l=1,\ldots,L_D,\quad j =1,\ldots,n_D,
\end{equation}
where $\mathbf{z}_{Dj}^{\top}$ is the $j$th row of the design matrix that contains the intercept, the continuous covariates that are modelled in a linear way (if any), the cubic B-splines basis representation for those modelled in a flexible way, the categorical covariates (if any), and their interaction(s) (if believed to exist). Also, $\boldsymbol{\beta}_{Dl}$ collects, for the $l$th component, the regression coefficients associated with the aforementioned covariates. For the covariate effects modelled using cubic B-splines, an important issue is the selection of the number and location of the knots at which to anchor the basis functions, as this has the potential to impact inferences, more so for the former than the latter. The selection of the number of knots can be assisted by a model selection criterion, for example, (the adaptation to the case of mixture models of) the deviance information criterion (DIC) \citep{Celeux2006}, the log pseudo marginal likelihood (LPML) \citep{Geisser1979}, and the widely applicable information criterion (WAIC) \citep{Gelman2014}. In turn, for the location of the interior knots themselves we follow \cite{Rosenberg1995} and use the quantiles of the covariate values.  

The regression coefficients and variances associated with each of the $L_D$ components are sampled from the conjugate centring distribution $(\boldsymbol{\beta}_{Dl},\sigma_{Dl}^{-2})\overset{\text{iid}}\sim\text{N}_{Q_D}(\mathbf{m}_D,\mathbf{S}_D)\Gamma(a_D,b_D)$, with conjugate hyperpriors $\mathbf{m}_D\sim\text{N}(\mathbf{m}_{D0},\mathbf{S}_{D0})$ and $\mathbf{S}_D^{-1}\sim\text{Wishart}(\nu_D,(\nu_D\Psi_D)^{-1})$ (a Wishart distribution with degrees of freedom $\nu_D$ and expectation $\Psi_D^{-1}$) and where $Q_D$ is the dimension of the vector $\mathbf{z}_{Dj}$ . Hyperparameters $\mathbf{m}_{D0}$ and $\Psi_D$ must be chosen to represent the prior belief about the regression coefficients associated to each mixture component and about their covariance matrix, respectively, whereas $\mathbf{S}_{D0}$ and $\nu_D$ are chosen to represent the confidence in the prior belief of $\mathbf{m}_{D0}$ and $\Psi_D$, respectively. As in the no-covariate case, by default in \code{cROC.bnp} test outcomes and covariates are standardised, which not only facilitates specification of the hyperparameter values but also improves the mixing of the Markov chain Monte Carlo (MCMC) chains. The default values are as follows:
\begin{equation*}	
\mathbf{m}_{D0}=\mathbf{0}_{Q_D}, \quad \mathbf{S}_{D0}=10I_{Q_D},\quad \nu_D=Q_D+2, \quad \Psi_D=I_{Q_D}, \quad a_D=2, \quad b_D=0.5.
\end{equation*}
When test outcomes and covariates are not standardised, the defaults are the following: 
\begin{equation*}
\mathbf{m}_{D0}=\widehat{\boldsymbol{\beta}}_D,\quad \mathbf{S}_{D0}=\widehat{\boldsymbol{\Sigma}}_D,\quad \nu_D=Q_D+2,\quad \Psi_D=30\widehat{\boldsymbol{\Sigma}}_D,\quad a_D=2, \quad b_D=\widehat{\sigma}_D^2/2, 
\end{equation*}
where $\widehat{\boldsymbol{\beta}}_D$ and $\widehat{\sigma}_D$ are the least squares estimates from fitting the linear model $y_{Dj}=\mathbf{z}_{Dj}^{\top}\boldsymbol{\beta}_D+\sigma_D\varepsilon_{Dj}$, where $E(\varepsilon_{Dj})=0$, $\text{var}(\varepsilon_{Dj})=1$, and $\widehat{\boldsymbol{\Sigma}}_D$ is the estimated covariance matrix of $\widehat{\boldsymbol{\beta}}_D$. With regard to the specification of $\alpha_D$ and $L_D$, as in the DPM model (no-covariate case) we set them, respectively, to $1$ and $10$. The blocked Gibbs sampler is used to simulate draws from the posterior distribution and details about it can be found, for instance, in the Supplementary Materials of \cite{Inacio17}.

Similarly to the analogous model for the no covariate case, at iteration $s$ of the Gibbs sampler procedure, the covariate-specific ROC curve is computed as
\begin{equation*} 
\text{ROC}^{(s)}(p\mid\mathbf{x})=1-F_D^{(s)}\left\{F_{\bar{D}}^{-1(s)}(1-p\mid\mathbf{x})\mid\mathbf{x}\right\},  \quad s=1,\ldots, S,
\end{equation*}
with
\begin{equation}\label{cdfdpm} 
F_D^{(s)}(y\mid\mathbf{x})=\sum_{l=1}^{L_D}\omega_{Dl}^{(s)}\Phi\left(y\mid \mathbf{z}^{\top}\boldsymbol{\beta}_{Dl}^{(s)},\sigma_{Dl}^{2(s)}\right),\quad
F_{\bar{D}}^{(s)}(y\mid\mathbf{x})=\sum_{k=1}^{L_{\bar{D}}}\omega_{\bar{D}k}^{(s)}\Phi\left(y\mid\mathbf{z}^{\top}\boldsymbol{\beta}_{\bar{D}k}^{(s)},\sigma_{\bar{D}k}^{2(s)}\right),
\end{equation}
and where the inversion is performed numerically. A point estimate for $\text{ROC}(p\mid\mathbf{x})$ can be obtained by computing the mean (or the median) of the ensemble $\{\text{ROC}^{(1)}(p\mid\mathbf{x}),\ldots,\text{ROC}^{(S)}(p\mid\mathbf{x})\}$, with pointwise credible bands derived from the percentiles of the ensemble. Although the results presented in \cite{Erkanli2006} can be extended to derive a closed-form expression for the covariate-specific AUC, for computational reasons, in \pkg{ROCnReg} the integral in \eqref{AUCConditional} is approximated using Simpson's rule, and the same applies for the partial areas. Conditionally on a specific covariate value, the computation of the Youden index and of the optimal threshold proceeds in a similar way as in the DPM model \citep[see][for details]{Inacio17}. As for the covariate-specific ROC curve, point and interval estimates can be obtained from the corresponding covariate-specific ensemble of each summary measure.
\subsection{Covariate-adjusted ROC curve}
All estimators for the covariate-adjusted ROC curve make use of Equation \eqref{aroc2} and rely on the following three steps
\begin{enumerate}
\item Estimation of the conditional distribution of test outcomes in the nondiseased group, $F_{\bar{D}}(y_{\bar{D}i}\mid\mathbf{x}_{\bar{D}i})$.
\item Computation of the placement value $U_D=1-F_{\bar{D}}(Y_D\mid\mathbf{X}_D)$ where, by a slight abuse of notation, we are designating it by the same letter used for the unconditional case.
\item Estimation of the cumulative distribution function of $U_D$. 
\end{enumerate}	
\subsubsection{Frequentist approaches}
The approaches used for estimation of the AROC curve proposed by \cite{Janes09a} and \cite{MX11a} only differ in Step 1. Specifically, once one has an estimate of the conditional CDF in the nondiseased group, say $\widehat{F}_{\bar{D}}(\cdot\mid\mathbf{x})$, Step 2 in the two approaches consists of trivially computing the diseased placement values as 
\begin{equation*}
\widehat{U}_{Dj}=1-\widehat{F}_{\bar{D}}(y_{Dj}\mid\mathbf{x}_{Dj}),\quad j = 1,\ldots, n_D.
\end{equation*}
Next, in Step 3, the AROC curve at a false positive fraction of $p$ is estimated via the empirical distribution
function of the placement values calculated in the previous step, $\{\widehat{U}_{Dj}\}_{j=1}^{n_D}$, that is,
\begin{equation*}
\widehat{\text{AROC}}(p)=\frac{1}{n_D}\sum_{j=1}^{n_D}I(\widehat{U}_{Dj}\leq p).
\end{equation*}
With regard to Step 1, both authors assume a location-scale regression model for the test outcomes in the nondiseased group and, as such and as explained in the previous section, the conditional CDF of the test results can be written using the CDF of the regression errors, i.e.,
\begin{equation*}
F_{\bar{D}}(y\mid\mathbf{x})=F_{\varepsilon_{\bar{D}}}\left(\frac{y-\mu_{\bar{D}}(\mathbf{x})}{\sigma_{\bar{D}}(\mathbf{x})}\right).
\end{equation*}
While \cite{Janes09a} assume a location-scale model of the form of \eqref{loc-scale-sp}, \cite{MX11a} rely on specification \eqref{locscaleker}. The estimation of the mean and variance functions follow exactly the same procedures as those described in the induced semiparametric linear model (for \citealt{Janes09a}) and induced kernel-based approach (for \citealt{MX11a}) for the covariate specific ROC curve (the only difference being that here we only need to perform the estimation for the nondiseased group). At last, and also as in the estimators for the covariate-specific ROC curve, $F_{\varepsilon_{\bar{D}}}$ can be either assumed to be the standard normal distribution or left unspecified and estimated empirically on the basis of the standardised residuals. In both cases, the AAUC and pAAUC can be computed as follows
\begin{align*}
\widehat{\text{AAUC}} &= \int_{0}^{1}\text{AROC}(p)\text{d}p=1-\frac{1}{n_D}\sum_{j=1}^{n_D}\widehat{U}_{Dj},\\
\widehat{\text{pAAUC}(u_1)} &= \int_{0}^{u_1}\text{AROC}(p)\text{d}p=u_1-\frac{1}{n_D}\sum_{j=1}^{n_D}\min\left\{u_1,\widehat{U}_{Dj}\right\},
\end{align*}
whereas the $\text{pAAUC}_{\text{TNF}}$ is computed as in Equation \eqref{pAUCConditional3} using numerical integration methods (function \code{integrate} in \proglang{R} package \pkg{stats}).

\subsubsection{Bayesian nonparametric approach}
Recently, \cite{Inacio18} proposed a Bayesian nonparametric estimator for the AROC curve that combines a dependent Dirichlet process mixture model and the Bayesian bootstrap. As in the Bayesian nonparametric approach for estimating the covariate-specific ROC curve, in Step 1, rather than assuming a location-scale regression model for the test outcomes in the nondiseased group, the entire conditional distribution is modelled using a single-weights dependent Dirichlet process mixture of normal distributions. Again, here, we also recommend to use cubic B-splines transformations of all continuous covariates. Using the same notation as before, we model the conditional density as
\begin{equation*}
F_{\bar{D}}(y_{\bar{D}i}\mid\mathbf{x}_{\bar{D}i})=\sum_{l=1}^{L_{\bar{D}}}\omega_{\bar{D}l}\Phi(y_{\bar{D}i}\mid\mathbf{z}_{\bar{D}i}^{\top}\boldsymbol{\beta}_{\bar{D}l},\sigma_{\bar{D}l}^2).
\end{equation*}
The same prior distributions and default values as in the \code{cROC.bnp} function are adopted for $\boldsymbol{\beta}_{\bar{D}l}$ and $\sigma_{\bar{D}l}^2$. Once Step 1 has been completed and given a posterior sample from the parameters of interest, the
corresponding realisation of the placement value of a diseased subject in the nondiseased population is easily computed as
\begin{equation*}
U_{Dj}^{(s)}=1-F_{\bar{D}}^{(s)}(y_{Dj}\mid\mathbf{x}_{Dj})=\sum_{l=1}^{L_{\bar{D}}}\omega_{\bar{D}l}^{(s)}\Phi\left(y_{Dj}\mid\mathbf{z}_{Dj}^{\top}\boldsymbol{\beta}_{\bar{D}l}^{(s)},\sigma_{\bar{D}l}^{2(s)}\right), \quad j =1,\ldots,n_D,\quad s=1,\ldots, S.
\end{equation*}
Finally, in Step 3, the cumulative distribution function of $\{U_{Dj}^{(s)}\}_{j=1}^{n_D}$ is estimated through the Bayesian bootstrap
\begin{equation*}
\text{AROC}^{(s)}(p)=\sum_{j=1}^{n_D}q_{j}^{(s)}I\left(U_{Dj}^{(s)}\leq p\right),\quad (q_{1}^{(s)},\ldots,q_{n_{D}}^{(s)})\sim\text{Dirichlet}(n_D; 1,\ldots,1).
\end{equation*}
As before, closed-form expressions do exist for the AAUC and pAAUC
\begin{align*}
\text{AAUC}^{(s)}&=\int_{0}^{1}\text{AROC}^{(s)}(p)\text{d}p=1-\sum_{j=1}^{n_D}q_j^{(s)}U_{Dj}^{(s)},\\
\text{pAAUC}^{(s)}(u_1)&=\int_{0}^{u_1}\text{AROC}^{(s)}(p)\text{d}p=u_1-\sum_{j=1}^{n_D}q_j^{(s)}\min\left\{u_1,U_{Dj}^{(s)}\right\},
\end{align*}
and the $\text{pAAUC}_{\text{TNF}}$ (Equation \eqref{pAUCConditional3}) is computed using numerical integration methods. With regard to the YI, it is obtained by directly plugging in $\text{AROC}^{(s)}(p)$ in Expression \eqref{YIAROC}.

A point estimate for $\text{AROC}(p)$ can be obtained by computing the mean (or the median) of the ensemble $\{\text{AROC}^{(1)}(p),\ldots,\text{AROC}^{(S)}(p)\}$, that is,
\begin{equation*}
\widehat{\text{AROC}}(p)=\frac{1}{S}\sum_{s=1}^{S}\text{AROC}^{(s)}(p), %\quad 0\leq p \leq 1,
\end{equation*}
and the percentiles of the ensemble can be used to provide pointwise credible bands/credible intervals. The same applies for the AAUC and pAAUC.
%% -- Illustrations ------------------------------------------------------------

%% - Virtually all JSS manuscripts list source code along with the generated
%%   output. The style files provide dedicated environments for this.
%% - In R, the environments {Sinput} and {Soutput} - as produced by Sweave() or
%%   or knitr using the render_sweave() hook - are used (without the need to
%%   load Sweave.sty).
%% - Equivalently, {CodeInput} and {CodeOutput} can be used.
%% - The code input should use "the usual" command prompt in the respective
%%   software system.
%% - For R code, the prompt "R> " should be used with "+  " as the
%%   continuation prompt.
%% - Comments within the code chunks should be avoided - these should be made
%%   within the regular LaTeX text.
\section{Package presentation and illustration} \label{sec:illustration}
This section describes the main functions in the \pkg{ROCnReg} package and illustrates their usage using endocrine data from a cross-sectional study carried out by the Galician Endocrinology and Nutrition Foundation (FENGA). A detailed description of this dataset can be found in \cite{Tome09}. It has also been previously analysed in \cite{MX11a, MX11b} and \cite{Inacio18}. For confidentiality reasons, the data used in this paper correspond to a synthetic dataset that was obtained by mimicking the original one and can be found in the \pkg{ROCnReg} package under the name \code{endosyn}. A summary of the data follows.

\begin{Schunk}
\begin{Sinput}
R> library("ROCnReg")
R> data("endosyn")
R> summary(endosyn)
\end{Sinput}
\begin{Soutput}
   cvd_idf            age          gender          bmi       
 Min.   :0.0000   Min.   :18.25   Men  :1317   Min.   :12.60  
 1st Qu.:0.0000   1st Qu.:29.57   Women:1523   1st Qu.:23.19  
 Median :0.0000   Median :39.28                Median :26.24  
 Mean   :0.2433   Mean   :41.43                Mean   :26.69  
 3rd Qu.:0.0000   3rd Qu.:50.84                3rd Qu.:29.74  
 Max.   :1.0000   Max.   :84.66                Max.   :46.20 
\end{Soutput}
\end{Schunk}
The dataset is comprised of $2840$ individuals ($1317$ men and $1523$ women, variable \code{gender}), with an \code{age} range between $18$ and $85$ years old. Variable \code{bmi} contains the body mass index (BMI) values, and  \code{cvd_idf} is the variable that indicates the presence (1) or absence (0) of two or more cardiovascular disease (CVD) risk factors. Following previous studies, the CVD risk factors considered include raised triglycerides, reduced HDL-cholesterol, raised blood pressure, and raised fasting plasma glucose. Note that from the $2840$ individuals, about $24\%$ present two or more CVD risk factors.

Using the \pkg{ROCnReg} package, in the subsequent sections we will illustrate how to ascertain, through the pooled ROC curve, the discriminatory capacity of the BMI (which acts as our diagnostic test in this example) in differentiating individuals with two or more CVD risk factors (those belonging to the diseased class $D$) from those having none or just one CVD risk factor (and that therefore belong to the nondiseased group $\bar{D}$). Also, in Section \ref{sec:croc_ilu} we will show how to evaluate, through the covariate-specific ROC curve, the possible modifying effect of \code{age} and \code{gender} on the discriminatory capacity of the BMI. Finally, Section \ref{sec:aroc_ilu} focuses on the covariate-adjusted ROC curve as a global summary measure of the BMI discriminatory ability, when taking the \code{age} and \code{gender} effects into account. In the Appendices we show the usage of the package for those methods not described in the main text.

\subsection{Pooled ROC curve}\label{sec:pooledroc_ilu}
The \pkg{ROCnReg} package allows estimating the pooled ROC curve by means of the four methods described in Section \ref{sec:methods}. Recall that function \code{pooledROC.emp} implements the empirical estimator, \code{pooledROC.kernel} the kernel-based approach, and \code{pooledROC.BB} and \code{pooledROC.dpm} correspond, respectively, to the Bayesian bootstrap estimator and the approach based on a Dirichlet process mixture (of normal distributions). The input arguments in the functions are method-specific (details can be found in the manual accompanying the package), but in all cases numerical and graphical summaries can be obtained by calling the functions \code{print.pooledROC}, \code{summary.pooledROC}, and \code{plot.pooledROC}, which can be abbreviated by \code{print}, \code{summary}, and \code{plot}.

By way of example, we present here the syntax using the \code{pooledROC.dpm} function. Recall that our aim is to ascertain, using the \code{endosyn} dataset, the discriminatory capacity of the BMI in differentiating individuals with two or more CVD risk factors from those having none or just one CVD risk factor.
\begin{Schunk}
\begin{Sinput}
R> set.seed(123, "L'Ecuyer-CMRG") # for reproducibility
R> pROC_dpm <- pooledROC.dpm(marker = "bmi", group = "cvd_idf", 
+    tag.h = 0, data = endosyn, standardise = TRUE, p = seq(0, 1, l = 101),
+    ci.level = 0.95, compute.lpml = TRUE, compute.WAIC = TRUE, compute.DIC = TRUE,
+    pauc = pauccontrol(compute = TRUE, focus = "FPF", value = 0.1), 
+    density = densitycontrol(compute = TRUE), prior.h = priorcontrol.dpm(), 
+    prior.d = priorcontrol.dpm(), 
+    mcmc = mcmccontrol(nsave = 8000, nburn = 2000, nskip = 1),
+    parallel = "snow", ncpus = 2, cl = NULL)
\end{Sinput}
\end{Schunk}
Before describing in detail the previous call, we first present the control functions that are used. In particular
\begin{CodeChunk}
\begin{CodeInput}
pauccontrol(compute = FALSE, focus = c("FPF", "TPF"), value = 1)
\end{CodeInput}
\end{CodeChunk}
can be used to indicate whether the pAUC should be computed (by default it is not computed), and in case it is computed (i.e., \code{compute = TRUE} ), whether the \code{focus} should be placed on restricted FPFs (pAUC, see (\ref{pAUC})) or on restricted TPFs ($\text{pAUC}_{\text{TPF}}$, see (\ref{pAUC_TPF})). In both cases, the upper bound $u_1$ (if focus is the FPF) or the lower bound $v_1$ (if focus is the TPF) should be indicated in the argument \code{value}. In addition to the pooled ROC curve, AUC, and pAUC (if required), the function \code{pooledROC.dpm} also allows computing the probability density function (PDF) of the test outcomes in both the diseased and nondiseased groups. In order to do so, we use
\begin{CodeChunk}
\begin{CodeInput}
densitycontrol(compute = FALSE, grid.h = NA, grid.d = NA)
\end{CodeInput}
\end{CodeChunk}
By default, PDFs are not returned by the function \code{pooledROC.dpm}, but this can be changed by setting \code{compute = TRUE}, and through \code{grid.h} and \code{grid.d} the user can specify a grid of test results where the PDFs are to be evaluated in, respectively, the nondiseased and diseased groups. Value \code{NA} signals auto initialisation, with default a vector of length $200$ in the range of the test results. Regarding the hyperparameters for the Dirichlet process mixture of normals model (used for the estimation of the PDFs/CDFs of the test outcomes in each group), they can be controlled using
\begin{CodeChunk}
\begin{CodeInput}
priorcontrol.dpm(m0 = NA, S0 = NA, a = 2, b = NA, alpha = 1, L = 10)
\end{CodeInput}
\end{CodeChunk}
A detailed description of these hyperparameters is found in Section \nameref{sec:methods}. Finally, to set the various parameters controlling the MCMC procedure (which in our case is simply a Gibbs sampler) we use
\begin{CodeChunk}
\begin{CodeInput}
mcmccontrol(nsave = 8000, nburn = 2000, nskip = 1)
\end{CodeInput}
\end{CodeChunk}
Here \code{nsave} is an integer value with the total number of scans to be saved, \code{nburn} is the number of burn-in scans, and \code{nskip} is the thinning interval. Unless due to memory usage reasons, we recommend not thinning and instead monitoring the effective sample size of the MCMC chain.

Coming back to the \code{pooledROC.dpm} function, through \code{marker} the user specifies the name of the variable containing the test results; in our case, these are the values of the BMI. The name of the variable that distinguishes diseased (two or more CVD risk factors, $D$) from nondiseased individuals (none or one CVD risk factor, $\bar{D}$) is represented by the argument \code{group}, and the value codifying nondiseased individuals in \code{group} is specified by \code{tag.h}. The \code{data} argument is a data frame containing the data and all needed variables. Setting \code{standardise = TRUE} (the default) will standardise (i.e., subtract the mean and divide by the standard deviation) the test outcomes. The set of FPFs at which to estimate the pooled ROC curve is specified in the argument \code{p}, and argument \code{ci.level} allows specifying the level for the credible intervals (by default: $0.95$). The LPML, WAIC, and DIC are computed by setting, respectively, the arguments \code{compute.lpml}, \code{compute.WAIC}, and \code{compute.DIC} to \code{TRUE}. Argument \code{pauc} is an (optional) list of values to replace the default values returned by the function \code{pauccontrol}. Here, we ask for the pAUC to be computed, with focus on restricted FPFs and upper bound $u_1 = 0.1$. Similarly, the argument \code{density} is an (optional) list of values to replace the default values returned by the function \code{densitycontrol}, as it is the argument \code{mcmc}. Through \code{prior.h} and \code{prior.d} arguments we specify the hyperparameters in the nondiseased and diseased classes, respectively. Again, both arguments are (optional) lists of values to replace the default values returned by the function \code{priorcontrol.dpm}. We shall remember that different hyperparameters' default values are set depending on whether test outcomes are standardised or not.  Finally, arguments \code{parallel}, \code{ncpus} and \code{cl} allow to perform parallel computations (based on the R-package \pkg{parallel}). In particular, through \code{parallel} the user specifies the type of parallel operation: either \code{"no"} (default), \code{"multicore"} (not available on Microsoft Windows operating systems) or \code{"snow"}. Argument \code{ncpus} is used to indicate the number of processes to be used in parallel operation (when \code{parallel = "multicore"} or \code{parallel = "snow"}), and \code{cl} is an optional parallel or snow cluster to be used when \code{parallel = "snow"}. If \code{cl} is not supplied (as in our example), a cluster on the local machine is created for the duration of the call.

A numerical summary of the fitted model can be obtained by calling the function \code{summary}, that provides, among other information, the estimated AUC (posterior mean) and $95\%$ credible interval (recall that, in the call to the function, we set \code{ci.level = 0.95}) and, if required, the LPML, WAIC, and DIC, separately, in the nondiseased (denoted here as \code{Group H}) and diseased (\code{Group D}) classes.

\begin{figure}[h!]
\begin{center}
\includegraphics[width=8cm]{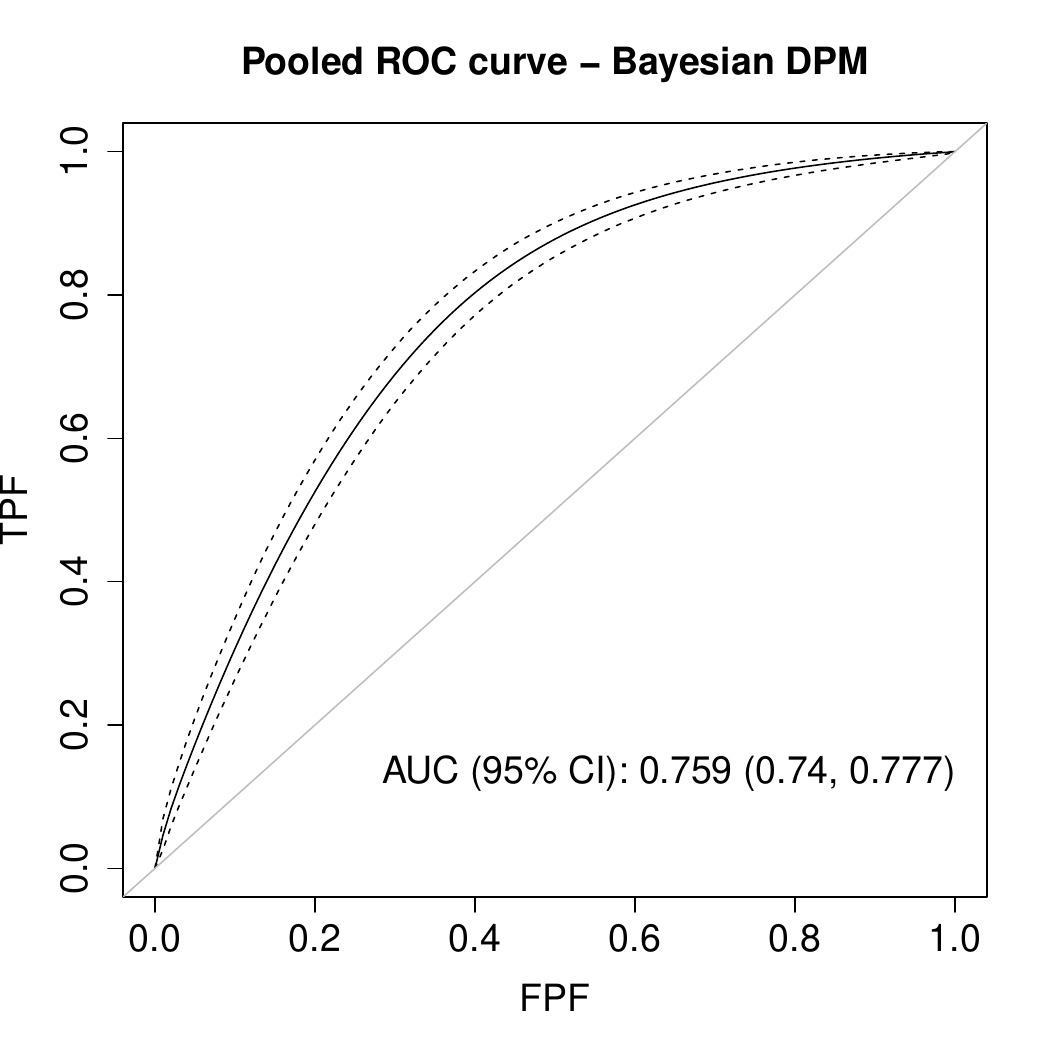}
\end{center}
\caption{Graphical results as provided by the \code{plot.pooledROC} function for an object of class \code{pooledROC.dpm}. Posterior mean and 95\% pointwise credible band for the pooled ROC curve and corresponding posterior mean and 95\% credible interval for the AUC.}
\label{pROC_dpm_plot}
\end{figure}

\begin{CodeChunk}
\begin{CodeInput}
R> summary(pROC_dpm)
\end{CodeInput}
\begin{CodeOutput}
Call:
pooledROC.dpm(marker = "bmi", group = "cvd_idf", tag.h = 0, data = endosyn, 
    standardise = TRUE, p = seq(0, 1, l = 101), ci.level = 0.95, 
    compute.lpml = TRUE, compute.WAIC = TRUE, compute.DIC = TRUE, 
    pauc = pauccontrol(compute = TRUE, focus = "FPF", value = 0.1), 
    density = densitycontrol(compute = TRUE), prior.h = priorcontrol.dpm(L = 10), 
    prior.d = priorcontrol.dpm(L = 10), mcmc = mcmccontrol(nsave = 8000, 
        nburn = 2000, nskip = 1), parallel = "snow", ncpus = 2)

Approach: Pooled ROC curve - Bayesian DPM
----------------------------------------------
Area under the pooled ROC curve: 0.759 (0.74, 0.777)*
Partial area under the pooled ROC curve (FPF = 0.1): 0.168 (0.139, 0.199)*
 * Credible level:  0.95

Model selection criteria:
                     Group H       Group D
WAIC               12490.485      4017.063
WAIC (Penalty)         8.431         5.468
LPML               -6245.247     -2008.541
DIC                12490.276      4016.920
DIC (Penalty)          8.326         5.396


Sample sizes:
                           Group H     Group D
Number of observations        2149         691
Number of missing data           0           0
\end{CodeOutput}
\end{CodeChunk}
To complement these numerical results, the \pkg{ROCnReg} package also provides graphical results that can be used to further explore the fitted model. Specifically, the function \code{plot} depicts the estimated pooled ROC curve and AUC (posterior means), jointly with \code{ci.level}$\times 100\%$ (pointwise) credible intervals (here $95\%$)
\begin{CodeChunk}
\begin{CodeInput}
R> plot(pROC_dpm, cex.main = 1.5, cex.lab = 1.5, cex.axis = 1.5, cex = 1.5)
\end{CodeInput}
\end{CodeChunk}
The result of the above code is shown in Figure \ref{pROC_dpm_plot}. 

\begin{figure}[ht!]
\begin{center}
\subfigure[DPM model with 10 mixture components in each group]{\includegraphics[width=5.5cm]{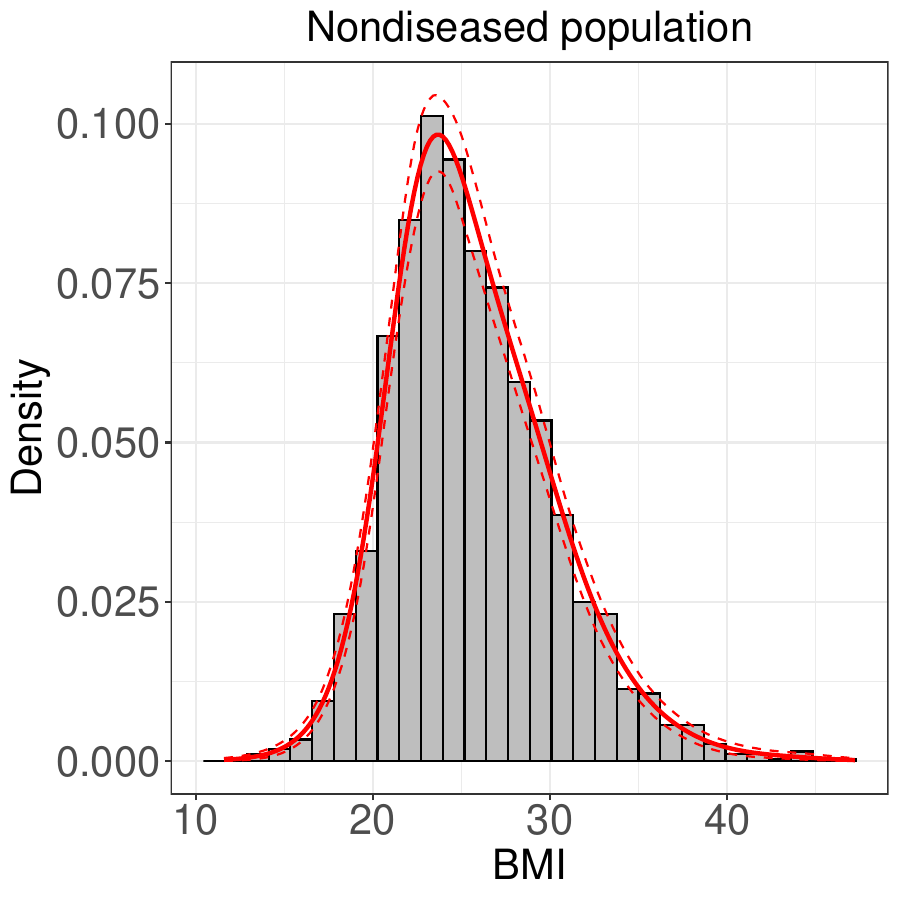}
\includegraphics[width=5.5cm]{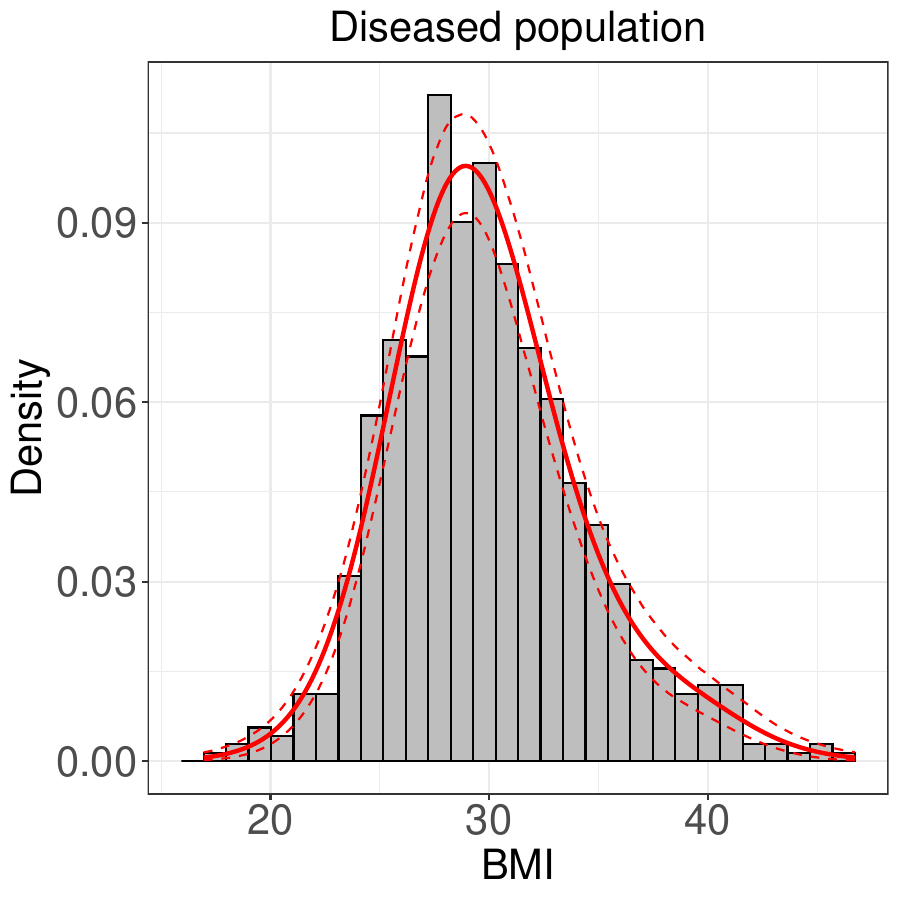}\label{pROC_dpm_densities_2}}
\subfigure[Normal model in each group]{\includegraphics[width=5.5cm]{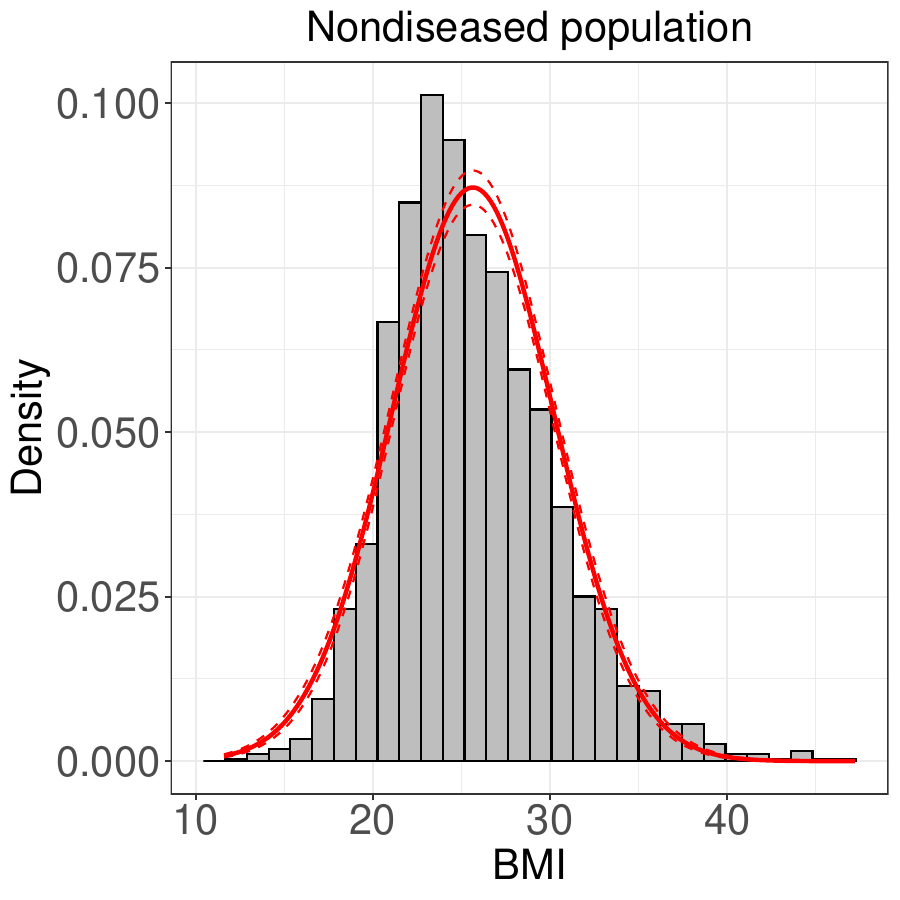}
\includegraphics[width=5.5cm]{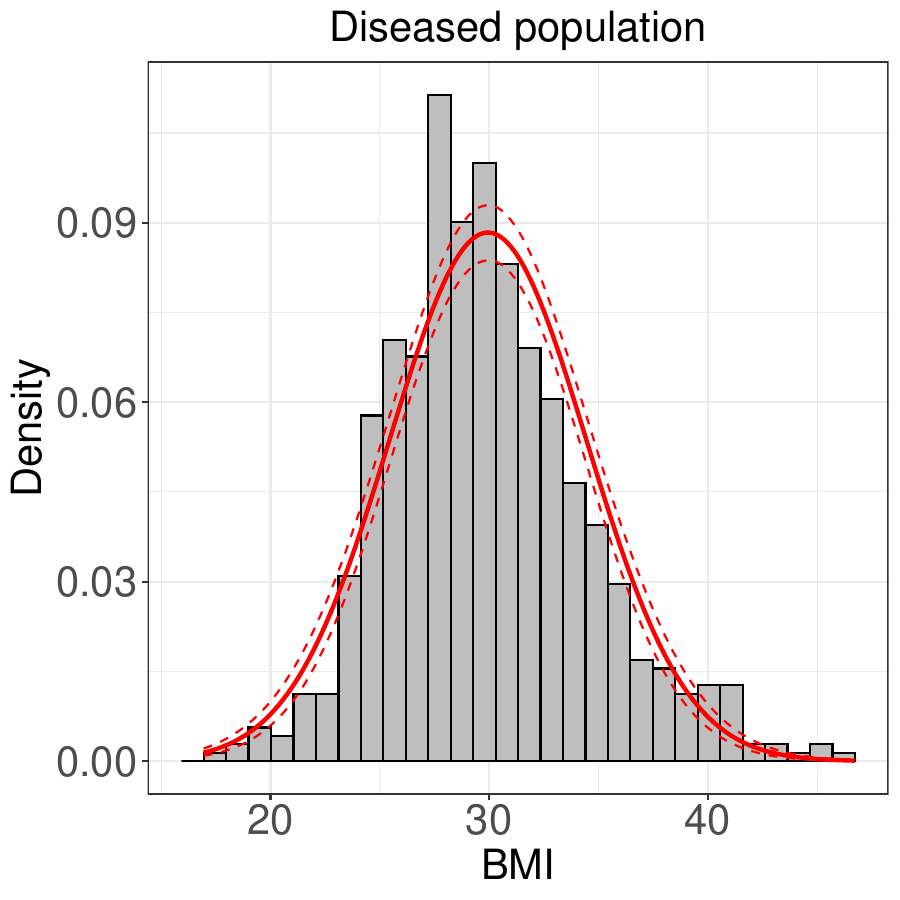}\label{pROC_normal_densities}}
\end{center}
\caption{Histogram of the (observed) BMI and posterior mean jointly along with 95\% pointwise credible bands (red lines) of the PDF of the BMI obtained using (a) a Dirichlet process mixture of normals model (object \code{pROC\_dpm}); and (b) a normal model (object \code{pROC\_normal}). Left: Nondiseased individuals (none or one CVD risk factor). Right: Diseased individuals (two or more CVD risk factors).}
\end{figure}

\begin{figure}[ht!]
\begin{center}
\includegraphics[width=14cm]{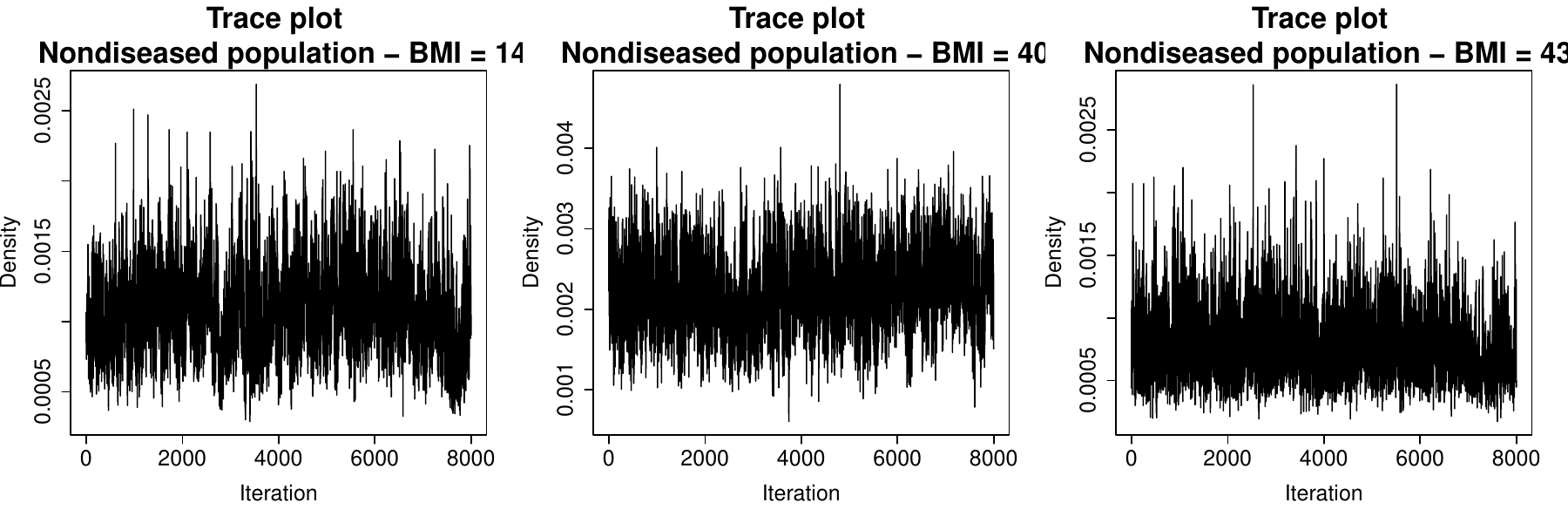}
\includegraphics[width=14cm]{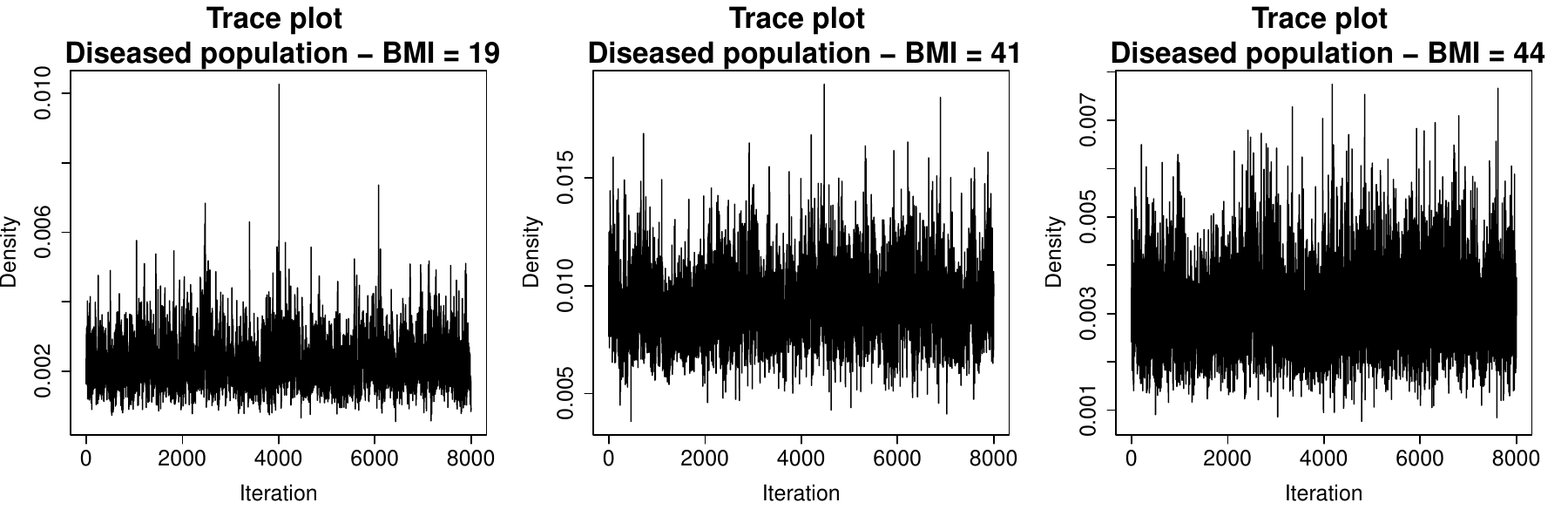}
\end{center}
\caption{Trace plots of the MCMC draws (after burn-in) of the PDFs of the BMI based on the model \code{pROC\_dpm}. Results are shown separately for the nondiseased and diseased populations and for different values of the BMI.}
\label{pROC_dpm_densities_tp}
\end{figure}

\begin{figure}[ht!]
\begin{center}
\includegraphics[width=6cm]{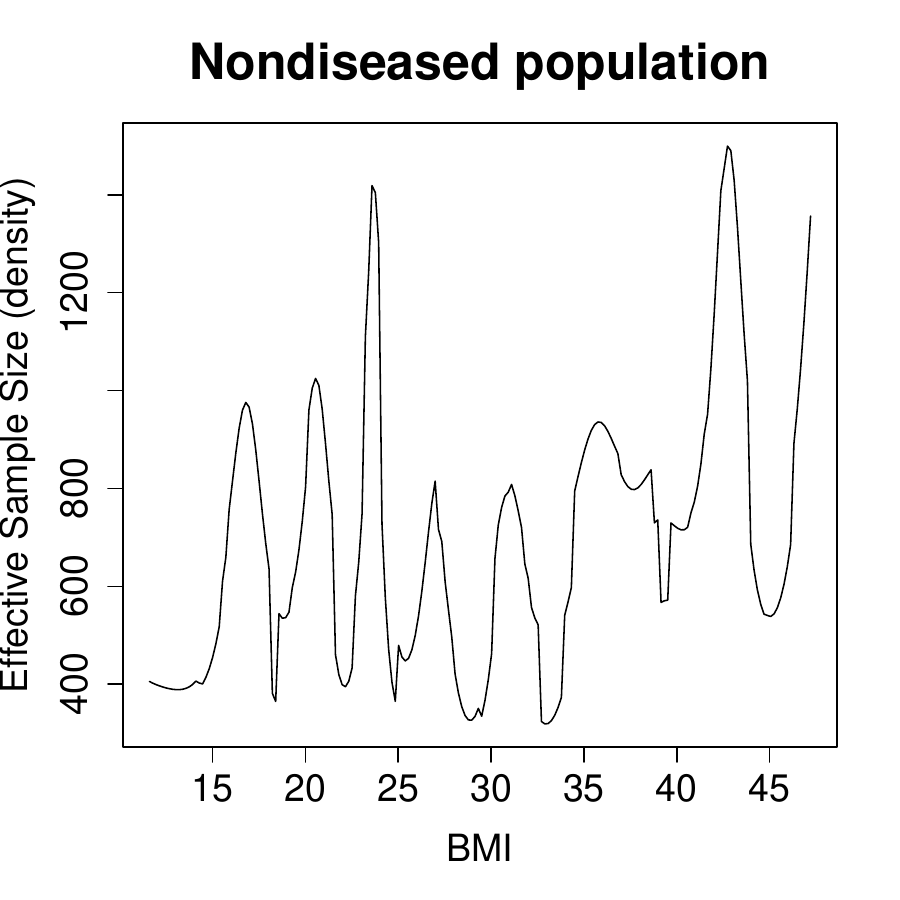}
\includegraphics[width=6cm]{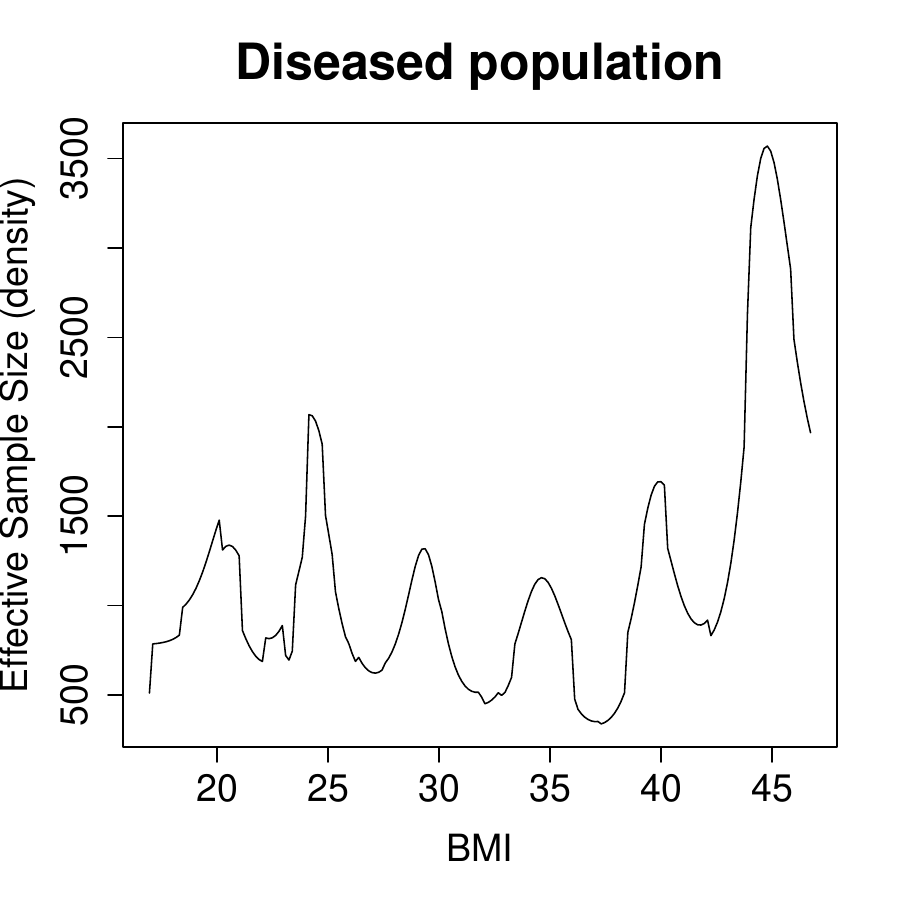}\\
\includegraphics[width=6cm]{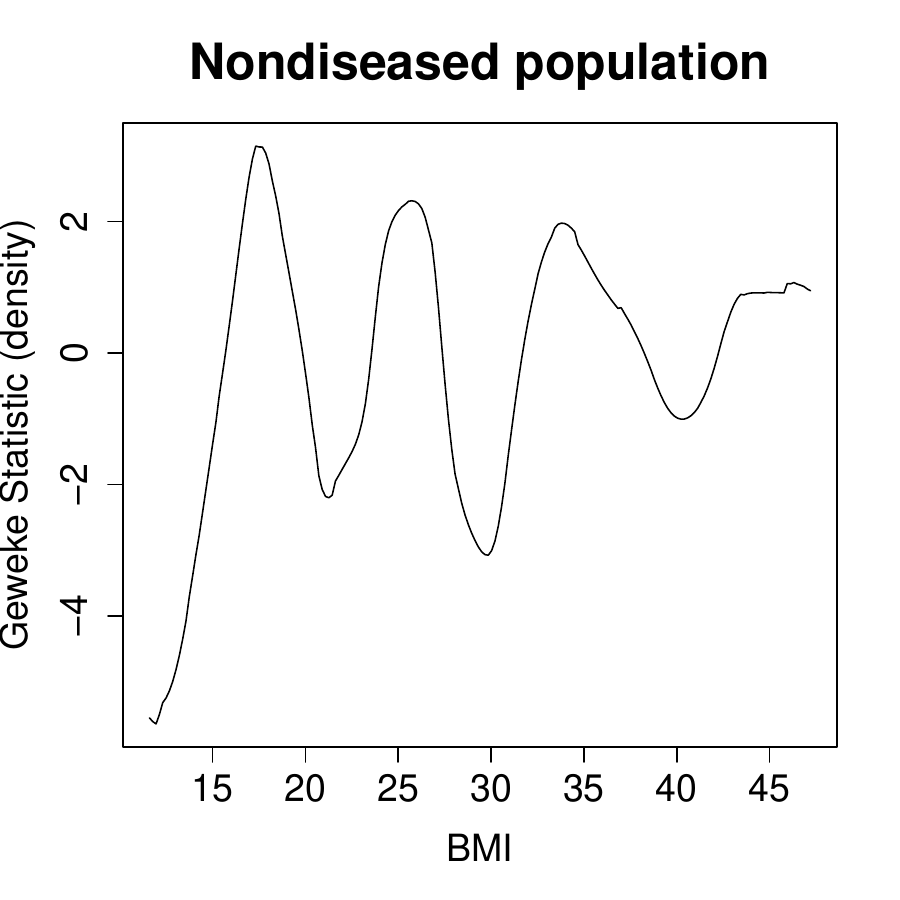}
\includegraphics[width=6cm]{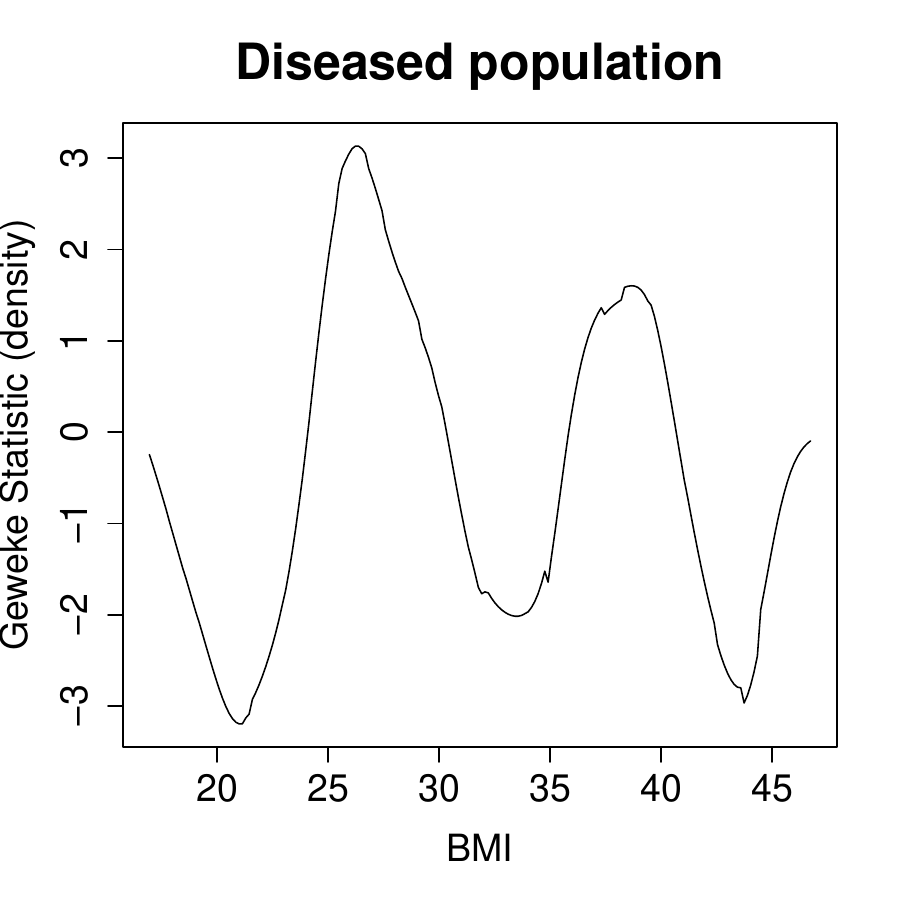}
\end{center}
\caption{Effective sample size and  Geweke statistic of the MCMC chains (after burn-in) and of the PDFs of the BMI based on model \code{pROC\_dpm} in the nondiseased and diseased populations. In both cases, results are shown along BMI values.}
\label{pROC_dpm_densities_es}
\end{figure}

By means of \code{density = densitycontrol(compute = TRUE)} in the call to the function, the estimates of the PDFs of the BMI in both classes are to be returned. This information can be accessed through component \code{dens} in object \code{pROC_dpm} (i.e., \code{pROC_dpm\$dens}), which is a list with elements \code{h} and \code{d} associated with the nodiseased and diseased groups, respectively. Each of the two elements is itself another list of two components: (1) \code{grid}, a vector that contains the grid of test results at which the PDFs have been evaluated (estimated); and (2) \code{dens}, a matrix with the PDFs at each iteration of the MCMC procedure. We can use these results to plot, e.g, the posterior mean (and 95\% pointwise credible bands) of the PDF of the BMI in the nondiseased and diseased populations (see Figure \ref{pROC_dpm_densities_2}, obtained using the \proglang{R} package \pkg{ggplot2} by \citealp{Wickham16}). As can be observed, the estimated densities obtained under the DPM method follow very closely the histograms of the data. Further, the estimated densities available in \code{dens} can be used, as advised by \cite{Gelman2013} (p.~553), to monitor convergence of the MCMC chains. The well-known label switching problem often leads to poor mixing of the chains of the component-specific parameters, but this may not impact convergence and mixing of the induced density/distribution of interest. For instance, Figure \ref{pROC_dpm_densities_tp} shows trace plots of the MCMC iterations (after burn-in) of the PDFs of the BMI in the two groups, for different (and randomly selected) values of the BMI, and Figure \ref{pROC_dpm_densities_es} depicts the corresponding effective sample sizes and and Geweke statistics (obtained using the \proglang{R} package \pkg{coda} by \citealp{coda06}). Note that all plots give evidence of a good mixing and do not suggest lack of convergence. For conciseness, the \proglang{R}-code for producing Figures \ref{pROC_dpm_densities_2}, \ref{pROC_dpm_densities_tp} and \ref{pROC_dpm_densities_es} is not provided here, but in the replication code that accompanies the paper (\url{https://bitbucket.org/mxrodriguez/rocnreg}). 

It is worth noting that the function \code{pooledROC.dpm} also allows fitting a normal distribution in each group; this is just a particular case (for which $L_D=L_{\bar{D}}=1$) of the more general DPM model. In order to fit such model, one simply needs to set $L=1$ in the \code{prior.d} and \code{prior.h} arguments. The code follows.
\begin{Schunk}
\begin{Sinput}
R> pROC_normal <- pooledROC.dpm(marker = "bmi", group = "cvd_idf", 
+    tag.h = 0, data = endosyn,
+    standardise = TRUE, p = seq(0, 1, l = 101), ci.level = 0.95, 
+    compute.lpml = TRUE, compute.WAIC = TRUE, compute.DIC = TRUE,
+    pauc = pauccontrol(compute = TRUE, focus = "FPF", value = 0.1), 
+    density = densitycontrol(compute = TRUE),
+    prior.h = priorcontrol.dpm(L = 1), prior.d = priorcontrol.dpm(L = 1),
+    mcmc = mcmccontrol(nsave = 8000, nburn = 2000, nskip = 1),
+    parallel = "snow", ncpus = 2, cl = NULL)
\end{Sinput}
\end{Schunk}
For the sake of space we omit from the summary the call to the function
\begin{CodeChunk}
\begin{CodeInput}
R> summary(pROC_normal)
\end{CodeInput}
\begin{CodeOutput}
Call: [...]

Approach: Pooled ROC curve - Bayesian DPM
----------------------------------------------
Area under the pooled ROC curve: 0.748 (0.728, 0.768)*
Partial area under the pooled ROC curve (FPF = 0.1): 0.224 (0.194, 0.253)*
 * Credible level:  0.95

Model selection criteria:
                     Group H       Group D
WAIC               12639.952      4049.004
WAIC (Penalty)         2.431         2.267
LPML               -6319.976     -2024.502
DIC                12639.505      4048.714
DIC (Penalty)          1.986         1.987


Sample sizes:
                           Group H     Group D
Number of observations        2149         691
Number of missing data           0           0
\end{CodeOutput}
\end{CodeChunk}
The fit of the DPM and normal models, in each group, can be compared on the basis of the WAIC, DIC, and/or the LPML. Remember that for the LPML, the higher its value, the better the model fit, while for the WAIC and DIC it is the other way around. By comparing these values, provided in the summary of each fitted model, we can conclude that the three criteria favour, in both the diseased and (especially in the) nondiseased groups, the more general DPM model. This is also corroborated by the plot of the fitted densities in each group shown in Figure \ref{pROC_normal_densities}.

%\begin{figure}[h!]
%\begin{center}
%\includegraphics[width=6cm]{densities_pooled_ROC_normal_h.pdf}
%\includegraphics[width=6cm]{densities_pooled_ROC_normal_d.pdf}
%\end{center}
%\caption{Histogram of the (observed) BMI and posterior mean jointly along with pointwise 95\% credible intervals (red lines) of the PDF of the BMI obtained using a normal model (object \code{pROC\_normal}). Left: Nondiseased individuals (none or one CVD risk factor). Right: Diseased individuals (two or more CVD risk factors).}
%\label{pROC_normal_densities}
%\end{figure}

\begin{figure}[h!]
\begin{center}
\includegraphics[width=8cm]{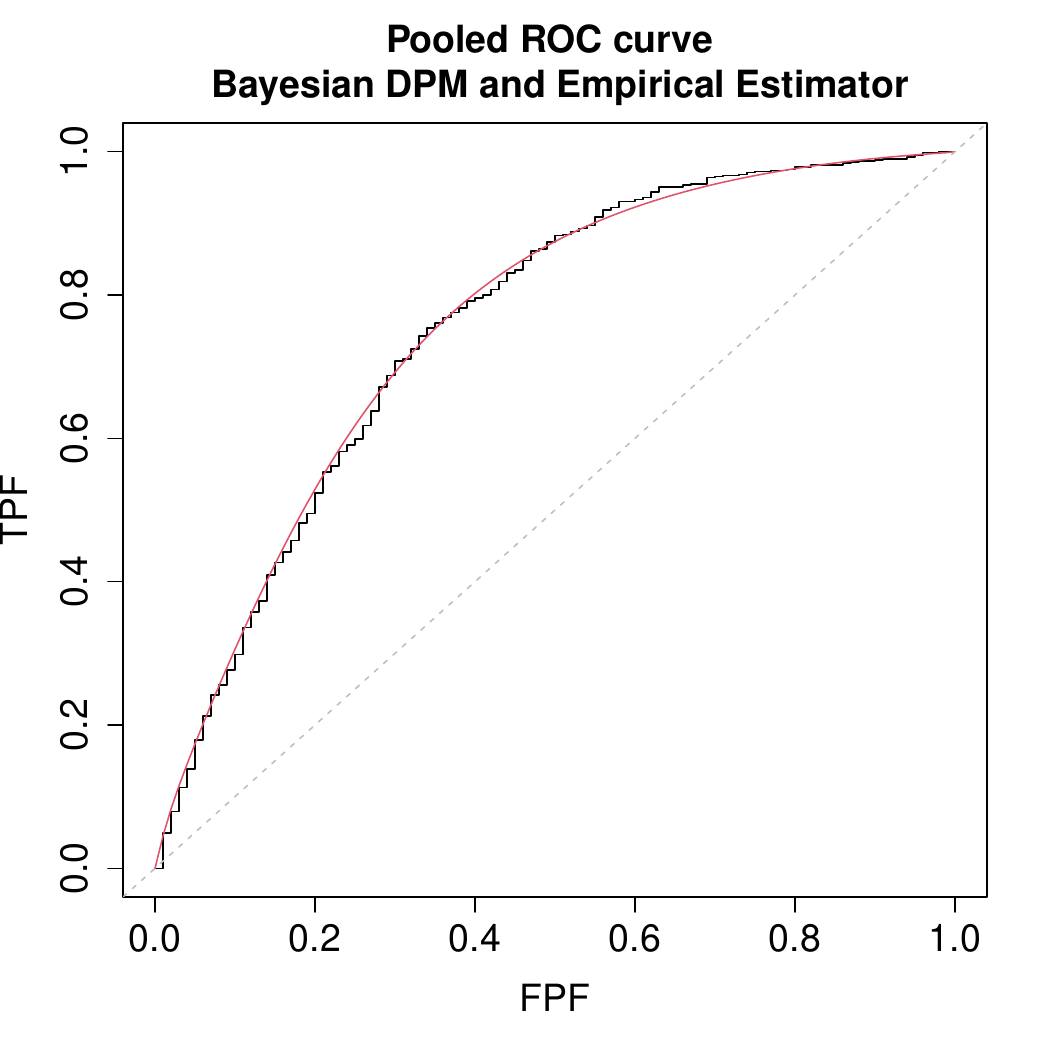}
\end{center}
\caption{Estimated ROC curve using the empirical approach (black line) and using the DPM method (red line--posterior mean).}
\label{pooled_ROC_dpm_emp}
\end{figure}

We now estimate the pooled ROC curve using the empirical estimator (function \code{pooledROC.emp}), and comparisons with the results obtained using the DPM approach are provided.
\begin{Schunk}
\begin{Sinput}
R> pROC_emp <- pooledROC.emp(marker = "bmi", group = "cvd_idf", 
+    tag.h = 0, data = endosyn, p = seq(0, 1, l = 101), 
+    pauc = pauccontrol(compute = TRUE, focus = "FPF", value = 0.1), B = 500,
+    ci.level = 0.95, parallel = "snow", ncpus = 2)
\end{Sinput}
\end{Schunk}
\begin{CodeChunk}
\begin{CodeInput}
R> summary(pROC_emp)
\end{CodeInput}
\begin{CodeOutput}
Call: [...]

Approach: Pooled ROC curve - Empirical
----------------------------------------------
Area under the pooled ROC curve: 0.76 (0.743, 0.778)*
Partial area under the pooled ROC curve (FPF = 0.1): 0.169 (0.14, 0.201)*
 * Confidence level:  0.95

Sample sizes:
                           Group H     Group D
Number of observations        2149         691
Number of missing data           0           0
\end{CodeOutput}
\end{CodeChunk}
Note that the posterior means for the AUC and pAUC obtained using the DPM method ($0.759$ and $0.168$, respectively) are very similar to the point estimates using the empirical approach ($0.760$ and $0.169$). This can also be observed when plotting the estimated ROC curves under the two methods (Figure \ref{pooled_ROC_dpm_emp}).

We finish this section by showing how to use \pkg{ROCnReg} to obtain an (optimal) threshold value which could be further used to `diagnose' an individual as diseased (two or more CVD risk factors) or nondiseased (none or only one CVD risk factor). To that aim, and for \code{pooledROC} objects (i.e., those obtained using functions \code{pooledROC.dpm}, \code{pooledROC.BB}, \code{pooledROC.emp}, and \code{pooledROC.kernel}), we use the function \code{compute.threshold.pooledROC}, which allows obtaining (optimal) threshold values using two criteria: the YI and the one that sets a target value for the FPF. For illustration, we show here the results using the YI criterion.
\begin{CodeChunk}
\begin{CodeInput}
R> th_pROC_dmp <- compute.threshold.pooledROC(pROC_dpm, criterion = "YI",
+    ci.level = 0.95, parallel = "snow", ncpus = 2)
R> th_pROC_dmp
\end{CodeInput}
\begin{CodeOutput}
$call
compute.threshold.pooledROC(object = pROC_dpm, criterion = "YI", 
    ci.level = 0.95, parallel = "snow", ncpus = 2)

$thresholds
     est       ql       qh 
26.46877 26.07129 26.85029 

$YI
      est        ql        qh 
0.4045776 0.3721684 0.4366298 

$FPF
      est        ql        qh 
0.3808336 0.3469580 0.4159478 

$TPF
      est        ql        qh 
0.7854112 0.7528575 0.8161865  	
\end{CodeOutput}
\end{CodeChunk}

The function returns the posterior mean (\code{est}) and \code{ci.level}$\times 100\%$ (here $95\%$ since \code{ci.level = 0.95}) credible interval (lower bound: \code{ql}, upper bound: \code{qh}) for the YI and associated threshold value, as well as for the FPF and TPF associated with this cutoff value. For our example, the (posterior mean of the) YI is $0.40$ and the YI-based threshold value is a BMI value of $26.5$, which falls in the nutritional status defined as pre-obesity by the World Health Organization. By using this YI-based threshold value, we would have a FPF of $0.38$ and a TPF of $0.79$.
\subsection{Covariate-specific ROC curve}\label{sec:croc_ilu}
We now turn our attention to the inclusion of covariates in ROC analysis. As shown in Table \ref{tab:overview} and Section \ref{sec:methods}, with \pkg{ROCnReg} the user can estimate the covariate-specific ROC curve by means of three approaches: function \code{cROC.sp} implements the frequentist parametric and semiparametric induced ROC regression estimator, \code{cROC.kernel} corresponds to the nonparametric, kernel-based, counterpart of \code{cROC.sp}, and \code{cROC.bnp} stands for the Bayesian approach based on a single-weights dependent Dirichlet process mixture of normal distributions. As for the functions in \pkg{ROCnReg} for estimating the pooled ROC curve, the input arguments are method-specific, and we refer the reader to the manual for details. For all methods, numerical and graphical summaries are obtained using functions \code{print.cROC}, \code{summary.cROC} and \code{plot.cROC}. Also, for objects of class \code{cROC.bnp}, \pkg{ROCnReg} provides the function \code{predictive.checks}, which implements tools for assessing model fit via posterior predictive checks.

Recall that, when including covariate information in ROC analysis, interest resides in evaluating if and how the discriminatory capacity of the test varies with such covariates. In particular, in our endocrine study we aim at evaluating the possible effect of both \code{age} and \code{gender} in the discriminatory capacity of the BMI. In what follows, we do that using the \code{cROC.bnp} function, and two different models are fitted. One which considers a normal distribution in each group and that incorporates the \code{age} effect in a linear way, and a second one which caps the maximum number of mixture components in each group at 10 (i.e., $L_{D}=L_{\bar{D}}=10$) and that models the \code{age} effect using cubic B-splines (and thus allows for a nonlinear effect of age). Following \cite{MX11a,MX11b}, both models consider the interaction between \code{age} and \code{gender}. For clarity, we first focus on the code that models \code{age} effect in a linear way, and use it to describe in detail the different arguments of the \code{cROC.bnp} function.

\begin{Schunk}
\begin{Sinput}
R> # Dataframe for predictions
R> agep <- seq(22, 80, l = 30)
R> endopred <- data.frame(age = rep(agep,2), 
+    gender = factor(rep(c("Women", "Men"), each = length(agep))))

R> set.seed(123, "L'Ecuyer-CMRG") # for reproducibility
R> cROC_bp <- cROC.bnp(formula.h = bmi ~ gender*age, 
+    formula.d = bmi ~ gender*age, group = "cvd_idf", tag.h = 0, 
+    data = endosyn, newdata = endopred,
+    standardise = TRUE, p = seq(0, 1, l = 101), ci.level = 0.95, 
+    compute.lpml = TRUE, compute.WAIC = TRUE, compute.DIC = TRUE, 
+    pauc = pauccontrol(compute = FALSE),
+    prior.h = priorcontrol.bnp(L = 1), prior.d = priorcontrol.bnp(L = 1),
+    density = densitycontrol(compute = TRUE),
+    mcmc = mcmccontrol(nsave = 8000, nburn = 2000, nskip = 1),
+    parallel = "snow", ncpus = 2)
\end{Sinput}
\end{Schunk}

As can be seen, many arguments coincide with those of the function \code{pooledROC.dpm} (described in Section \ref{sec:pooledroc_ilu}). We thus focus here on those that are specific to \code{cROC.bnp}. The arguments \code{formula.h} and \code{formula.d} are \code{formula} objects specifying the model for the regression functions (see Equation \eqref{mu_ddp}) in, respectively, the nondiseased and diseased classes. They are similar to the formula used with the \code{glm} function, except that nonlinear functions (modelled by means of cubic B-splines) can be added using function \code{f} (an example will follow later in this section). Note that in both cases, the left-hand side of the formulas should include the name of the test/marker (in our case \code{bmi}). In our application, and for both groups, the model for the component's means includes, in addition to the linear effect of \code{age} and \code{gender}, the (linear) interaction between these two covariates (i.e., \code{gender*age} $\equiv$ \code{gender + age + gender:age}). Through the \code{newdata} argument the user can specify a new data frame containing the values of the covariates at which the covariate-specific ROC curve and AUC (and also pAUC and PDFs, if required) are to be computed. Finally, \code{prior.h} (the same holds for \code{prior.d}) is an (optional) list of values to replace the defaults returned by \code{priorcontrol.bnp}  
\begin{Schunk}
\begin{Sinput}
priorcontrol.bnp(m0 = NA, S0 = NA, nu = NA, Psi = NA, a = 2, b = NA, 
    alpha = 1, L = 10)
\end{Sinput}
\end{Schunk}
which allows setting the hyperparameters for the single-weights dependent Dirichlet process mixture of normals model (see Section \ref{sec:methods} and the manual accompanying the package for more details). In our example, we only modified the upper bound for the number of components in the mixture model (by default $L = 10$) and set it to $1$. With this configuration, the model for the covariate-specific ROC curve can be regarded as a Bayesian counterpart of the induced ROC approach proposed by \cite{Faraggi03} and we denote it as the Bayesian normal linear model (for the test outcomes).

\begin{figure}[h!]
\begin{center}
\includegraphics[width=14cm]{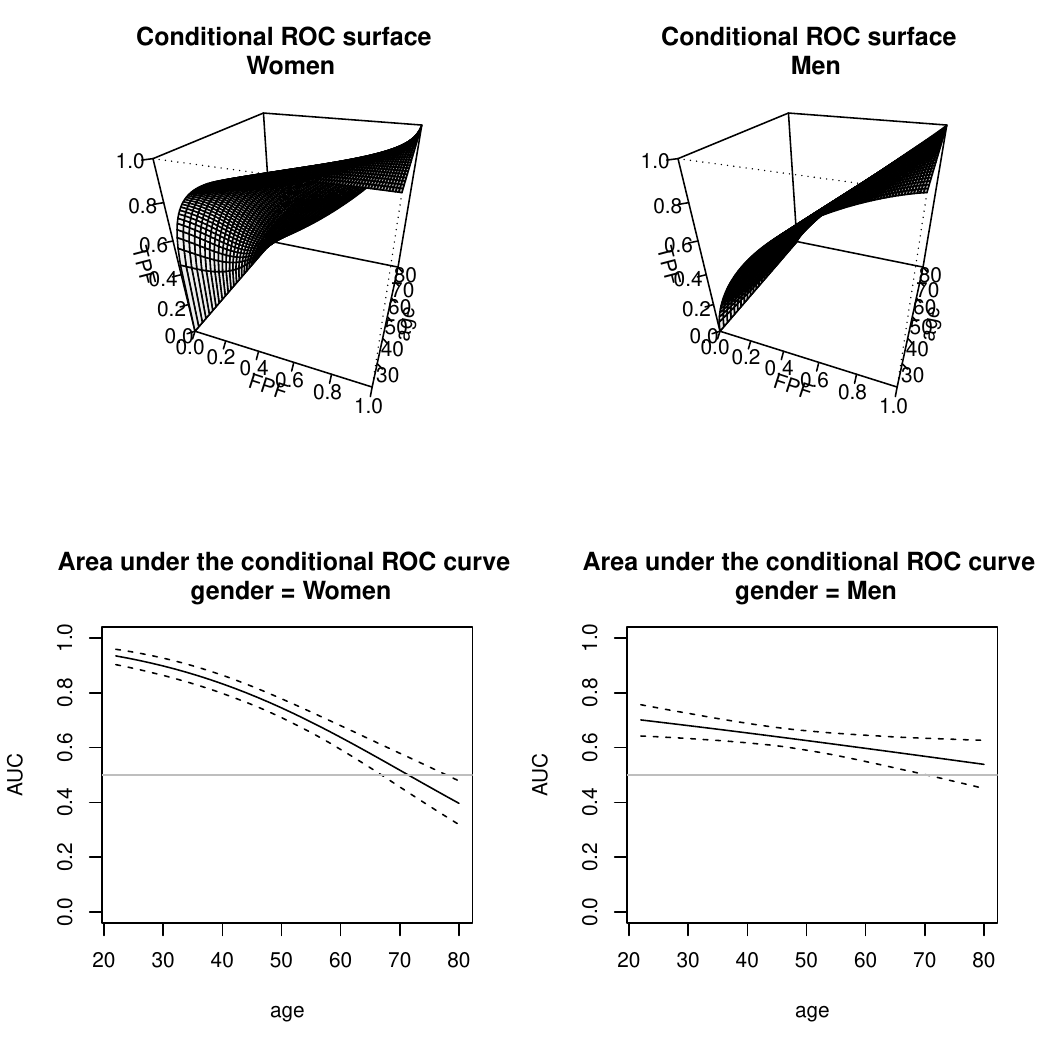}
\end{center}
\caption{Graphical results as provided by the \code{plot.cROC} function for an object of class \code{cROC.bnp}. Results for the model that includes the linear interaction between \code{age} and \code{gender} and one mixture component. Top row: Posterior mean of the covariate-specific ROC curve along age, separately for men and women. Bottom row: Posterior mean and 95\% pointwise credible band for the covariate-specific AUC along age, separately for men and women.}
\label{cROC_bp_plot}
\end{figure}

In this case, the \code{summary} of the fitted model provides the following information. %For the sake of page space, in what follows we omit from the summary the call to the function.
\begin{Schunk}
\begin{Sinput}
R> summary(cROC_bp)
\end{Sinput}
\begin{Soutput}
Call: [...]

Approach: Conditional ROC curve - Bayesian nonparametric
----------------------------------------------------------

Parametric coefficients
Group H:
                   Post. mean    Post. quantile 2.5%    Post. quantile 97.5%
(Intercept)           26.1459                25.8765                 26.4096
genderWomen           -0.9160                -1.2726                 -0.5680
age                    1.1949                 0.9180                  1.4690
genderWomen:age        1.1948                 0.8455                  1.5394


Group D:
                   Post. mean    Post. quantile 2.5%    Post. quantile 97.5%
(Intercept)           29.1865                28.7625                 29.6115
genderWomen            2.0826                 1.3705                  2.7665
age                    0.6578                 0.2162                  1.0904
genderWomen:age       -0.7711                -1.4655                 -0.0956


ROC curve:
                   Post. mean    Post. quantile 2.5%    Post. quantile 97.5%
(Intercept)           -0.6959                -0.8177                 -0.5776
genderWomen           -0.6863                -0.8695                 -0.5046
age                    0.1229                 0.0045                  0.2415
genderWomen:age        0.4499                 0.2745                  0.6245
b                      0.9391                 0.8824                  0.9975


Model selection criteria:
                     Group H       Group D
WAIC               12174.986      4007.980
WAIC (Penalty)         6.283         5.646
LPML               -6087.492     -2003.990
DIC                12173.664      4007.329
DIC (Penalty)          4.994         5.053


Sample sizes:
                           Group H     Group D
Number of observations        2149         691
Number of missing data           0           0
\end{Soutput}
\end{Schunk}
The first aspect to note is that, in this case, the \code{summary} function does not provide the estimated AUC as there is one (possibly different) AUC for each combination of covariate values. Also, given that: (1) only one component has been considered for modelling the CDFs of test results in the diseased and nodiseased groups, and (2) covariate effects have been modelled in a linear way, the \code{summary} function provides the posterior mean (and quantiles) of the (parametric) coefficients associated with the regression functions (see Equation \eqref{loc-scale-sp}) and with the covariate-specific ROC curve (see Equation \eqref{a-b-sp}). We note that since in the call to the function we have specified \code{standardise = TRUE} (and consequently both the test outcomes and covariates are standardised), the regression coefficients are on the scale of the standardised covariates. If we focus on the coefficients for the covariate-specific ROC curve, it seems that the discriminatory capacity of the BMI decreases with age, with the decrease being more pronounced in women (note that the expression of the covariate-specific ROC curve in \eqref{cROC-a-b-sp} implies that positive coefficients correspond to a decrease in discriminatory capacity). These results are possibly better judged by plotting the estimated covariate-specific ROC curves and associated AUCs. This can be done using the \code{plot} function. For the covariate-specific ROC curve, the depicted graphics will depend on the number and nature of the covariates included in the analyses. In particular, for our application, we obtain, separately for men and women, the covariate-specific ROC curves (and AUCs) along age. These are shown in Figure \ref{cROC_bp_plot}, obtained using the code
\begin{Schunk}
\begin{Sinput}
R> op <- par(mfrow = c(2,2))
R> plot(cROC_sp, ask = FALSE)
R> par(op)
\end{Sinput}
\end{Schunk}

Although in this example we have modelled the \code{age} effect linearly and only one mixture component was considered, \pkg{ROCnReg} also allows for modelling the effect of continuous covariates in a nonlinear way, either using cubic B-spline basis expansions (through the function \code{cROC.bnp}) or  kernel-based smoothers (via the function \code{cROC.kernel}). Also, as noted before, using only one mixture component for the single-weights dependent Dirichlet process mixture of normals model (function \code{cROC.bnp}) is equivalent to consider a (Bayesian) normal model, which might be too restrictive for most data applications. In what follows, we provide more flexibility to the model for the covariate-specific ROC curve by means of (1) increasing the number of mixture components, and (2) modelling the \code{age} effect in a nonlinear way (recall our considerations in Section \nameref{sec:cROC} about the lack of flexibility of the single-weights dependent Dirichlet process mixture of normals model when covariates effects on the components' means are modelled linearly). The former is done by modifying the value of \code{L} in the arguments \code{prior.h} and \code{prior.d}, with $10$ being the default value. Regarding the latter, this is done by making use of the function \code{f} when specifying the component's mean functions through \code{formula.h} and \code{formula.d}. In particular, in our application we are interested in modelling the factor-by-curve interaction between \code{age} and \code{gender} (i.e., we model the \code{age} effect `separately' for men and women). This is done using, e.g., \code{bmi \~\ gender + f(age, by = gender, K = c(3,5))}. Through the argument \code{K} we indicate the number of internal knots that are used for constructing the cubic B-spline basis that is used to approximate the nonlinear effect of \code{age} (with the quantiles of \code{age} used to anchor the knots). Note that we can specify a different number of internal knots for men and women (\code{K = c(3,5)}), where the order of vector \code{K} should match the ordering of the levels of the factor \code{gender}. We also note that to assist in the selection of the number of interior knots (in \pkg{ROCnReg} the location is always based on the quantiles of the corresponding covariates), the user can make use of the WAIC, DIC, and/or LPML. For instance, for this application, we fitted different models with different number of internal knots and we have chosen the model that provided the lowest WAIC (this was done in both the nondiseased and diseased populations and we remark that the number of knots does not need to be the same in the two populations). The final model is shown below.

\begin{Schunk}
\begin{Sinput}
R> # Levels of gender, and its ordering. 
R> # Needed if we want to specify different
R> # number of knots for men and women
R> levels(endosyn$gender)
\end{Sinput}
\begin{Soutput}
[1] "Men"   "Women"
\end{Soutput}
\begin{Sinput}
R> set.seed(123, "L'Ecuyer-CMRG") # for reproducibility
R> cROC_bnp <- cROC.bnp(
+    formula.h = bmi ~ gender + f(age, by = gender, K = c(0,0))
+    formula.d = bmi ~ gender + f(age, by = gender, K = c(4,4)),
+    group = "cvd_idf", tag.h = 0, data = endosyn, newdata = endopred,
+    standardise = TRUE, p = seq(0, 1, l = 101), ci.level = 0.95, 
+    compute.lpml = TRUE, compute.WAIC = TRUE, compute.DIC = TRUE, 
+    pauc = pauccontrol(compute = FALSE),
+    prior.h = priorcontrol.bnp(L = 10), prior.d = priorcontrol.bnp(L = 10),
+    density = densitycontrol(compute = TRUE),
+    mcmc = mcmccontrol(nsave = 8000, nburn = 2000, nskip = 1),
+    parallel = "snow", ncpus = 2)

R> summary(cROC_bnp)
\end{Sinput}
\begin{Soutput}
Call: [...]

Approach: Conditional ROC curve - Bayesian nonparametric
----------------------------------------------------------

Model selection criteria:
                     Group H       Group D
WAIC               11833.000      3909.828
WAIC (Penalty)        31.236        38.583
LPML               -5916.766     -1955.449
DIC                11829.750      3904.532
DIC (Penalty)         29.611        35.934


Sample sizes:
                           Group H     Group D
Number of observations        2149         691
Number of missing data           0           0
\end{Soutput}
\end{Schunk}
\begin{Schunk}
\begin{Sinput}
R> op <- par(mfrow = c(2,2))
R> plot(cROC_sp, ask = FALSE)
R> par(op)
\end{Sinput}
\end{Schunk}
\begin{figure}[h!]
\begin{center}
\includegraphics[width=14cm]{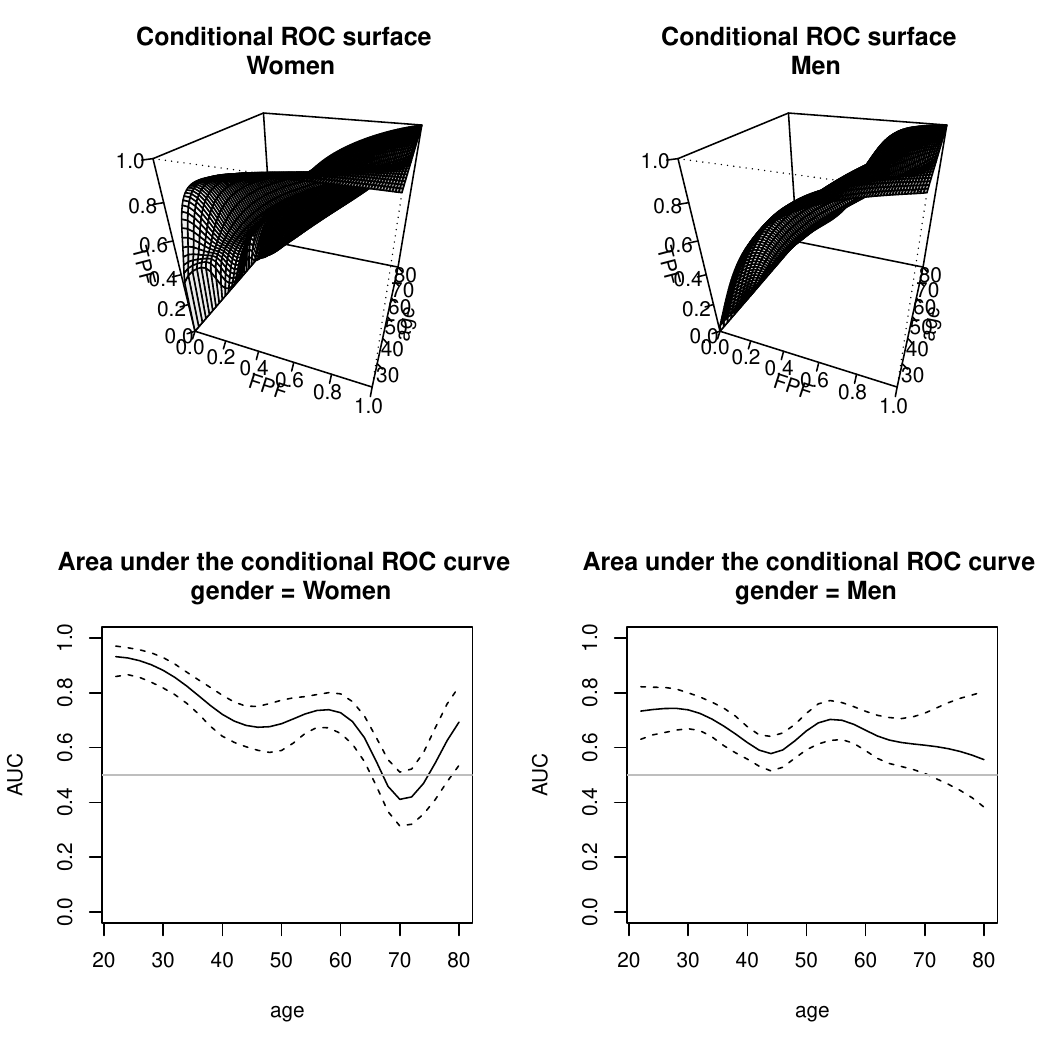}
\end{center}
\caption{Graphical results as provided by the \code{plot.cROC} function for an object of class \code{cROC.bnp}. Results for the model that includes the factor-by-curve interaction between \code{age} and \code{gender} and 10 mixture components. Top row: Posterior mean of the covariate-specific ROC curve along age, separately for men and women. Bottom row: Posterior mean and 95\% pointwise credible band for the covariate-specific AUC along age, separately for men and women.}
\label{cROC_bnp_plot}
\end{figure}

%The graphical results are shown in Figure \ref{cROC_bnp_plot}. Note that, especially for women, age displays a marked nonlinear effect. Recall that for objects of class \code{cROC.bnp}, and if required in the call to the function, the \code{summary} function provides, separately for the diseased and nondiseased/healthy groups, the WAIC, LPML, and DIC. Note that, in both cases, the three criteria support the use of the more flexible model that uses B-splines and 10 mixture components for modelling the distribution of the BMI (model \code{cROC_bnp}) over the more restrictive Bayesian normal linear model (model \code{cROC_bp}). Since WAIC, LPML, and DIC are relative criteria, posterior predictive checks are also available in \pkg{ROCnReg} through the function \code{predictive.checks}. Specifically, the function generates replicated datasets from the posterior predictive distribution in both class $D$ and $\bar{D}$ and compares them to the test values (BMI values in our application) using specific statistics. For the choice of such statistics we follow \cite{Gabry2019}, who suggest choosing statistics that are `orthogonal' to the model parameters. Since we are using a location-scale mixture of normals model for the test outcomes, we use here the skewness and kurtosis and check how well the posterior predictive distribution captures these two quantities.

\begin{figure}[h!]
\begin{center}
\subfigure[Model including the linear interaction between \code{age} and \code{gender} and one component]{\includegraphics[width=15cm]{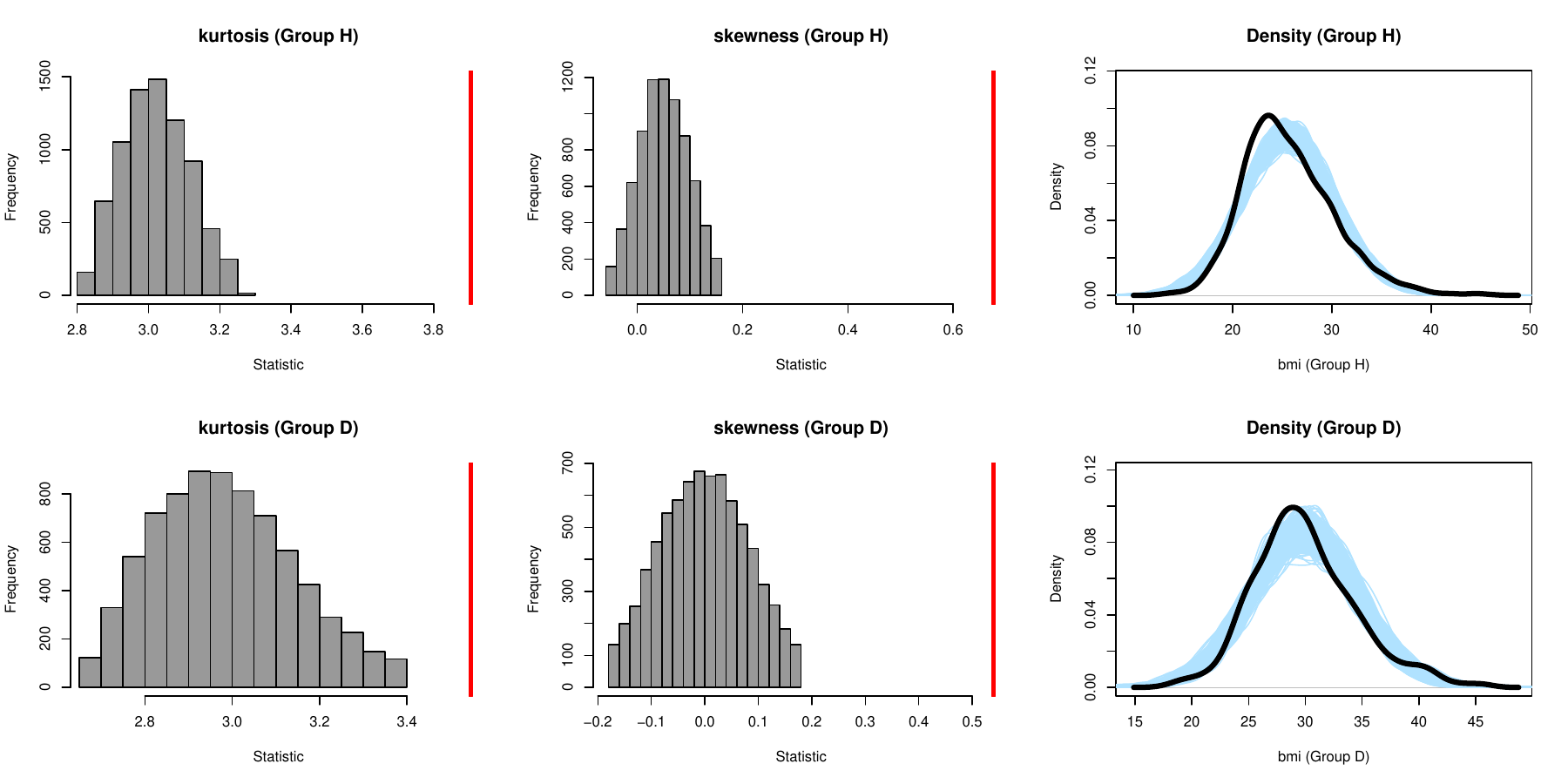}}
\subfigure[Model including the factor-by-curve interaction between \code{age} and \code{gender} and 10 mixture components]{\includegraphics[width=15cm]{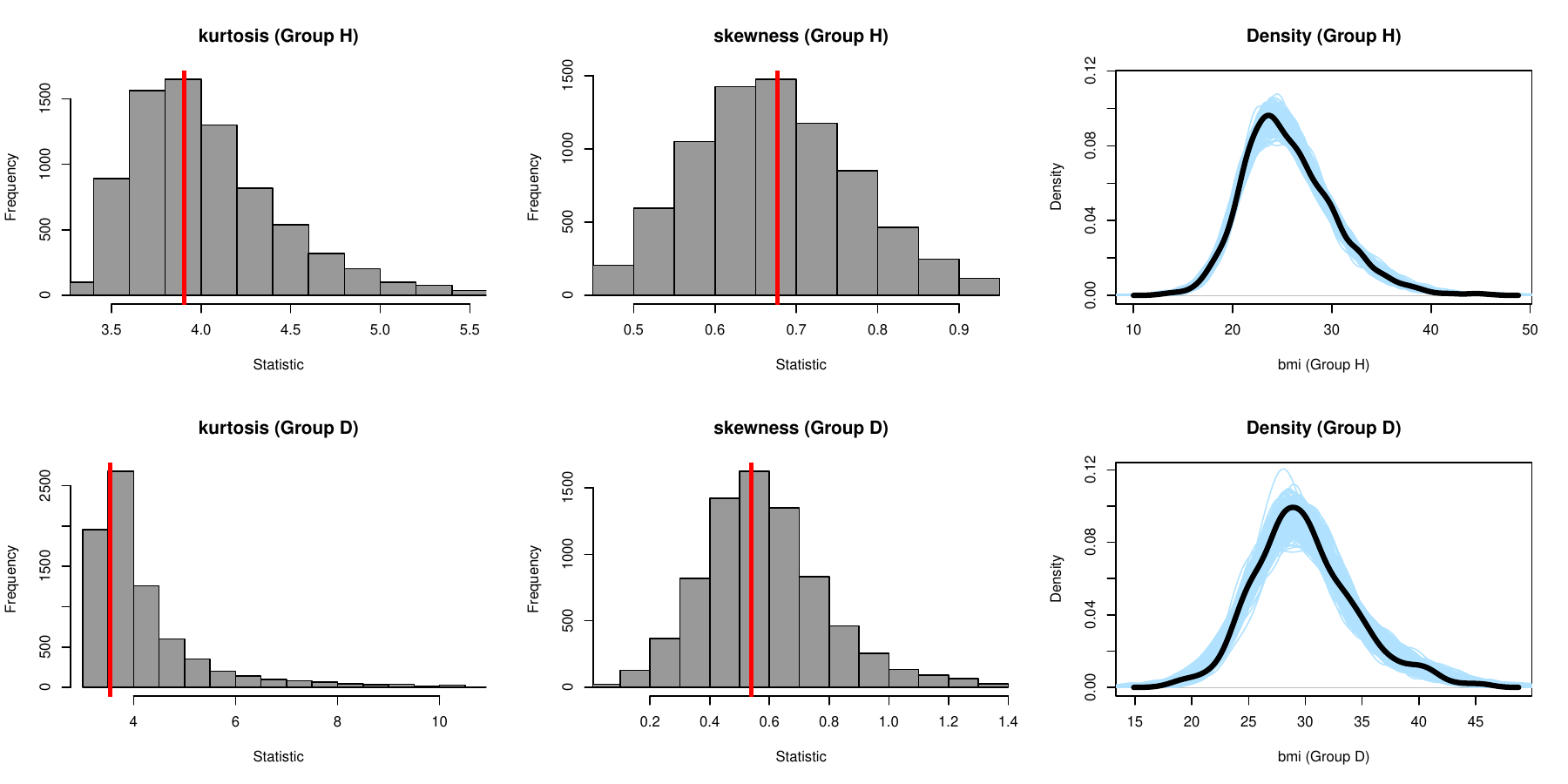}}
\end{center}
\caption{Graphical results as provided by the \code{predictive.checks} function for an object of class \code{cROC.bnp}. Histograms of the statistics \textit{skewness} and \textit{kurtosis} computed from $8000$ draws from the posterior predictive distribution in the diseased and nondiseased group. The red line is the estimated statistic from the observed BMI values. The right-hand side plots show the kernel density estimate of the observed BMI (solid black line), jointly with the kernel density estimates for $500$ simulated datasets drawn from the posterior predictive distributions.}
\label{cROC_bnp_pred_checks}
\end{figure}

The graphical results are shown in Figure \ref{cROC_bnp_plot}. Note that, especially for women, age displays a marked nonlinear effect. Recall that for objects of class \code{cROC.bnp}, and if required in the call to the function, the \code{summary} function provides, separately for the diseased and nondiseased groups, the WAIC, LPML, and DIC. Note that, in both cases, the three criteria support the use of the more flexible model that uses cubic B-splines and 10 mixture components for modelling the distribution of the BMI (model \code{cROC$\_$bnp}) over the more restrictive Bayesian normal linear model (model \code{cROC$\_$bp}). Because the WAIC, LPML, and DIC are relative criteria, posterior predictive checks are also available in \pkg{ROCnReg} through the function \code{predictive.checks}. Specifically, the function generates replicated datasets from the posterior predictive distribution in both class $D$ and $\bar{D}$ and compares them to the test values (BMI values in our application) using specific statistics. For the choice of such statistics we follow \cite{Gabry2019}, who suggest choosing statistics that are `orthogonal' to the model parameters. Since we are using a location-scale mixture of normals model for the test outcomes, we use here the skewness and kurtosis and check how well the posterior predictive distribution captures these two quantities.
\begin{Schunk}
\begin{Sinput}
R> op <- par(mfrow = c(2,3))
R> pc_cROC_bp <- predictive.checks(cROC_bp, 
+    statistics = c("kurtosis", "skewness"), devnew = FALSE)
R> par(op)

R> op <- par(mfrow = c(2,3))
R> pc_cROC_bnp <- predictive.checks(cROC_bnp, 
+    statistics = c("kurtosis", "skewness"), devnew = FALSE)
R> par(op)
\end{Sinput}
\end{Schunk}
Results are shown in Figure \ref{cROC_bnp_pred_checks}. As can be seen, the model that includes the factor-by-curve interaction between \code{age} and \code{gender} and 10 mixture components performs quite well in capturing both quantities, while the Bayesian normal linear model fails to do so. Also shown in Figure \ref{cROC_bnp_pred_checks} (and provided by function \code{predictive.checks}) are the kernel density estimates of $500$ randomly selected datasets drawn from the posterior predictive distribution, in each group, compared to the kernel density estimate of the observed BMI (in each group). Again, the more flexible model, as opposed to the Bayesian normal linear model, is able to simulate data that are very much similar to the observed BMI values.

As for the pooled ROC curve (Section \ref{sec:pooledroc_ilu}), \pkg{ROCnReg} also provides a function that allows obtaining (optimal) threshold values for the covariate-specific ROC curve. For illustration, instead of the threshold values based on the Youden index, we now use the criterion that sets a target value for the FPF. The code for model \code{cROC_bnp}, when setting the $\text{FPF} = 0.3$, is as follows. 
\begin{Schunk}
\begin{Sinput}
R> th_fpf_cROC_bnp <- compute.threshold.cROC(cROC_bnp, 
+    criterion = "FPF", FPF = 0.3, newdata = endopred,
+    ci.level = 0.95, parallel = "snow", ncpus = 2)
R> names(th_fpf_cROC_bnp)
\end{Sinput}
\begin{Soutput}
[1] "newdata"    "thresholds" "TPF"    "FPF"    "call"
\end{Soutput}
\end{Schunk}
%In addition to the data frame \code{newdata} containing the covariate values at which the thresholds are computed, function \code{compute.threshold.cROC} returns the covariate-specific \code{thresholds} corresponding to the specified FPF as well as the covariate-specific \code{TPF} attached to these thresholds. In both cases, the function returns the posterior mean and \code{ci.level}$\times 100\%$ (here $95\%$) pointwise credible intervals. Although \pkg{ROCnReg} does not provide a function for plotting the results obtained using \code{compute.threshold.cROC}, graphical results can be easily obtained. For simplicity, we only show here the code for the covariate-specific threshold values (\code{thresholds}), but a similar code can be used to plot the covariate-specific TPFs (\code{TPF}). Both plots are shown in Figure \ref{cROC_bnp_th}. As can be observed, for a FPF of $0.3$, the BMI age-specific thresholds tend to increase with age both for men and women, although for the latter there is a slight decrease after an age of about 70 years old. The age-specific TPFs corresponding to the thresholds for which the FPF is $0.3$ show a nonlinear behaviour and these are in general higher for women than for men (of the same age).

\begin{figure}[h!]
\begin{center}
\subfigure[BMI threshold values]{\includegraphics[width=12cm]{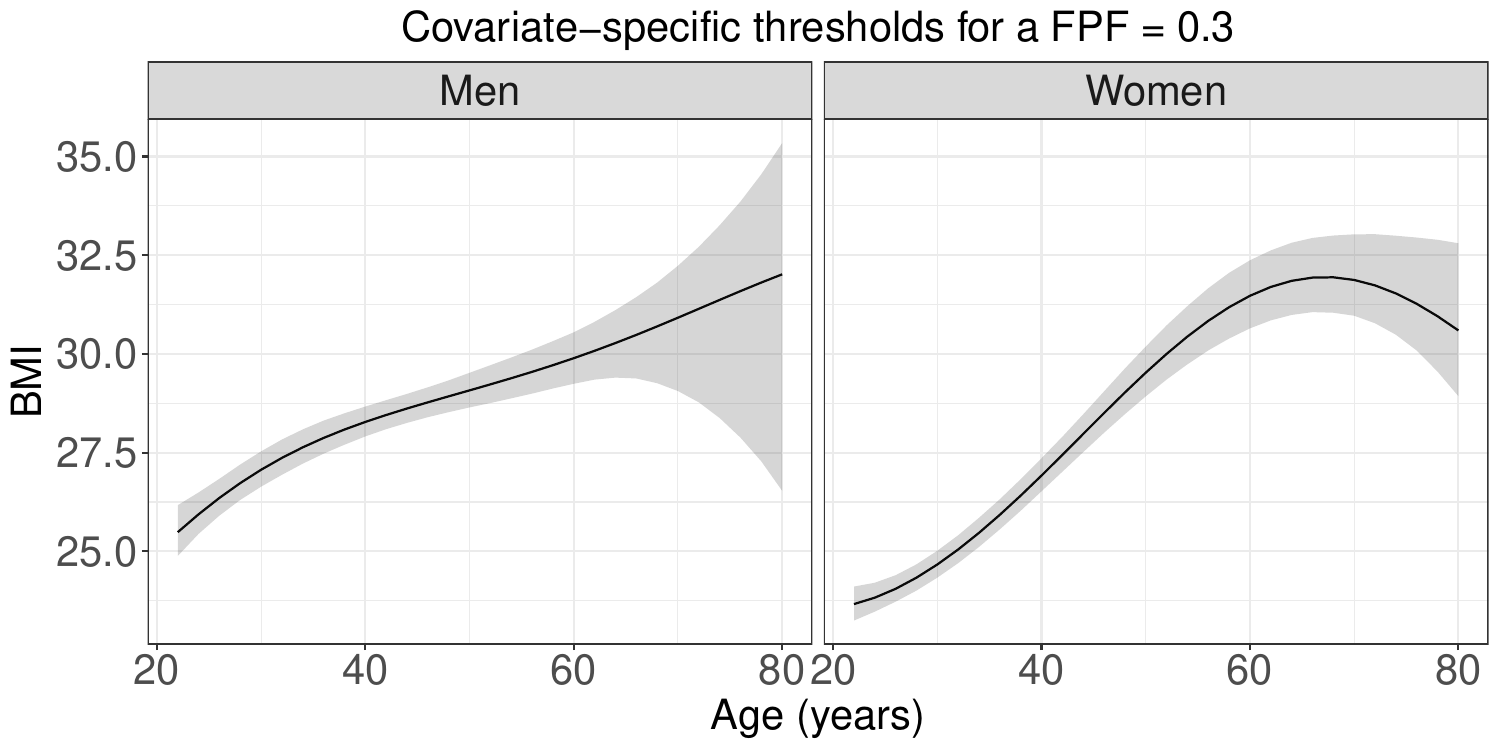}}
\subfigure[True positive fractions]{\includegraphics[width=12cm]{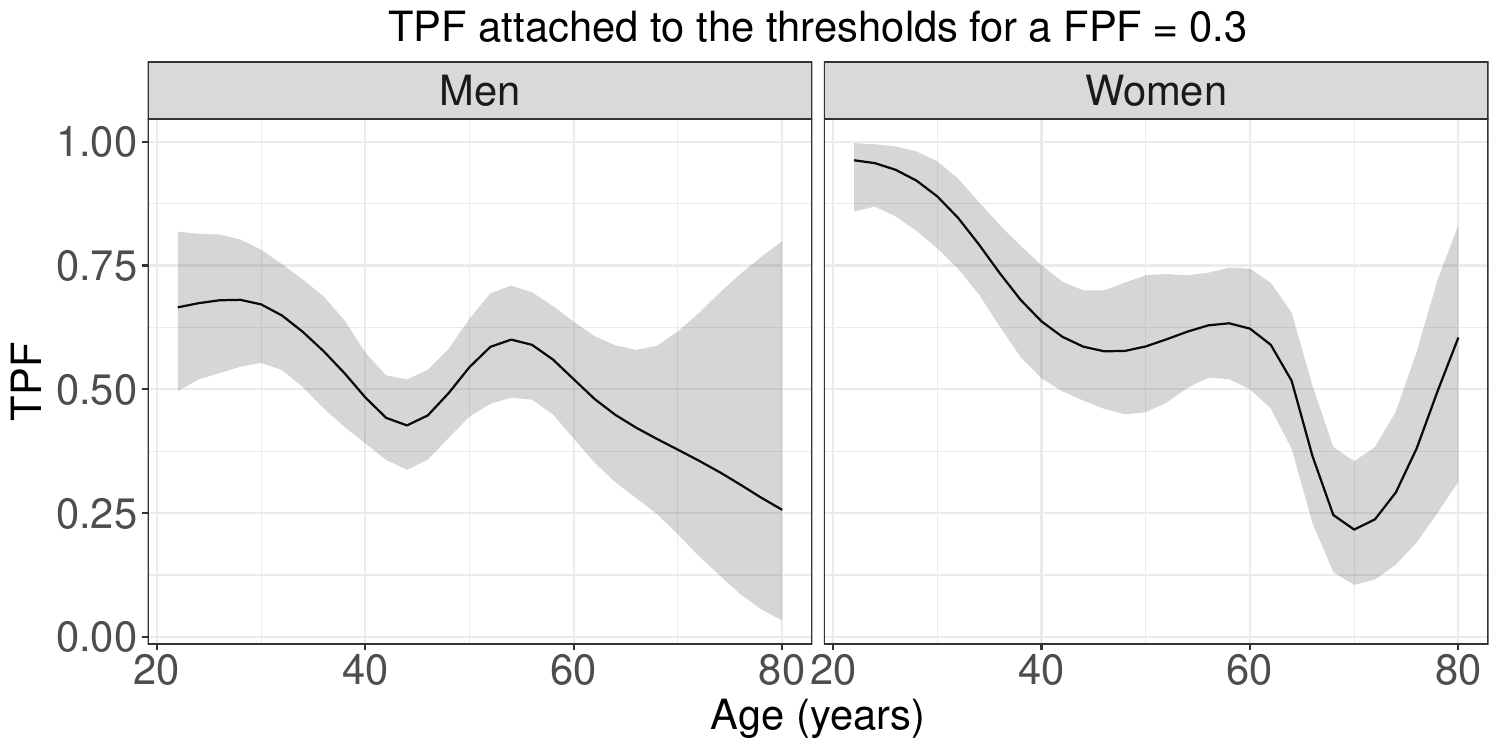}}
\end{center}
\caption{Top row: Posterior mean (solid black line) and 95\% pointwise credible band for the BMI threshold values, along age, corresponding to a FPF of 0.3. Bottom row: Posterior mean (solid black line) and 95\% pointwise credible band of the TPFs, along age, corresponding to the BMI threshold values for which $\text{FPF}=0.3$.}
\label{cROC_bnp_th}
\end{figure}

In addition to the data frame \code{newdata} containing the covariate values at which the thresholds are computed, the function \code{compute.threshold.cROC} also returns the covariate-specific \code{thresholds} corresponding to the specified FPF as well as the covariate-specific \code{TPF} attached to these thresholds. In both cases, the function returns the posterior mean and the \code{ci.level}$\times 100\%$ (here $95\%$) pointwise credible intervals. Although \pkg{ROCnReg} does not provide a function for plotting the results obtained using \code{compute.threshold.cROC}, graphical results can be easily obtained. For simplicity, we only show here the code for the covariate-specific threshold values (\code{thresholds}), but a similar code can be used to plot the covariate-specific TPFs. Both plots are shown in Figure \ref{cROC_bnp_th}. As can be observed, for a FPF of $0.3$, the BMI age-specific thresholds tend to increase with age both for men and women, although for the latter there is a slight decrease after an age of about 70 years old. The age-specific TPFs corresponding to the thresholds for which the FPF is $0.3$ show a nonlinear behaviour and these are in general higher for women than for men (of the same age).

\begin{Schunk}
\begin{Sinput}
df <- data.frame(age = th_fpf_cROC_bnp$newdata$age,
+    gender = th_fpf_cROC_bnp$newdata$gender,  
+    y = th_fpf_cROC_bnp$thresholds[[1]][,"est"], 
+    ql = th_fpf_cROC_bnp$thresholds[[1]][,"ql"],  
+    qh = th_fpf_cROC_bnp$thresholds[[1]][,"qh"])
          
R> g0 <- ggplot(df, aes(x = age, y = y, ymin = ql, ymax = qh)) +
+    geom_line() + 
+    geom_ribbon(alpha = 0.2) + 
+    labs(title = "Covariate-specific thresholds for a FPF = 0.3", 
+    x = "Age (years)", y = "BMI") + 
+    theme(strip.text.x = element_text(size = 20), 
+    plot.title = element_text(hjust = 0.5, size = 20), 
+    axis.text = element_text(size = 20), 
+    axis.title = element_text(size = 20)) +
+    facet_wrap(~gender)
R> print(g0)
\end{Sinput}
\end{Schunk}

%Finally, we mention that for conciseness we have not shown here how to perform convergence diagnostics of the MCMC chains for models fitted using \code{cROC.bnp}. In very much the same way as shown in Section \ref{sec:pooledroc_ilu} for the object \code{pROC_dpm}, using the information contained in component \code{dens} in the list of returned values (if required), one can produce trace plots of the conditional densities at some sampled values, as well as, obtain the corresponding effective sample sizes. Some results are provided in Appendix \ref{AppB}, and the associated code can be found in the \proglang{R} replication code that accompanies this paper.

For conciseness we have not shown here how to perform convergence diagnostics of the MCMC chains for models fitted using the function \code{cROC.bnp}. In very much the same way as shown in the previous section for the object \code{pROC_dpm}, using the information contained in component \code{dens} in the list of returned values (if required), one can produce trace plots of the conditional densities at some sampled values, as well as obtain the corresponding effective sample sizes and Geweke statistics. Some results are provided in the Appendices, and the associated code can be found in the replication code that accompanies this paper.
\subsection{Covariate-adjusted ROC curve}\label{sec:aroc_ilu}
In this section we illustrate how to conduct inference about the covariate-adjusted ROC curve using  \pkg{ROCnReg}. Similarly to the covariate-specific ROC curve, three approaches are available, namely, function \code{AROC.sp} implements the frequentist approaches that postulate that test outcomes in the nondiseased group follow a linear model with the CDF of the error term being either a standard normal distribution or estimated via the empirical CDF of the standardised residuals, \code{AROC.kernel} corresponds to the kernel-based counterpart, and \code{AROC.bnp} implements the Bayesian (nonparametric) approach based on a single-weights dependent Dirichlet process mixture of normal distributions and the Bayesian bootstrap.

\begin{figure}[h!]
\begin{center}
\subfigure[AROC curve]{\includegraphics[width=7.5cm]{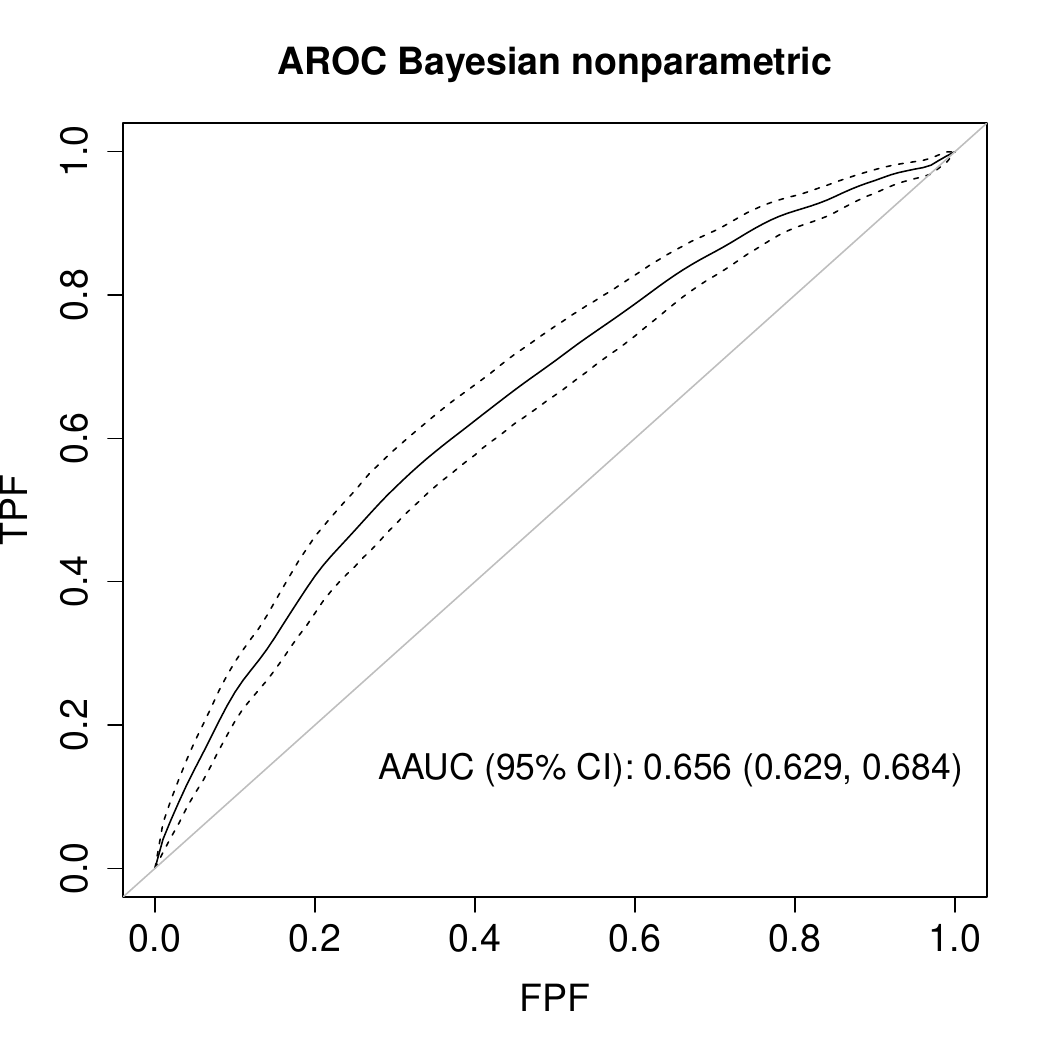}\label{AROC_bnp}}
\subfigure[AROC vs Poooled ROC curve]{\includegraphics[width=7.5cm]{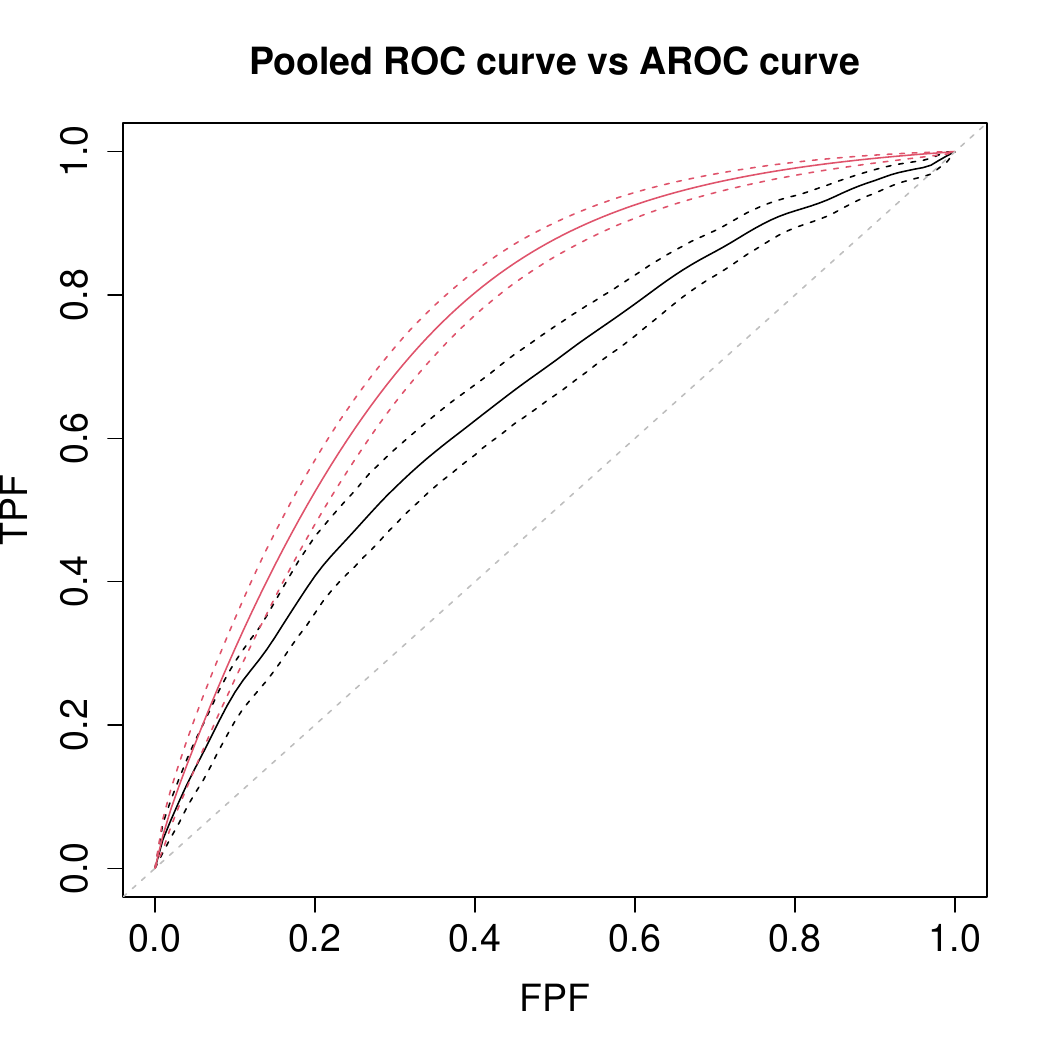}\label{AROC_bnp_pooled}}
\end{center}
\caption{(a) Age and gender-adjusted ROC curve: posterior mean and 95\% pointwise credible band. (b) Age and gender-adjusted ROC curve (in black) and pooled ROC curve (estimated using a DPM of normals model) (in red). Solid lines represent the posterior means and dashed lines the  95\% pointwise credible bands.}
\label{AROC_bnp_complete}
\end{figure}

Recall that the AROC curve is a global summary measure of diagnostic accuracy that takes covariate information into account. In the context of our endocrine application we seek to study the overall discriminatory capacity of the BMI for detecting the presence of CVD risk factors when adjusting for age and gender. Here we focus on how to estimate the AROC curve using the \code{AROC.bnp} function. The function syntax is exactly similar to the one of \code{cROC.bnp}, with the only difference being that we only need to specify the arguments related to the nondiseased population. The code and respective summary follow.
\begin{Schunk}
\begin{Sinput}
R> set.seed(123, "L'Ecuyer-CMRG") # for reproducibility
R> AROC_bnp <- AROC.bnp(
+    formula.h = bmi ~ gender + f(age, by = gender, K = c(0,0)), 
+    group = "cvd_idf", tag.h = 0, data = endosyn, standardise = TRUE, 
+    p = seq(0, 1, l = 101), ci.level = 0.05, 
+    compute.lpml = TRUE, compute.WAIC = TRUE, compute.DIC = TRUE, 
+    pauc = pauccontrol(compute = FALSE), prior = priorcontrol.bnp(L = 10),
+    density = densitycontrol.aroc(compute = FALSE),
+    mcmc = mcmccontrol(nsave = 8000, nburn = 2000, nskip = 1),
+    parallel = "snow", ncpus = 2)

R> summary(AROC_bnp)
\end{Sinput}
\begin{Soutput}
Call: [...]

Approach: AROC Bayesian nonparametric
----------------------------------------------
Area under the covariate-adjusted ROC curve: 0.656 (0.629, 0.684)*
 * Credible level:  0.95

Model selection criteria:
                     Group H
WAIC               11833.000
WAIC (Penalty)        31.236
LPML               -5916.766
DIC                11829.750
DIC (Penalty)         29.611


Sample sizes:
                           Group H     Group D
Number of observations        2149         691
Number of missing data           0           0
\end{Soutput}
\end{Schunk}
The area under the AROC curve is $0.656$ (95\% credible interval: $(0.629, 0.684)$) thus revealing a reasonable good ability of the BMI to detect the presence of CVD risk factors when teasing out the age and gender effects. As for the pooled ROC curve and the covariate-specific ROC curve, a \code{plot} function is also available (result in Figure \ref{AROC_bnp}).
\begin{Schunk}
\begin{Sinput}
R> plot(AROC_bnp, cex.main = 1.5, cex.lab = 1.5, cex.axis = 1.5, cex = 1.3) 
\end{Sinput}
\end{Schunk}
We finish with a comparison of the AROC curve with the pooled ROC curve that was obtained earlier by using a DPM model with 10 components in each group. In Figure \ref{AROC_bnp_pooled} we show the plots of the two curves and, as can be noticed, the pooled ROC curve lies well above the AROC curve, thus evidencing the need for incorporating covariate information into the analysis.
\begin{Schunk}
\begin{Sinput}
R> plot(AROC_bnp$p, AROC_bnp$ROC[,1], 
+    type = "l", xlim = c(0,1), ylim = c(0,1),
+    xlab = "FPF", ylab = "TPF", 
+    main = "Pooled ROC curve vs AROC curve",
+    cex.main = 1.5, cex.lab = 1.5, cex.axis = 1.5, cex = 1.5)
R> lines(AROC_bnp$p, AROC_bnp$ROC[,2], col = 1, lty = 2)
R> lines(AROC_bnp$p, AROC_bnp$ROC[,3], col = 1, lty = 2)
R> lines(pROC_dpm$p, pROC_dpm$ROC[,1], col = 2)
R> lines(pROC_dpm$p, pROC_dpm$ROC[,2], col = 2, lty = 2)
R> lines(pROC_dpm$p, pROC_dpm$ROC[,3], col = 2, lty = 2)
R> abline(0, 1, col = "grey", lty = 2)
\end{Sinput}
\end{Schunk}

%We end remarking that in the Supplementary Materials accompanying this paper we show the usage of the package for those methods not described in the main manuscript.
\subsection{Computational aspects}\label{sec:comp_times}
We finish this section with some comments on computational aspects. In our experience, the methods with the largest computing times are those implemented in \code{cROC.bnp} when $L_{\bar{D}} > 1$ and in \code{cROC.kernel} when confidence bands are to be constructed. In the first case, the main reason behind the computational burden is the need to invert $F_{\bar{D}}\left(\cdot \mid \mathbf{x}\right)$ in order to obtain the covariate-specific ROC curve (see equation (\ref{ROCConditional})). Note that when $L_{\bar{D}}>1$, the  conditional distribution function in the nondiseased group is given by a mixture of normal distributions and the corresponding quantile function needs to be computed for each covariate(s) value we might be interested in and for each iteration of the Gibbs sampler procedure. Regarding the \texttt{cROC.kernel} function, the computing time is mainly driven by the number of bootstrap samples used for constructing the confidence bands. In Table~\ref{tab:computingtimes} we show the time, in seconds, needed for fitting the pooled, the covariate-specific, and the covariate-adjusted ROC curve using the Bayesian nonparametric and the kernel approaches for the synthetic endocrine data and when both parallel (with 2 and 4 processes) and no parallel options are used. We note that for the Bayesian approaches we also computed the densities/conditional densities, as well as the WAIC, LPML, and DIC, which further increase the computing time (in the case of the AROC curve these were only computed in the nondiseased population). With respect to the kernel-based approach (in this case the fit is done separately for men and women and the corresponding results are presented in the Supplementary Material) we have used $500$ bootstrap samples to construct the confidence bands. As it can be appreciated, for these two intense tasks, using 4 processes drastically improves the computation time. All computations were performed in a iMac with 3.6GHz quad core Intel i7 processor and 32GB RAM running under a macOS Catalina 10.15.5 operating system.

\begin{table}[ht!]
\centering
\begin{tabular}{ccccc}
\hline
& No parallel & Snow (2 cores) & Snow (4 cores) \\ \hline
\code{pooledROC.dpm} & 138 & 118& 111\\ \hline
\code{pooledROC.kernel} & 376& 196& 105 \\ \hline
\code{cROC.bnp} & 2052& 1117& 680 \\ \hline
\multirow{2}*{\code{cROC.kernel}} & Men: 1159 &  Men: 528&   Men: 279 \\ 
& Women: 1885 & Women: 916 & Women: 466\\ \hline
\code{AROC.bnp} & 126& 115& 112 \\ \hline
\multirow{2}*{\code{AROC.kernel}} & Men: 847 & Men: 404& Men: 214  \\
& Women: 1707 & Women: 833 & Women: 438\\\hline
\end{tabular}
\caption{\label{tab:computingtimes} Time in seconds (rounded to the nearest second) needed to fit the pooled, the covariate-specific, and the covariate-adjusted ROC curve for the Bayesian nonparametric and kernel approaches.}
\end{table}
\section{Summary and future plans} \label{sec:summary}
In this paper we have introduced the capabilities of the \proglang{R} package \pkg{ROCnReg} for conducting inference about the pooled ROC curve, the covariate-specific ROC curve, and the covariate-adjusted ROC curve and their associated summary indices. As we have illustrated, the current version of the package provides several options for estimating ROC curves, both under frequentist and Bayesian paradigms, either parametrically, semiparametrically, or nonparametrically. To the best of our knowledge, this is the first software package implementing Bayesian inference for ROC curves. Several additions/extensions are planned in the future and these, among others, include:
\begin{itemize}
\item Implement the most time-consuming parts in \proglang{C} or \proglang{C++}.
\item Incorporate methods for non-binary disease status (e.g., no disease, mild disease, severe disease). That is, implement ROC surface models.
\item Implement new (optimal) threshold criteria (e.g., YI including costs).
\end{itemize}
\section*{Computational details}
The results in this paper were obtained using \proglang{R}~4.0.3 with the \pkg{ROCnReg}~1.0-5 package. The \pkg{ROCnReg} package has multiple dependencies: \pkg{graphics}, \pkg{grDevices}, \pkg{parallel}, \pkg{splines}, \pkg{stats}, \pkg{moments} \citep{moments_package}, \pkg{nor1mix} \citep{nor1mix_package}, \pkg{Matrix} \citep{Matrix_package}, \pkg{spatstat} \citep{spatstat_package}, \pkg{np} \citep{Hayfield08}, \pkg{lattice} \citep{lattice_package}, \pkg{MASS} \citep{MASS_package} and \pkg{pbivnorm} \citep{pbivnorm_package}. \proglang{R} itself and all packages used are available from the Comprehensive \proglang{R} Archive Network (CRAN) at \url{https://CRAN.R-project.org/}.
\section*{Acknowledgments}
MX Rodr\'iguez-\'Alvarez was funded by project MTM2017-82379-R (AEI/FEDER, UE), by the Basque Government through the BERC 2018-2021 program and Elkartek project 3KIA (KK-2020/00049), and by the Spanish Ministry of Science, Innovation, and Universities (BCAM Severo Ochoa accreditation SEV-2017-0718). %The work of V In\'acio was partially supported by FCT (Funda\c c\~ao para a Ci\^encia e a Tecnologia, Portugal), through the projects PTDC/MAT-STA/28649/2017 and \linebreak UID/MAT/00006/2020. 

\bibliography{refs}

\newpage
\begin{appendix}
\section{Further computational tools for pooled ROC curve}\label{AppA}	
\subsection{Informal model diagnostics for the Bayesian methods: quantile residuals}
We illustrate the use of quantile residuals \citep{Dunn96} which can be helpful for models fitted using the function \code{pooledROC.dpm}, for instance, in deciding if $L_{D}=1$ or $L_{D}>1$ (the same obviously applies to the nondiseased group). Quantile residuals are based on the well-known fact that for a continuous random variable, say $Y_D$, with CDF given by $F_D$, one has that $F_D(Y_D)\sim\text{U}(0,1)$. As a consequence, quantile residuals defined by $\widehat{r}_{Dj}=\Phi^{-1}\{\widehat{F}_{D}(y_{Dj})\}$, for $j=1,\ldots,n_D$, should follow, approximately, a standard normal distribution if a correct model has been specified. A quantile-quantile (QQ) plot can then be used to determine deviations of the quantile residuals from the standard normal distribution. Below we provide the code to construct, for the diseased population, such QQ plot using output from the object \code{pROC\_dpm} (that assumed $L_{D}=L_{\bar{D}}=10$) obtained using the function \code{pooledROC.dpm}. The code for the nondiseased population follows in a similar manner, and is provided in the \proglang{R} replication code that accompanies this paper.

\begin{figure}[h!]
\begin{center}	
\subfigure[DPM model with 10 mixture components in each group]{\includegraphics[width=7cm]{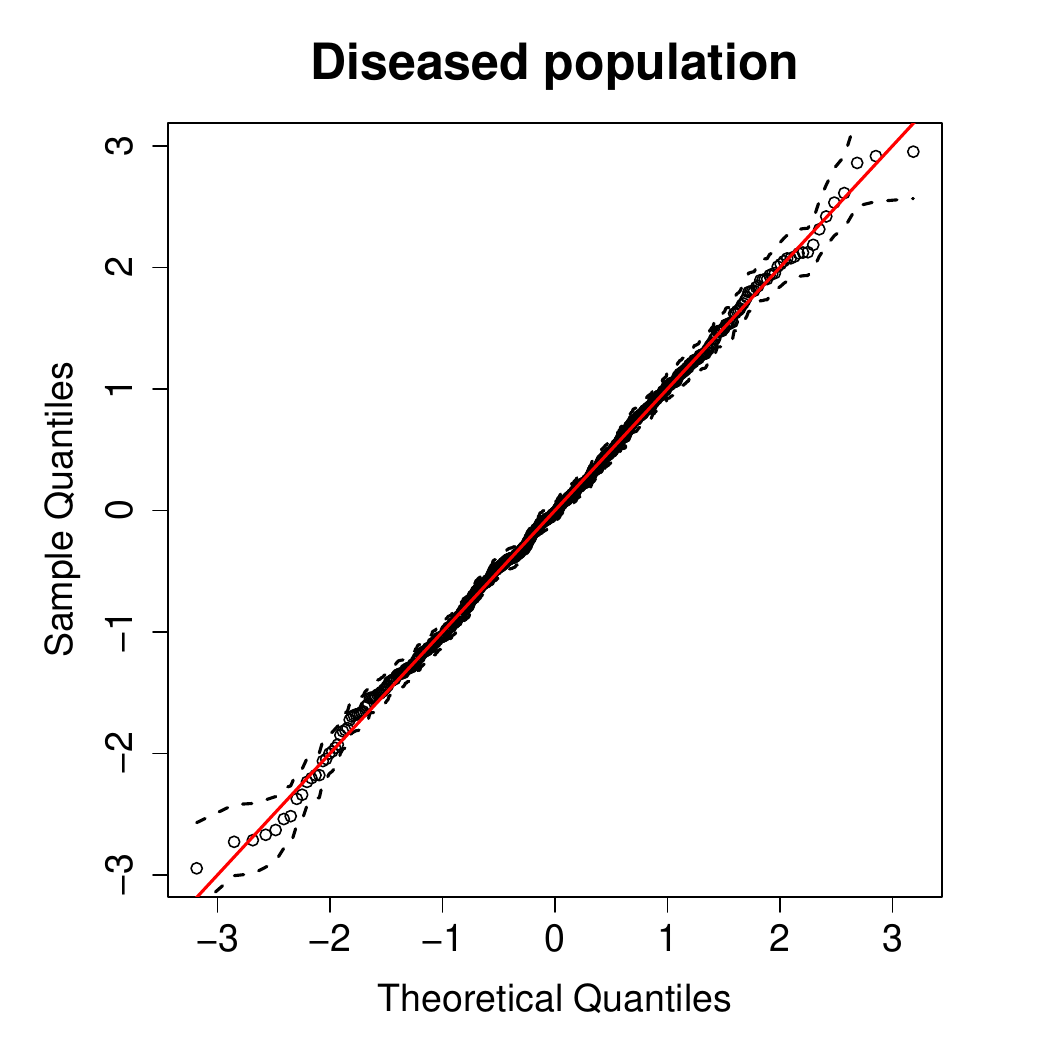}
\includegraphics[width=7cm]{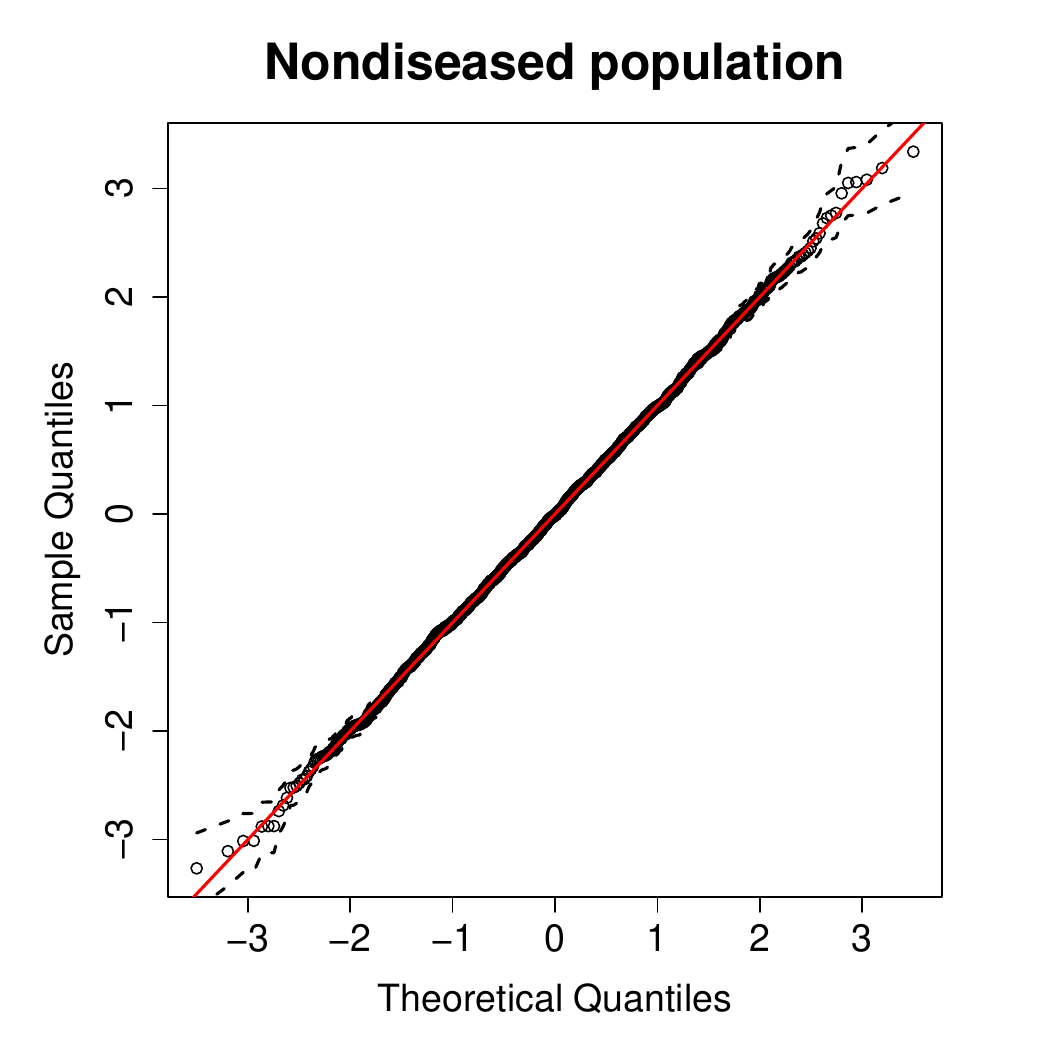}\label{qqplots_dpm}}
\subfigure[Normal model in each group]{\includegraphics[width=7cm]{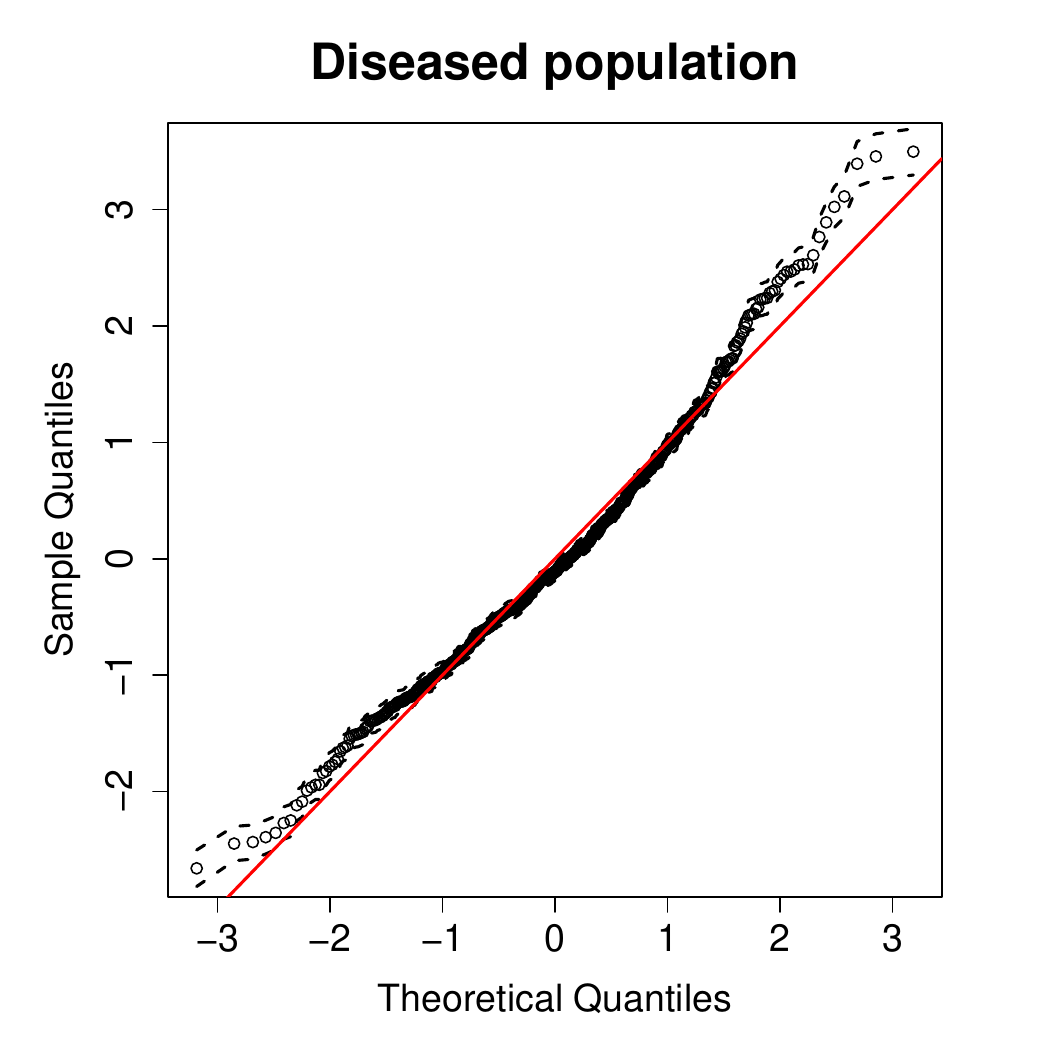}
\includegraphics[width=7cm]{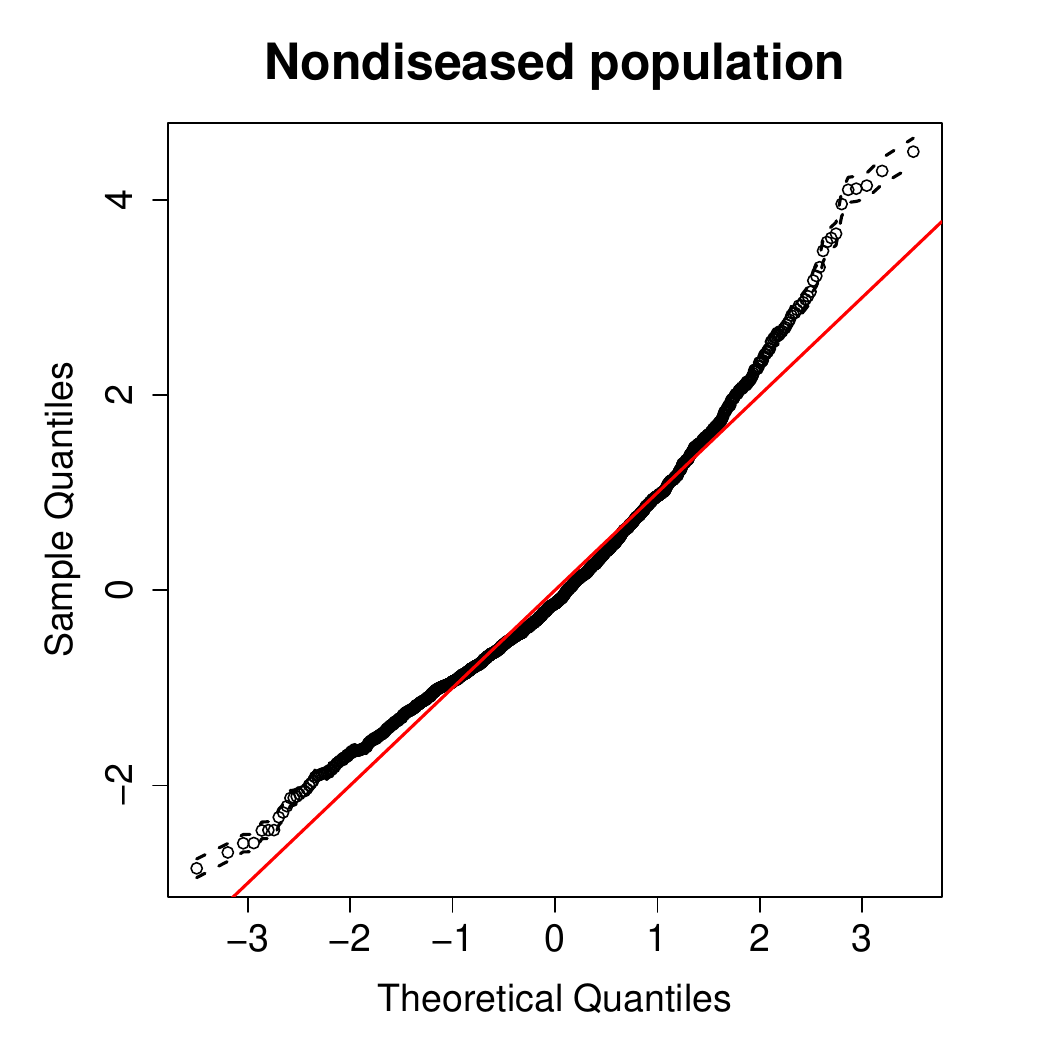}\label{qqplots_dpm_normal}}
\caption{Quantile residuals of the BMI data versus the theoretical quantiles of the standard normal distribution. The circles represent the posterior mean quantiles over all posterior samples, while the dashed lines represent the corresponding 95\% credible bands. Top row: DPM model with 10 components in each group (model \code{pROC\_dpm} in the main manuscript). Bottom row: normal model in each group (model \code{pROC\_normal} in the main manuscript).}
\label{qqplots_dpm_global}
\end{center}
\end{figure}

\begin{Schunk}
\begin{Sinput}
R> library("nor1mix")
R> traj <- matrix(0, nrow = pROC_dpm$mcmc$nsave, 
+    ncol = length(pROC_dpm$marker$d))
R> lgrid <- length(pROC_dpm$marker$d)
R> grid <- qnorm(ppoints(lgrid))
R> for (l in 1:pROC_dpm$mcmc$nsave) {
+    aux <- norMix(mu = pROC_dpm$fit$d$mu[l,], 
+    sigma = pROC_dpm$fit$d$sd[l,],
+    w = pROC_dpm$fit$d$probs[l,])
+    traj[l, ] <- quantile(qnorm(pnorMix(pROC_dpm$marker$d, aux)), 
+    ppoints(lgrid), type = 2)
+ }
R> l.band <- apply(traj, 2, quantile, prob = 0.025)
R> trajhat <- apply(traj, 2, mean)
R> u.band <- apply(traj, 2, quantile, prob = 0.975)
		
R> op <- par(pty = "s")
R> plot(grid, trajhat, xlab = "Theoretical Quantiles",
+    ylab = "Sample Quantiles", main = "Nondiseased population",
+    cex.main = 2, cex.lab = 1.5, cex.axis = 1.5)
R> lines(grid, l.band, lty = 2, lwd = 2)
R> lines(grid, u.band, lty = 2, lwd = 2)
R> abline(a = 0, b = 1, col ="red", lwd = 2)
R> par(op)
\end{Sinput}
\end{Schunk}
\end{appendix}
 The resulting QQ plots are shown in Figure \ref{qqplots_dpm} and show virtually no deviations from the standard normal distribution quantiles, thus revealing a good fit of the DPM model that assumes 10 mixture components in both the diseased and nondiseased groups. In contrast, those QQ plots obtained when fitting a normal model in each group (i.e., $L_{D}=L_{\bar{D}} = 1$; model \code{pROC\_normal} in the main manuscript), clearly show some deviations from the assumed normal distribution quantiles (see Figure \ref{qqplots_dpm_normal}). The code used follows. 
  
\begin{Schunk}
\begin{Sinput}
R> traj_normal <- matrix(0, nrow = pROC_normal$mcmc$nsave,
+    ncol = length(pROC_normal$marker$d))
R> lgrid_normal <- length(pROC_normal$marker$d)
R> grid_normal <- qnorm(ppoints(lgrid_normal))
R> for (l in 1:pROC_normal$mcmc$nsave) {
+    traj_normal[l, ] <- quantile(qnorm(pnorm(pROC_normal$marker$d, 
+    pROC_normal$fit$d$mu[l], pROC_normal$fit$d$sd[l])),
+    ppoints(lgrid_normal), type = 2)
+ }
R> l.band_normal <- apply(traj_normal, 2, quantile, prob = 0.025)
R> trajhat_normal <- apply(traj_normal, 2, mean)
R> u.band_normal <- apply(traj_normal, 2, quantile, prob = 0.975)

R> op <- par(pty = "s")
R> plot(grid, trajhat, xlab = "Theoretical Quantiles",
+    ylab = "Sample Quantiles", main = "Nondiseased population",
+    cex.main = 2, cex.lab = 1.5, cex.axis = 1.5)
R> lines(grid, l.band, lty = 2, lwd = 2)
R> lines(grid, u.band, lty = 2, lwd = 2)
R> abline(a = 0, b = 1, col ="red", lwd = 2)
R> par(op)
\end{Sinput}
\end{Schunk}

\subsection{Bayesian bootstrap and kernel estimators of the pooled ROC curve}
\begin{figure}[h!]
\begin{center}
\includegraphics[width=8cm]{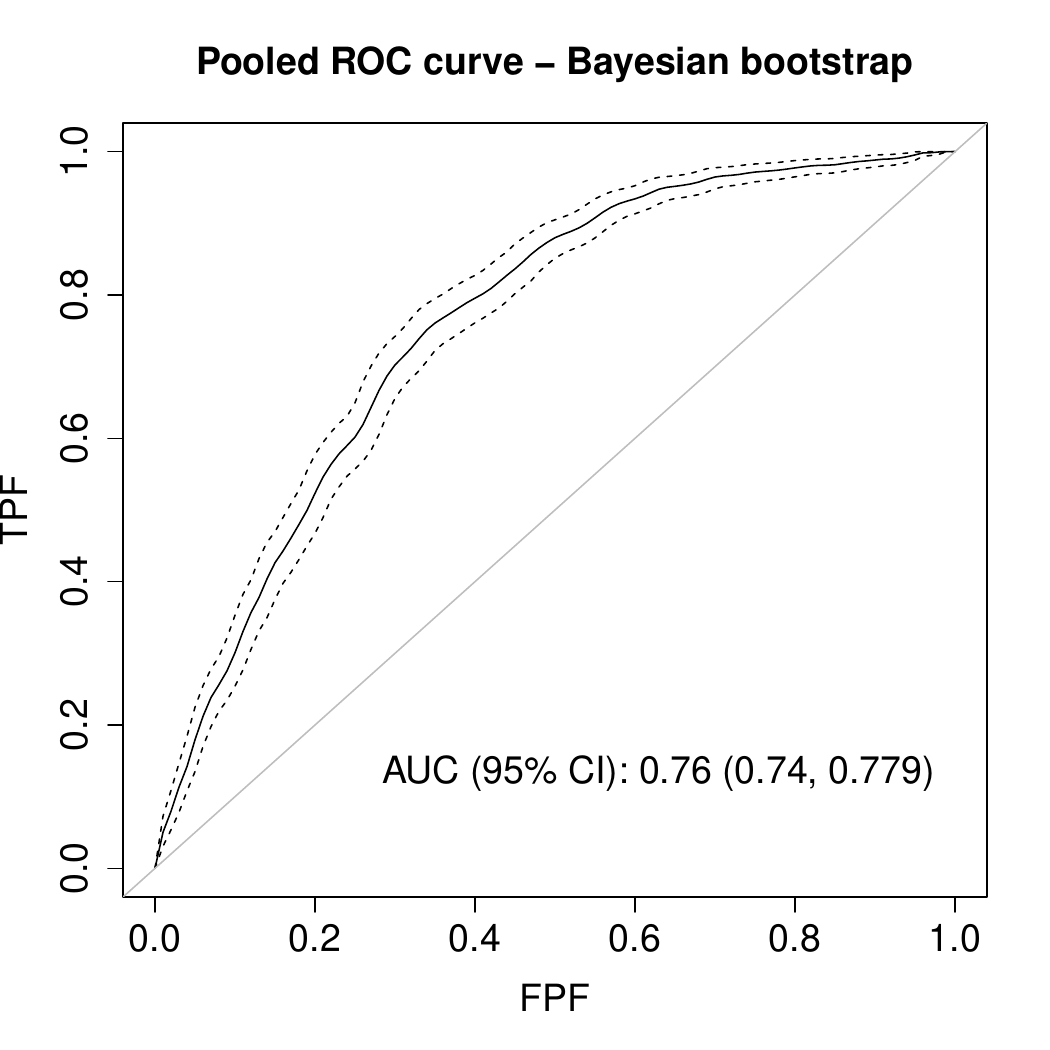}
\end{center}
\caption{Graphical results as provided by the \code{plot.pooledROC} function for an object of class \code{pooledROC.BB}. Posterior mean and 95\% pointwise credible band of the pooled ROC curve and corresponding AUC (posterior mean and 95\% credible interval).}
\label{pROC_BB_plot}
\end{figure}

The following code is used to fit the Bayesian bootstrap ROC curve estimator for the pooled ROC curve (function \code{pooledROC.BB}). The number of iterations considered is \code{B = 5000} and, for the sake of illustration, we also compute the partial area under the curve corresponding to true positive fractions (TPF) or sensitivities between $0.8$ and $1$.
\begin{Schunk}
\begin{Sinput}
R> set.seed(123, "L'Ecuyer-CMRG") # for reproducibility
R> pROC_BB <- pooledROC.BB(marker = "bmi", group = "cvd_idf", 
+    tag.h = 0, data = endosyn, p = seq(0, 1, l = 101), ci.level = 0.95,
+    pauc = pauccontrol(compute = TRUE, focus = "TPF", value = 0.8), 
+    B = 5000, parallel = "snow", ncpus = 2)
		
R> summary(pROC_BB)
\end{Sinput}
\begin{Soutput}
Call: [...]
		
Approach: Pooled ROC curve - Bayesian bootstrap
----------------------------------------------
Area under the pooled ROC curve: 0.76 (0.74, 0.779)*
Partial area under the pooled ROC curve (FPF = 0.1): 0.17 (0.14, 0.201)*
 * Credible level:  0.95

Sample sizes:
                           Group H     Group D
Number of observations        2149         691
Number of missing data           0           0
\end{Soutput}
\end{Schunk}
Note that all partial areas' values returned are normalised and, as such, what is being reported, in this case, is 
\[
\text{pAUC}_{\text{TPF}}(0.8)/(1-0.8).
\]
The estimated pooled ROC curve and AUC (posterior means), jointly with 95\% credible intervals, are obtained as follows
\begin{Schunk}
\begin{Sinput}
R> plot(pROC_BB, cex.main = 1.5, cex.lab = 1.5, cex.axis = 1.5, cex = 1.5)
\end{Sinput}
\end{Schunk}
The result is shown in Figure \ref{pROC_BB_plot}.

\begin{figure}[h!]
\begin{center}
\includegraphics[width=8cm]{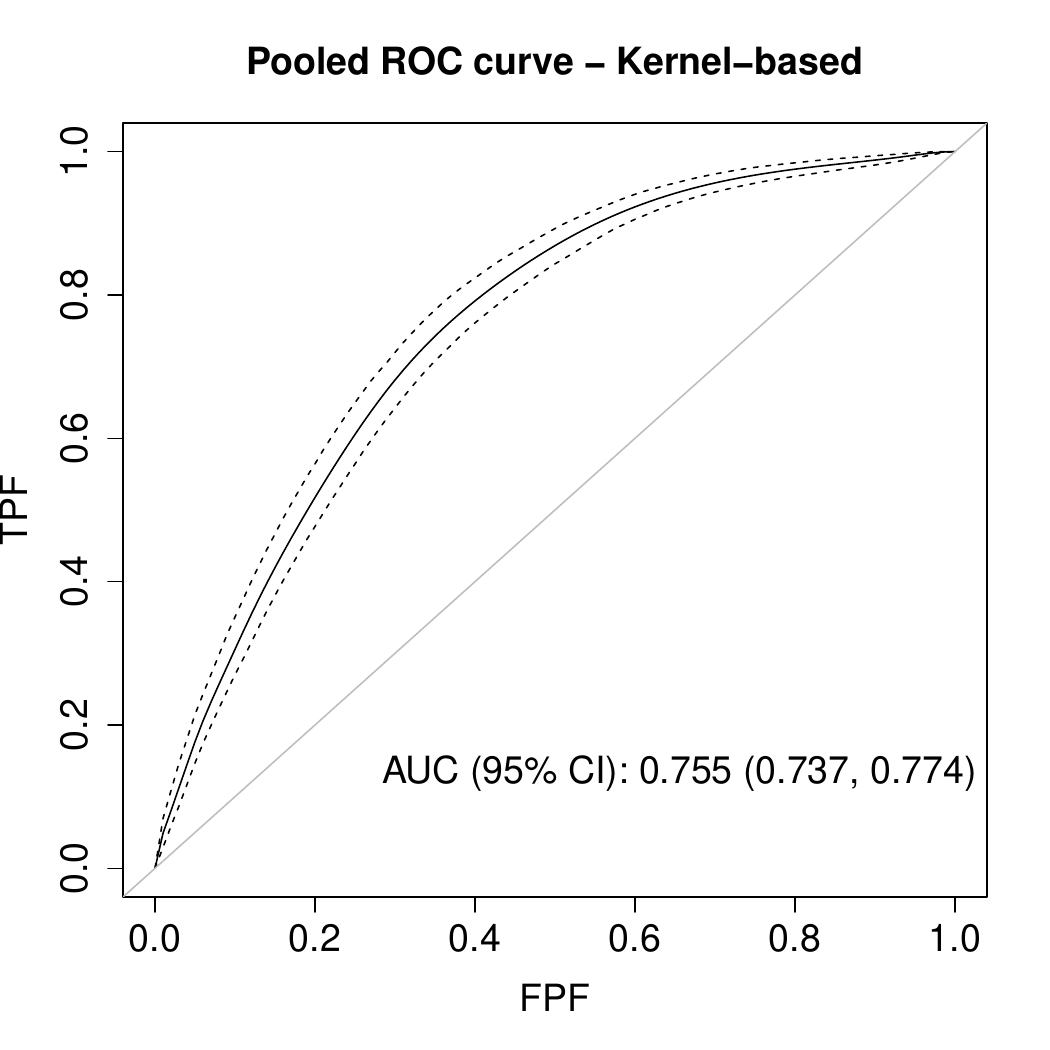}
\end{center}
\caption{Graphical results as provided by the \code{plot.pooledROC} function for an object of class \code{pooledROC.kernel}. Estimate and 95\% pointwise bootstrap confidence interval of the pooled ROC curve and corresponding AUC.}
\label{pROC_kernel_plot}
\end{figure}

\begin{figure}[h!]
\begin{center}
\includegraphics[width=8cm]{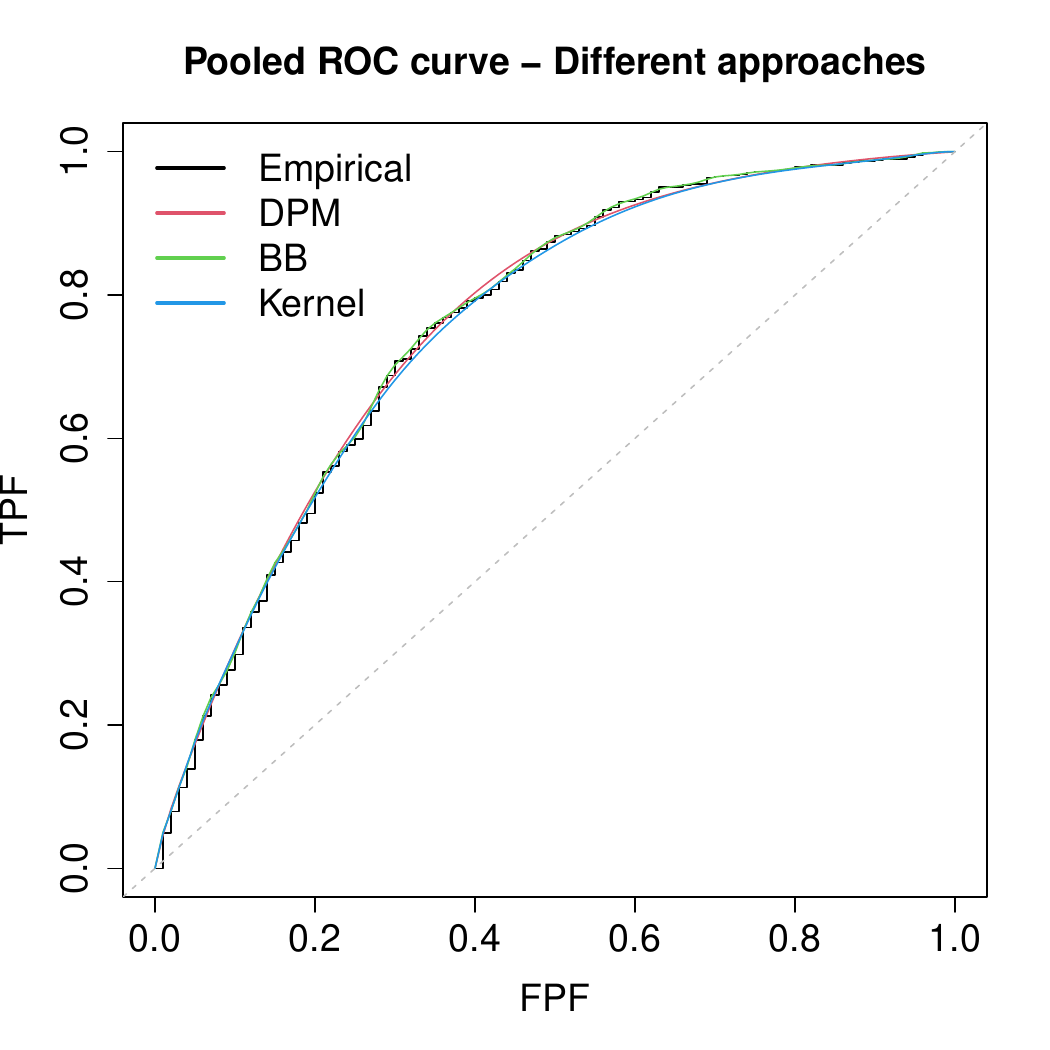}
\end{center}
\caption{ROC curve estimated using the different approaches implemented in \pkg{ROCnReg}. `Empirical' stands for the empirical estimator, `DPM' for the Dirichlet process mixture of normal distributions estimator, `BB' for the Bayesian bootstrap approach, and `Kernel' for the kernel estimator.}
\label{pROC_kernel_comp}
\end{figure}

\noindent We shall present now the syntax associated to the kernel estimator of the pooled ROC curve (function \code{pooledROC.kernel}). In terms of arguments, \code{bw} specifies how the bandwidth should be computed, with \code{SRT} standing for Silverman's rule of thumb and \code{UCV} for least squares cross-validation. Additionally, here \code{B} stands for the number of bootstrap replications used to compute the confidence intervals/bands. The syntax follows. 
\begin{Schunk}
\begin{Sinput}
R> set.seed(123, "L'Ecuyer-CMRG") # for reproducibility
R> pROC_kernel <- pooledROC.kernel(marker = "bmi", group = "cvd_idf", 
+    tag.h = 0, data = endosyn, p = seq(0, 1, l = 101), bw = "SRT",
+    B = 500, ci.level = 0.95, method = "coutcome",
+    pauc = pauccontrol(compute = TRUE, focus = "TPF", value = 0.8),
+    parallel = "snow", ncpus = 2)
		
R> summary(pROC_kernel)
\end{Sinput}
\begin{Soutput}
Call: [...]
		
Approach: Pooled ROC curve - Kernel-based
----------------------------------------------
Area under the pooled ROC curve: 0.755 (0.737, 0.774)*
Partial area under the specificity pooled ROC curve (Se = 0.8): 0.408 (0.373, 0.446)*
 * Confidence level:  0.95

                Group H     Group D
Bandwidths:       0.867       1.019



Bandwidth Selection Method: Silverman's rule-of-thumb

Sample sizes:
                           Group H     Group D
Number of observations        2149         691
Number of missing data           0           0
\end{Soutput}
\end{Schunk}
Graphical results are obtained with the following code, and present in Figure \ref{pROC_kernel_plot}
\begin{Schunk}
\begin{Sinput}
R> plot(pROC_kernel, cex.main = 1.5, cex.lab = 1.5, cex.axis = 1.5, 
+    cex = 1.5)
\end{Sinput}
\end{Schunk}
We finish this section with a comparison of the estimated pooled ROC curves obtained using all methods incorporated in \pkg{ROCnReg} (Figure \ref{pROC_kernel_comp}). 
\begin{Schunk}
\begin{Sinput}
R> plot(pROC_emp$p, pROC_emp$ROC[,1], type = "s", xlim = c(0,1),
+    ylim = c(0,1), xlab = "FPF", ylab = "TPF", 
+    main = "Pooled ROC curve - Different approaches",
+    cex.main = 1.5, cex.lab = 1.5, cex.axis = 1.5)
+    lines(pROC_dpm$p, pROC_dpm$ROC[,1], col = 2)
+    lines(pROC_BB$p, pROC_BB$ROC[,1], col = 3)
+    lines(pROC_kernel$p, pROC_kernel$ROC[,1], col = 4)
+    abline(0, 1, col = "grey", lty = 2)
+    legend("topleft", legend = c("Empirical", "DPM", "BB", "Kernel"), 
+    lty = 1, col = 1:4, bty = "n", lwd = 2, cex = 1.5)
\end{Sinput}
\end{Schunk}

\section{Further computational tools for the covariate-specific ROC curve}\label{AppB}
\subsection{Convergence assessments for the Bayesian methods: trace plots, effective sample sizes, and Geweke statistics}
\begin{figure}[h!]
\begin{center}
\includegraphics[width=13cm]{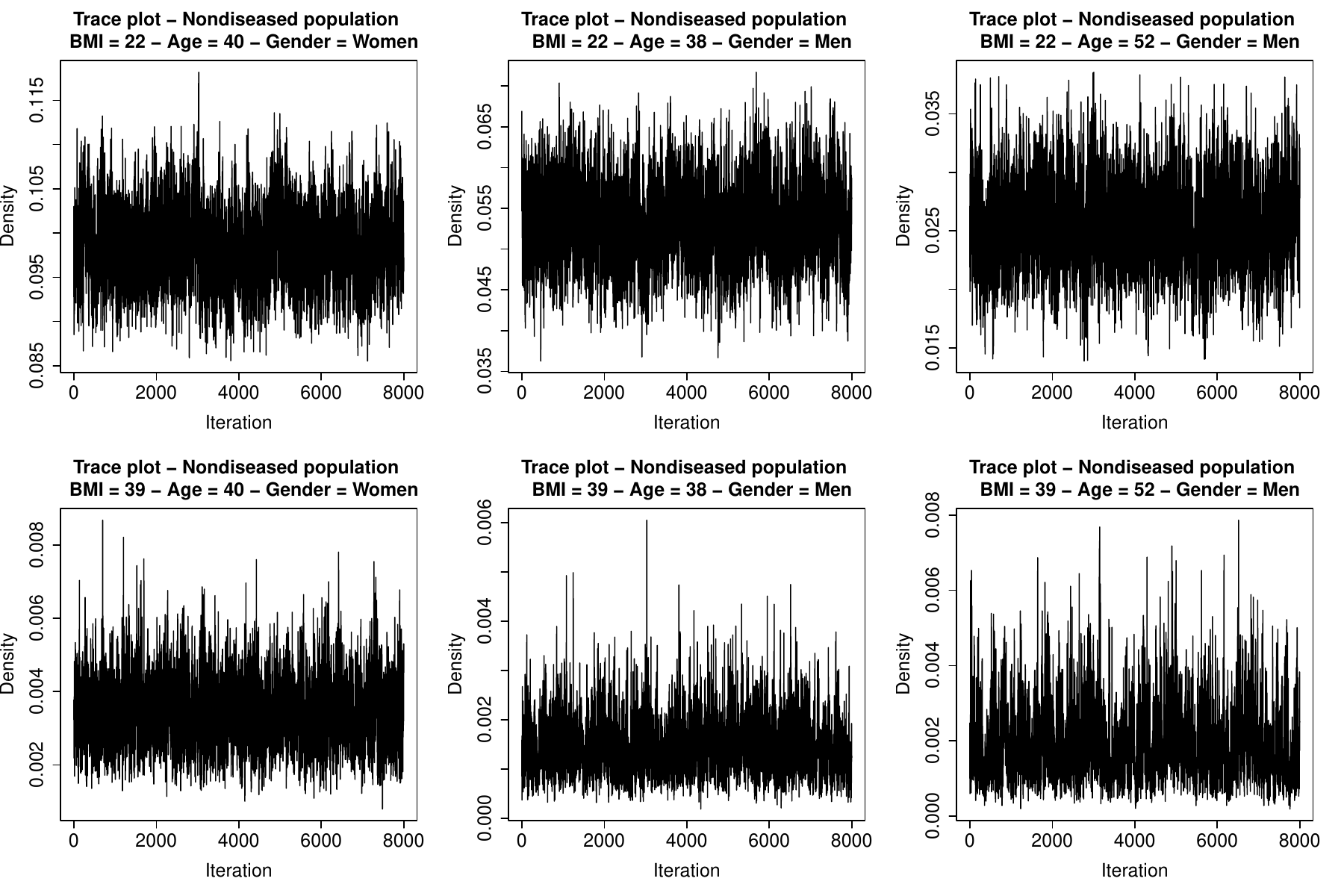}
\includegraphics[width=13cm]{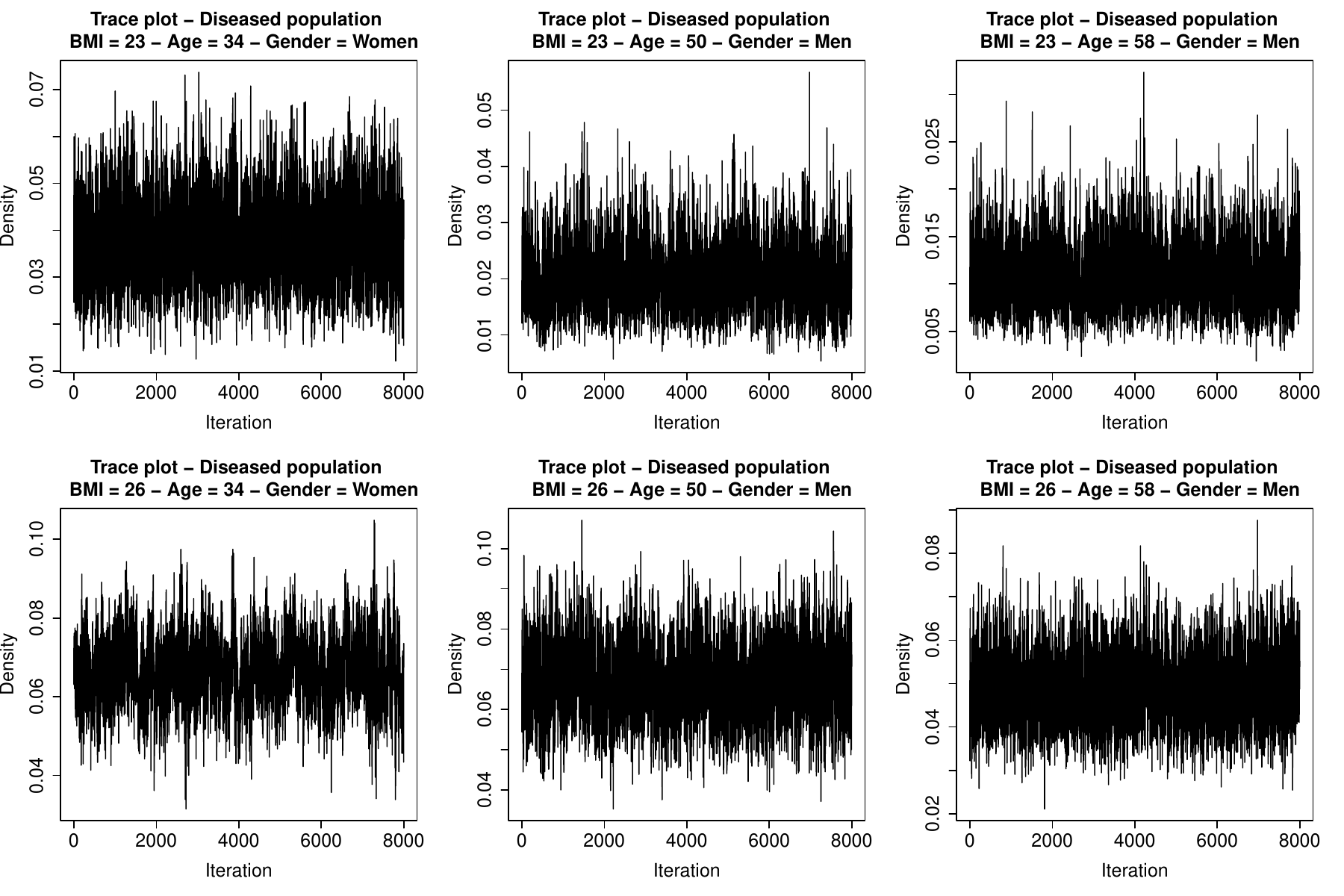}
\end{center}
\caption{Trace plots of the MCMC draws (after burn-in) of the conditional PDFs of BMI based on model \code{cROC\_bnp}. Results are shown separately for the nondiseased and diseased population, for different combinations of \code{age} and \code{gender} (covariates) and for different values of the BMI.}
\label{cROC_bnp_densities_tp}
\end{figure}

\begin{figure}[h!]
\begin{center}
\includegraphics[width=15cm]{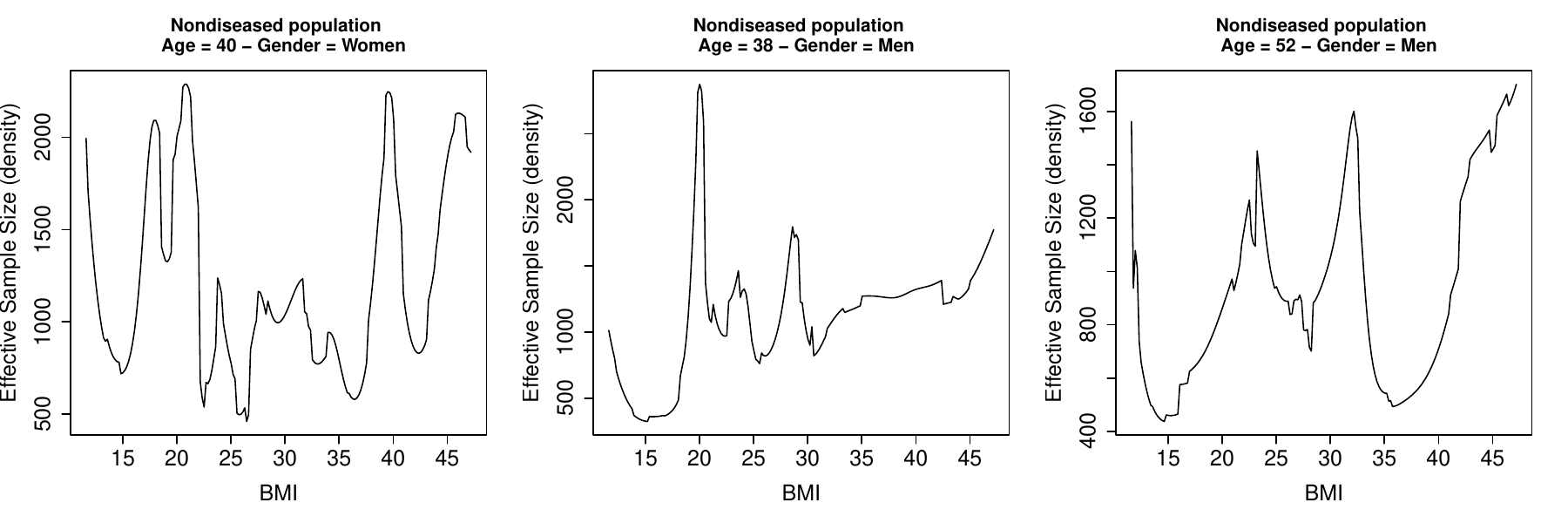}
\includegraphics[width=15cm]{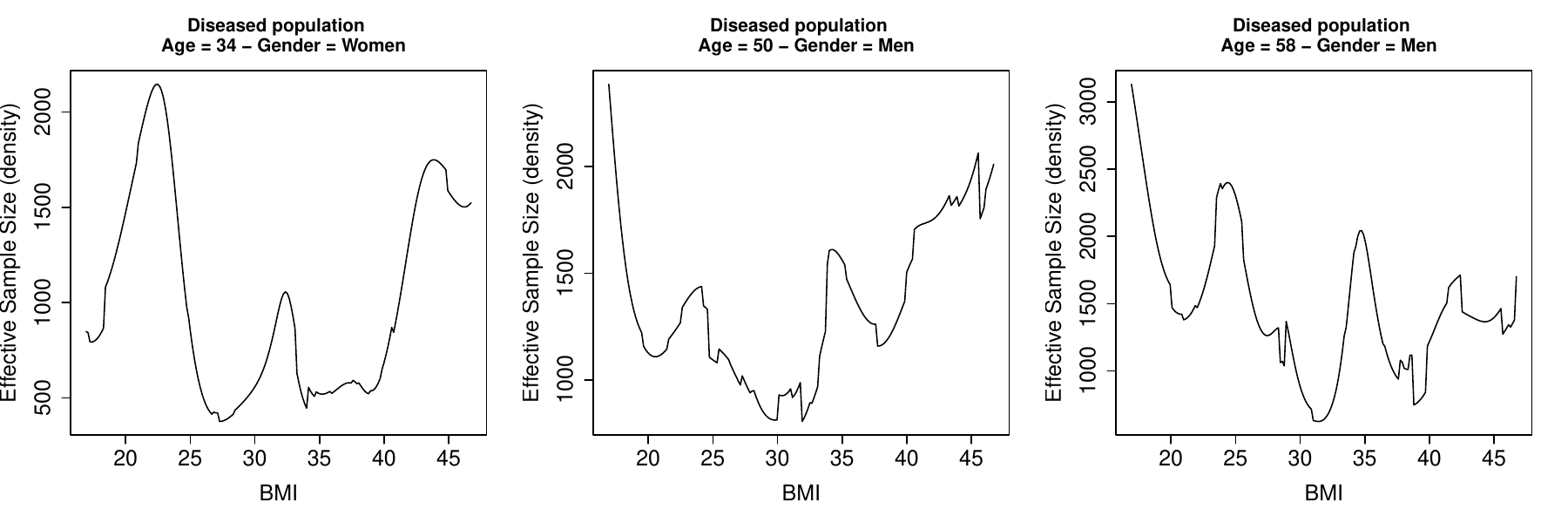}
\end{center}
\caption{Effective sample size of the MCMC chains (after burn-in) of the conditional PDFs of BMI based on model \code{pROC\_dpm}. Results are shown separately for the nondiseased and diseased population and for different combinations of \code{age} and \code{gender}. In all cases, results are shown along BMI values.}
\label{cROC_bnp_densities_es}
\end{figure}

\begin{figure}[h!]
\begin{center}
\includegraphics[width=15cm]{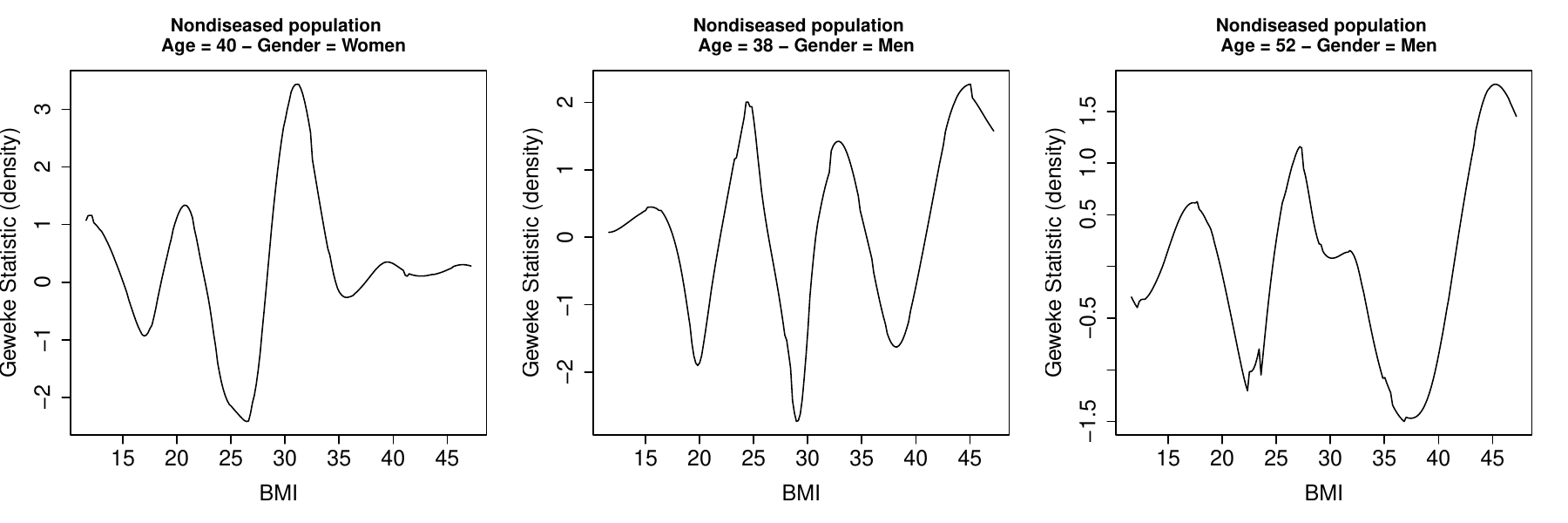}
\includegraphics[width=15cm]{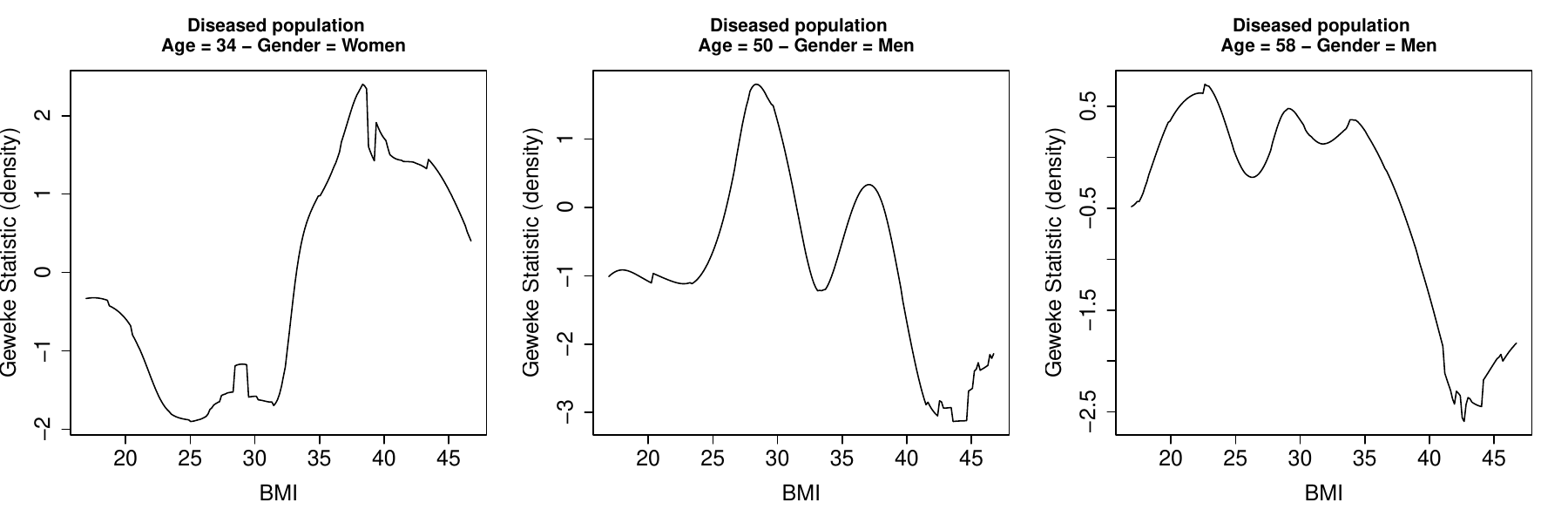}
\end{center}
\caption{Geweke statistic of the MCMC chains (after burn-in) of the conditional PDFs of BMI based on model \code{cROC\_bnp}. Results are shown separately for the nondiseased and diseased population and for different combinations of \code{age} and \code{gender}. In all cases, results are shown along BMI values.}
\label{cROC_bnp_densities_geweke}
\end{figure}

We start this section by including, for model \code{cROC\_bnp} in Section \ref{sec:croc_ilu}, some trace plots of the MCMC draws (after burn-in) of the conditional PDFs of BMI (Figure \ref{cROC_bnp_densities_tp}) and corresponding effective sample sizes and Geweke statistics (Figure \ref{cROC_bnp_densities_es} and \ref{cROC_bnp_densities_geweke}, respectively). All plots show good mixing of the MCMC chains and do not suggest lack of convergence. For conciseness, the \proglang{R}-code for producing Figures \ref{cROC_bnp_densities_tp}, \ref{cROC_bnp_densities_es}, and \ref{cROC_bnp_densities_geweke} is not provided here, but in the \proglang{R} replication code that accompanies this paper. 

\subsection{Informal model diagnostics for the Bayesian methods: quantile residuals and conditional densities}
\begin{figure}[h!]
\begin{center}	
	\subfigure[BNP model with 10 mixture components and a factor by curve interaction in each group]
	{\includegraphics[width=7cm]{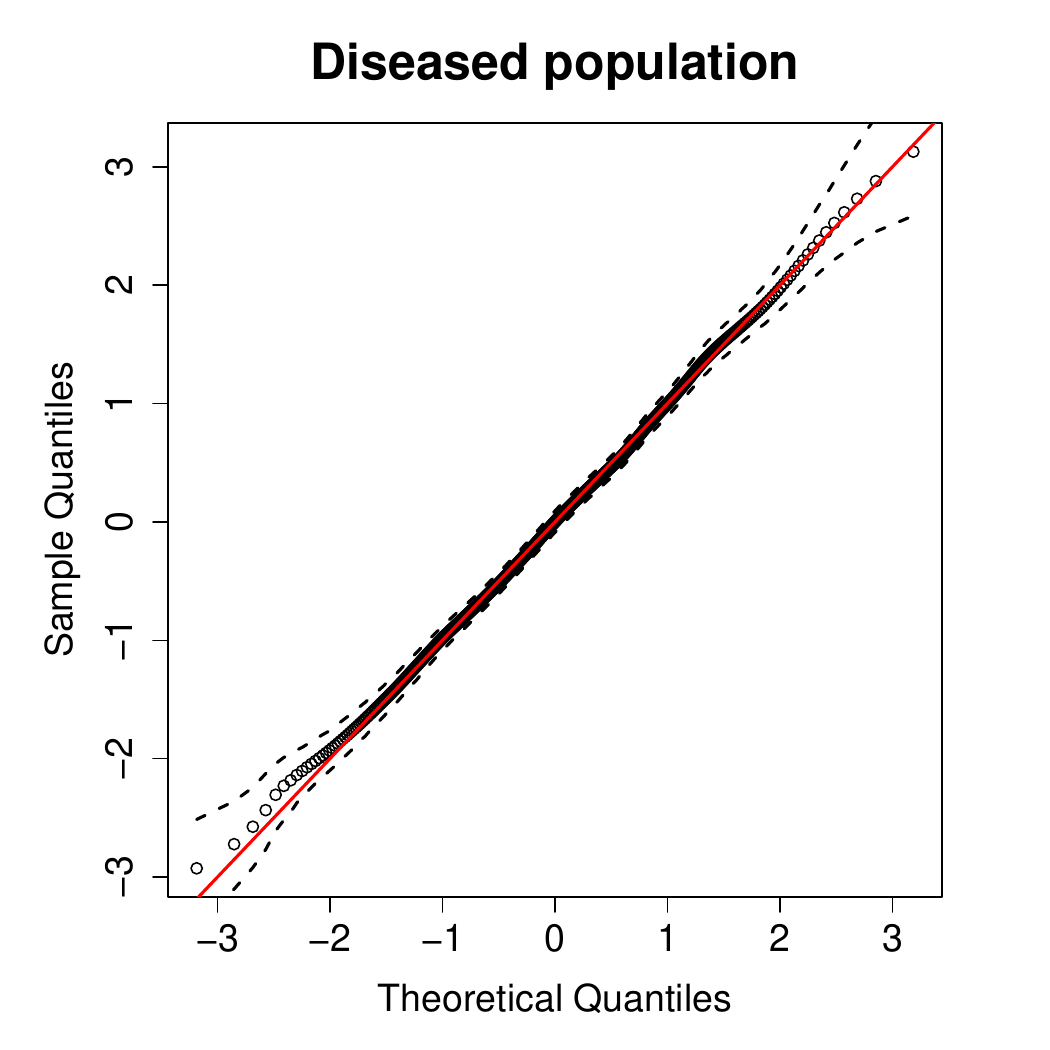}
	\includegraphics[width=7cm]{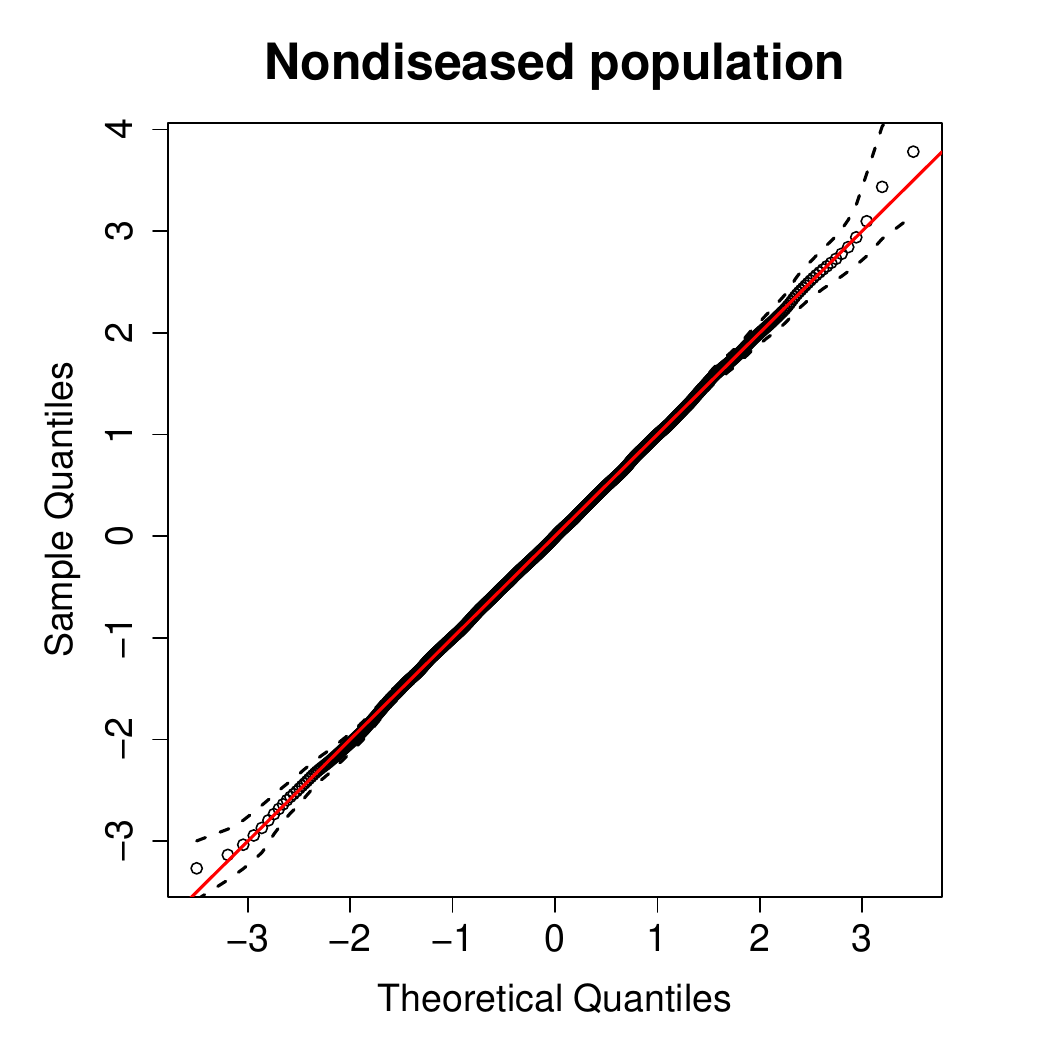}}%\label{qqplots_dpm}}
\\
		\subfigure[Bayesian normal linear model in each group]{\includegraphics[width=7cm]{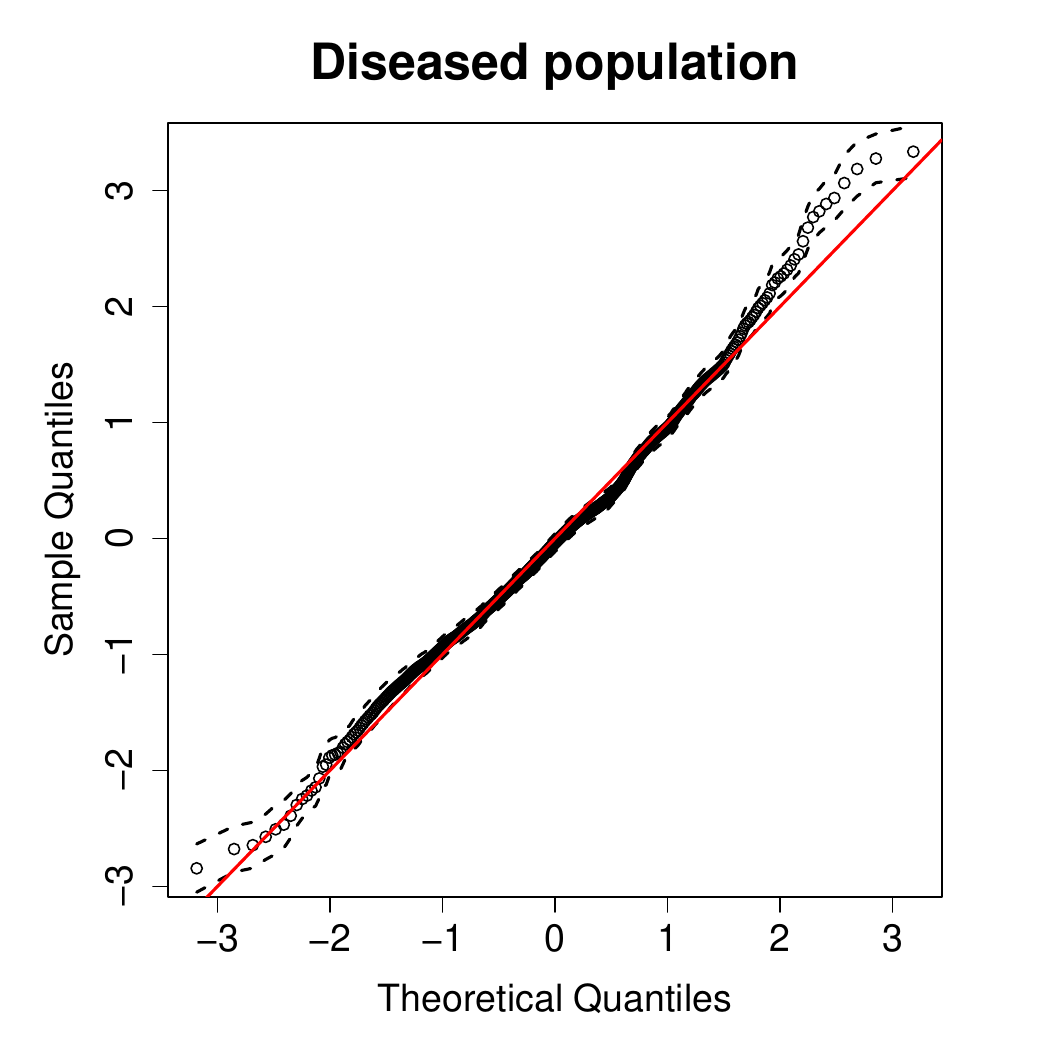}
			\includegraphics[width=7cm]{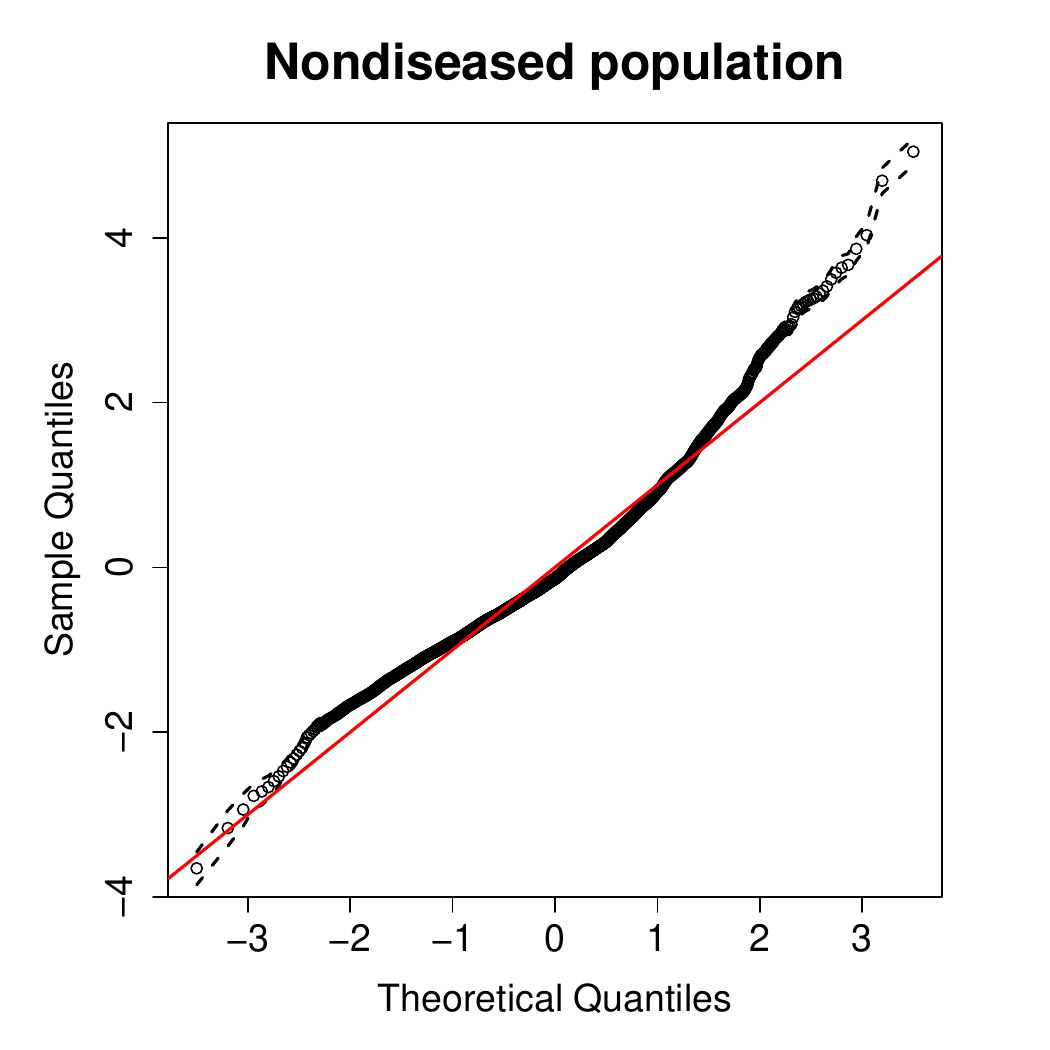}}%\label{qqplots_dpm_normal}}
		\caption{Quantile residuals of the BMI data versus the theoretical quantiles of the standard normal distribution. The circles represent the posterior mean quantiles over all posterior samples, while the dashed lines represent the corresponding 95\% credible bands. Top row: BNP model with 10 mixture components and a factor by curve interaction in each group (model \code{cROC\_bnp} in the main manuscript). Bottom row: Bayesian normal linear model in each group (model \code{cROC\_bp} in the main manuscript).}
		\label{qqplots_covsl}
	\end{center}
\end{figure}

\begin{figure}[h!]
	\begin{center}
\includegraphics[width=14cm]{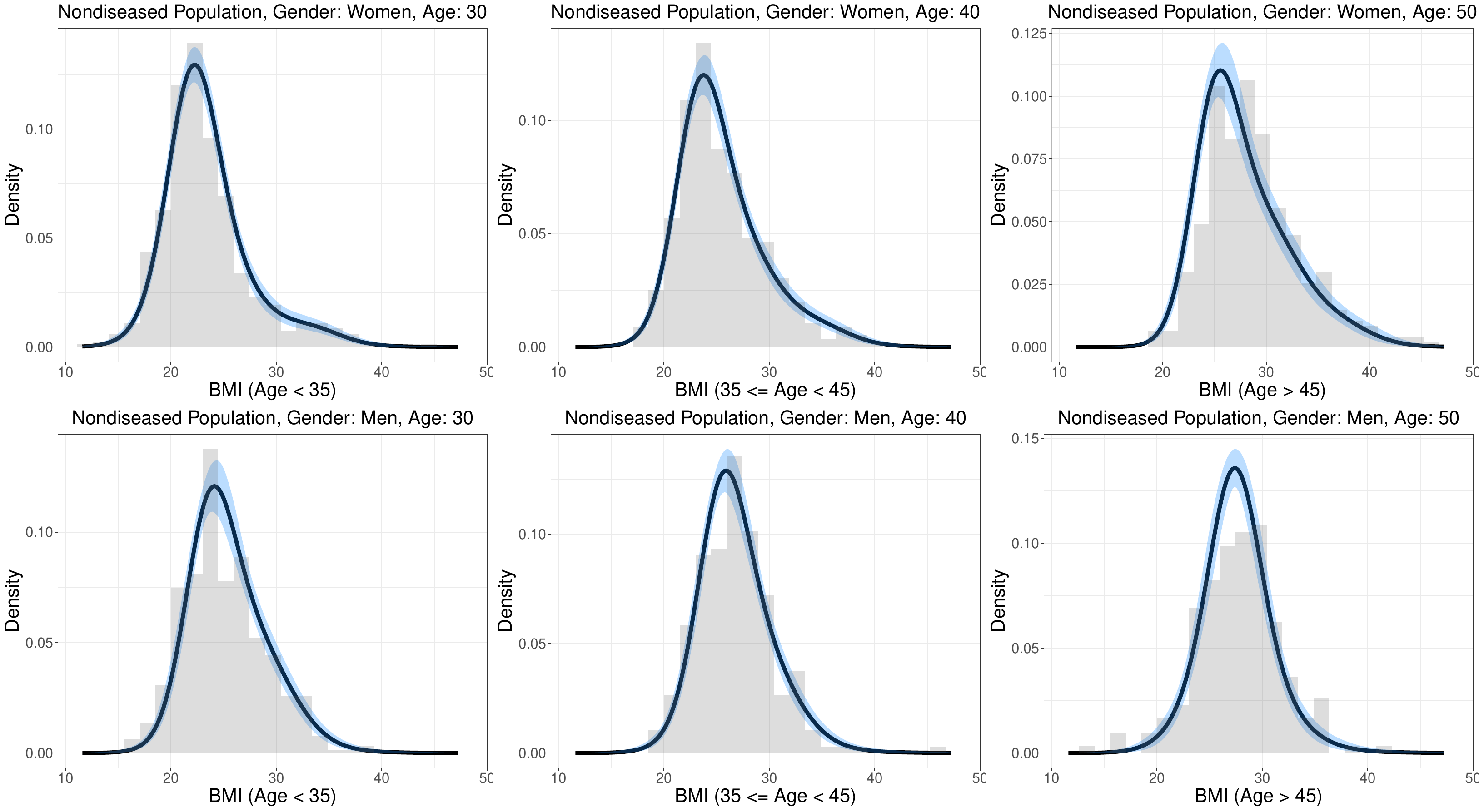}
\end{center}
\caption{Histograms of the BMI for selected age intervals and for men and women, along with the pointwise mean (continuous blue lines) and $95\%$ credible bands (shaded blue areas) for the conditional distribution of BMI given age and gender, in the nondiseased group, under the BNP model with 10 mixture components and a factor by curve interaction (model \code{cROC\_bnp} in the main manuscript). The ages of 30, 40, and 50 correspond, respectively and approximately, to the 25th, 50th, and 75th percentiles of the age variable in the whole dataset.}
\label{cond_hists_h}
\end{figure}

\begin{figure}[h!]
	\begin{center}
\includegraphics[width=14cm]{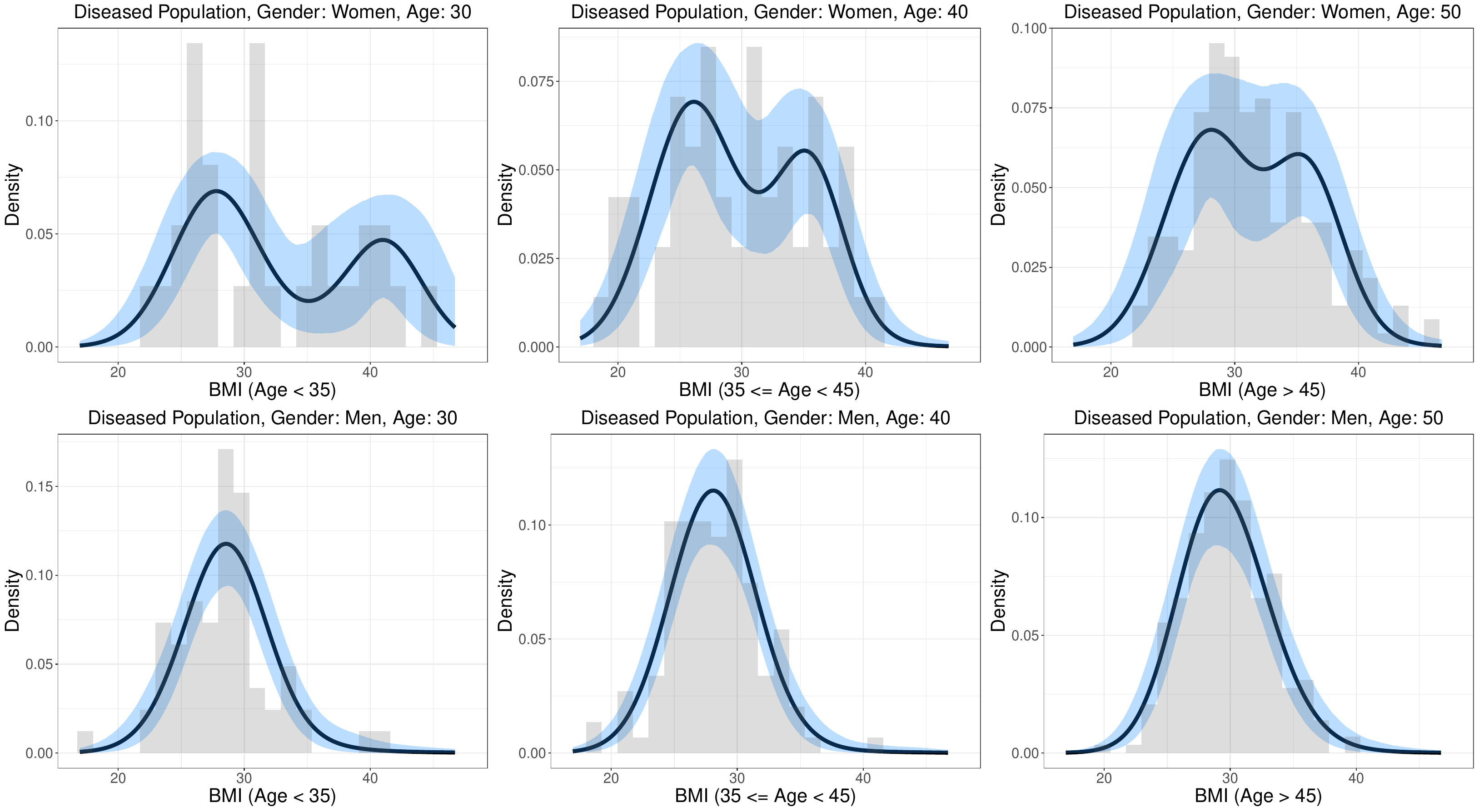}
\end{center}
	\caption{Histograms of the BMI for selected age intervals and for men and women, along with the pointwise mean (continuous blue lines) and $95\%$ credible bands (shaded blue areas) for the conditional distribution of BMI given age and gender, in the diseased group, under the BNP model with 10 mixture components and a factor by curve interaction (model \code{cROC\_bnp} in the main manuscript). The ages of 30, 40, and 50 correspond, respectively and approximately, to the 25th, 50th, and 75th percentiles of the age variable in the whole dataset.}
	\label{cond_hists_d}
\end{figure}

As in the no covariates case, we also present the quantile residuals for both the Bayesian normal linear model (designated as model \code{cROC_bp} in Section ``Package presentation and illustration'' of the main manuscript) and for the Bayesian nonparametric model with 10 mixture components and a factor by curve interaction (model \code{cROC\_bnp} in the main text). The results are shown in Figure \ref{qqplots_covsl} and, as for the unconditional case, they show virtually no deviation from the quantiles of the standard normal distribution, in both the diseased and nondiseased groups, for the Bayesian nonparametric method. Further, in Figures \ref{cond_hists_h} and \ref{cond_hists_d} we also show the histograms of the BMI for selected age intervals and for both men and women, along with the estimated (by model \code{cROC\_bnp}) conditional distribution of BMI given age and gender, in the nondiseased and diseased groups. As it can be appreciated, the estimated conditional densities follow quite nicely histograms of the BMI for the different age intervals and genders. For conciseness, the \proglang{R}-code for producing these figures is not provided here, but in the \proglang{R} replication code that accompanies the paper.

\subsection{Frequentist estimators of the covariate-specific ROC curve}
\begin{figure}[h!]
\begin{center}
\includegraphics[width=14cm]{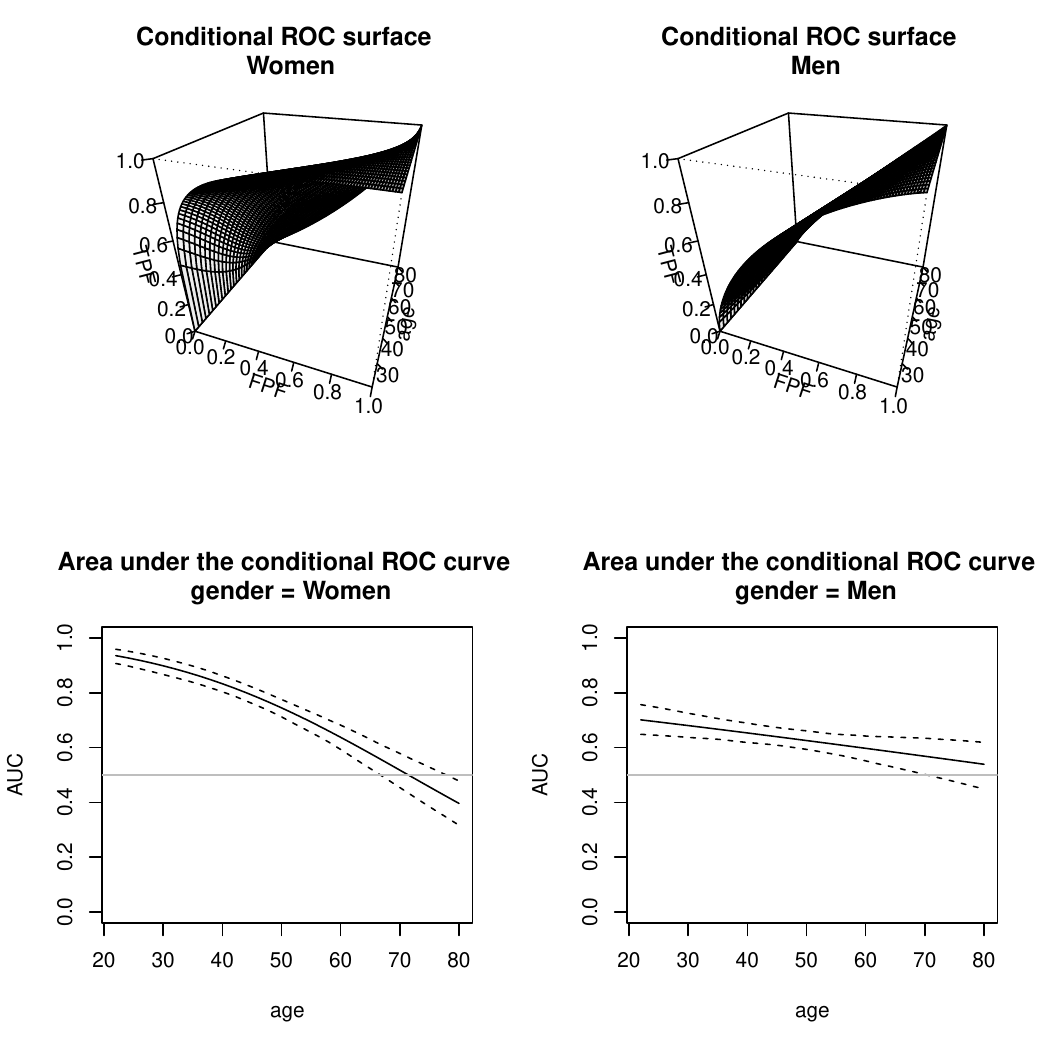}
\end{center}
\caption{Graphical results as provided by the \code{plot.cROC} function for an object of class \code{cROC.sp}. Results for a model that includes the linear interaction between \code{age} and \code{gender}. Top row: Estimate of the covariate-specific ROC curve along age, separately for men and women. Bottom row: Estimate and 95\% pointwise bootstrap confidence interval of the covariate-specific AUC along age, separately for men and women.}
\label{cROC_sp_plot}
\end{figure}

We now turn our attention on how to estimate the covariate-specific ROC curve using the induced (semiparametric) linear model (function \code{cROC.sp}). As for the Bayesian linear model described in the main manuscript, for both nondiseased and diseased groups, the model for the regression functions includes, in addition to the linear effect of \code{age} and \code{gender}, the (linear) interaction between the two (i.e., \code{gender*age} $\equiv$ \code{gender + age + gender:age}). Also, by specifying \code{est.cdf = "normal"}, we assume that the error term in both groups follows a standard normal distribution. Finally, uncertainty estimation for this method is based on the bootstrap, and through argument \code{B = 500}, we indicate the number of resamples. As usual, numeric and graphical summaries are obtained using, respectively, functions \code{summary} and \code{plot} (see Figure \ref{cROC_sp_plot}).

\begin{Schunk}
\begin{Sinput}
R> set.seed(123, "L'Ecuyer-CMRG") # for reproducibility
R> cROC_sp <- cROC.sp(formula.h = bmi ~ gender*age, 
+    formula.d = bmi ~ gender*age, group = "cvd_idf", tag.h = 0,
+    data = endosyn, newdata = endopred, est.cdf = "normal", 
+    p = seq(0, 1, l = 101), B = 500, ci.level = 0.95, 
+    parallel = "snow", ncpus = 2)

R> summary(cROC_sp)
\end{Sinput}
\begin{Soutput}
Call: [...]
		
Approach: Conditional ROC curve - semiparametric
----------------------------------------------------------

Parametric coefficients
Group H:
                   Estimate    Quantile 2.5%    Quantile 97.5%
(Intercept)         22.7670          21.9331           23.5509
genderWomen         -4.2942          -5.2247           -3.3091
age                  0.0885           0.0680            0.1091
genderWomen:age      0.0885           0.0645            0.1124


Group D:
                   Estimate    Quantile 2.5%    Quantile 97.5%
(Intercept)         26.9171          25.5389           28.2935
genderWomen          4.7363           2.2900            7.1651
age                  0.0440           0.0161            0.0719
genderWomen:age     -0.0515          -0.0954           -0.0075


ROC curve:
                   Estimate    Quantile 2.5%    Quantile 97.5%
(Intercept)         -0.9482          -1.3376           -0.6040
genderWomen         -2.0633          -2.6498           -1.4612
age                  0.0102           0.0026            0.0183
genderWomen:age      0.0320           0.0191            0.0442
b                    0.9378           0.8728            1.0044


Model selection criteria:
          Group H      Group D
AIC     12173.673     4007.215
BIC     12202.037     4029.905


Sample sizes:
                           Group H     Group D
Number of observations        2149         691
Number of missing data           0           0
\end{Soutput}
\end{Schunk}
\begin{Schunk}
\begin{Sinput}
R> op <- par(mfrow = c(2,2))
R> plot(cROC_sp, ask = FALSE)
R> par(op)
\end{Sinput}
\end{Schunk}

\begin{figure}[h!]
\begin{center}
\includegraphics[width=14cm]{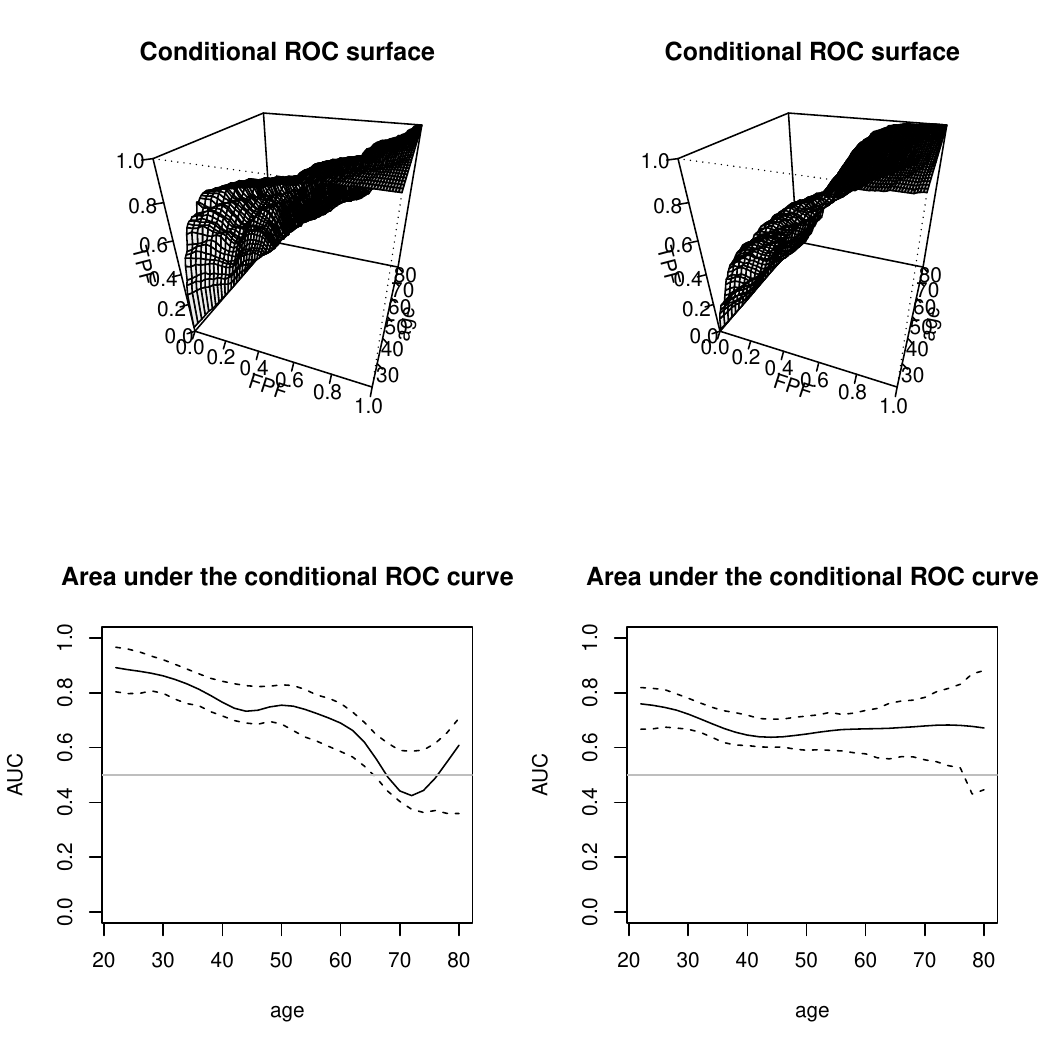}
\end{center}
\caption{Graphical results as provided by the \code{plot.cROC} function for an object of class \code{cROC.kernel}. Top row: Estimate of the covariate-specific ROC curve along age, separately for men (left) and women (right). Bottom row: Estimate and 95\% pointwise bootstrap confidence interval of the covariate-specific AUC along age, separately for men (left) and women (right). Results in this case where obtained separately for men and women.}
\label{cROC_kernel_plot}
\end{figure}

\noindent When it comes to estimating the covariate-specific ROC curve using the induced kernel-based approach (function \code{cROC.kernel}), we should keep in mind that it can only deal with one continuous covariate. As a consequence, and for our endocrine study, we evaluate \code{age} effect separately for men and women, i.e., we fit two different models. The code follows. Note that in contrast with the other functions for estimating the covariate-specific ROC curve, function \code{cROC.kernel} expects as arguments \code{marker} and \code{covariate}, where the user specifies, respectively, the name of the variables that contain the test results (in our example \code{bmi}) and the covariate (\code{age}). Uncertainty estimation for this method is also based on the bootstrap, and through argument \code{B = 500}, we indicate the number of resamples. Numeric and graphical summaries are obtained using, respectively, functions \code{summary} and \code{plot}. The graphical results, for both men and women, are shown in Figure \ref{cROC_kernel_plot}.      
\begin{Schunk}
\begin{Sinput}
R> # For prediction
R> agep <- seq(22, 80, l = 30)
R> endopred_ker <- data.frame(age = agep)
		
R> set.seed(123, "L'Ecuyer-CMRG") # for reproducibility
R> cROC_kernel_men <- cROC.kernel(marker = "bmi", covariate = "age", 
+    group = "cvd_idf", tag.h = 0, data = subset(endosyn, gender == "Men"),
+    newdata = endopred_ker, p = seq(0, 1, l = 101), B = 500, ci.level = 0.95, 
+    parallel = "snow", ncpus = 2)

R> summary(cROC_kernel_men)
\end{Sinput}
\begin{Soutput}
Call: [...]
		
Approach: Conditional ROC curve - Kernel-based
----------------------------------------------------------

Regression functions:

                Group H      Group D
Bandwidth:     5.767820     6.477821

Kernel Estimator: Local-Constant
Bandwidth Selection Method: Least Squares Cross-Validation
Continuous Kernel Type: Second-Order Gaussian

Variance functions:

                Group H       Group D
Bandwidth:     6.489771     18.567326

Kernel Estimator: Local-Constant
Bandwidth Selection Method: Least Squares Cross-Validation
Continuous Kernel Type: Second-Order Gaussian

Sample sizes:
                           Group H     Group D
Number of observations         899         418
Number of missing data           0           0
\end{Soutput}
\end{Schunk}

\begin{Schunk}
\begin{Sinput}
R> cROC_kernel_women <- cROC.kernel(marker = "bmi", covariate = "age",
+    group = "cvd_idf", tag.h = 0, data = subset(endosyn, gender == "Women"),
+    newdata = endopred_ker, p = seq(0, 1, l = 101), B = 500, ci.level = 0.95, 
+    parallel = "snow", ncpus = 2)

R> summary(cROC_kernel_women)
\end{Sinput}
\begin{Soutput}
Call: [...]
		
Approach: Conditional ROC curve - Kernel-based
----------------------------------------------------------

Regression functions:

                Group H      Group D
Bandwidth:     3.993242     4.308757

Kernel Estimator: Local-Constant
Bandwidth Selection Method: Least Squares Cross-Validation
Continuous Kernel Type: Second-Order Gaussian

Variance functions:

                 Group H       Group D
Bandwidth:     18.250813     10.490069

Kernel Estimator: Local-Constant
Bandwidth Selection Method: Least Squares Cross-Validation
Continuous Kernel Type: Second-Order Gaussian

Sample sizes:
                           Group H     Group D
Number of observations        1250         273
Number of missing data           0           0
\end{Soutput}
\end{Schunk}

\begin{Schunk}
\begin{Sinput}
R> op <- par(mfcol = c(2,2))
R> plot(cROC_kernel_women, ask = FALSE)
R> plot(cROC_kernel_men, ask = FALSE)
R> par(op)
\end{Sinput}
\end{Schunk}

\section{Frequentist estimators of the AROC curve}\label{AppC}	
We finish this document by presenting the code for estimating the covariate-adjusted ROC curve (AROC curve) using the induced semiparametric linear model (function \code{AROC.sp}) and the kernel-based approach (function \code{AROC.kernel}). We avoid giving many details, and simply present the code for fitting the models and obtaining the numerical and graphical summaries. It is important to note that, since the kernel-based approach only deals with one continuous covariate, the AROC curve in this case is estimated separately in men and women. This is to be differentiated from the AROC curve obtained by including both \code{age} and \code{gender}, which reflects the discriminatory capacity solely due to the \code{bmi} while teasing out both \code{age} and \code{gender} effects.
\begin{Schunk}
\begin{Sinput}
R> # Induced semiparametric linear model
R> set.seed(123, "L'Ecuyer-CMRG") # for reproducibility
R> AROC_sp <- AROC.sp(formula.h = bmi ~ gender*age, group = "cvd_idf", 
+    tag.h = 0, data = endosyn, est.cdf = "normal", p = seq(0, 1, l = 101),
+    B = 500, ci.level = 0.95, parallel = "snow", ncpus = 2)
		
R> summary(AROC_sp)
\end{Sinput}
\begin{Soutput}
Call: [...]
		
Approach: AROC semiparametric
----------------------------------------------
Area under the covariate-adjusted ROC curve: 0.638 (0.612, 0.664)*
 * Confidence level:  0.95

Parametric coefficients (Group H):
                   Estimate    Quantile 2.5%    Quantile 97.5%
(Intercept)         22.7670          22.0029           23.6441
genderWomen         -4.2942          -5.4177           -3.3017
age                  0.0885           0.0658            0.1080
genderWomen:age      0.0885           0.0637            0.1147


Model selection criteria:
          Group H
AIC     12173.673
BIC     12202.037


Sample sizes:
                           Group H     Group D
Number of observations        2149         691
Number of missing data           0           0
\end{Soutput}
\end{Schunk}

\begin{Schunk}
\begin{Sinput}
R> plot(AROC_sp, cex.main = 1.5, cex.lab = 1.5, cex.axis = 1.5, cex = 1.3)
\end{Sinput}
\end{Schunk}
\begin{figure}[h!]
\begin{center}
\includegraphics[width=8cm]{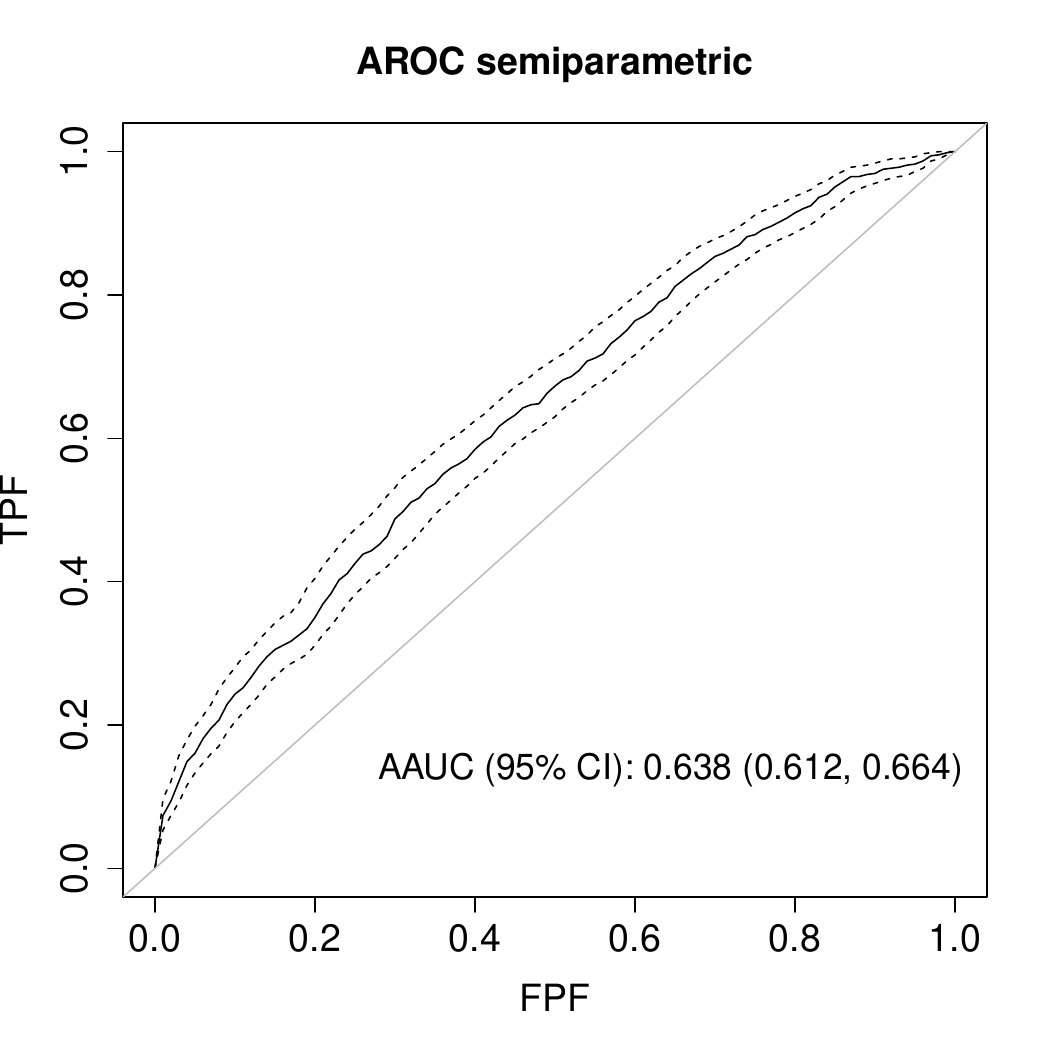}
\end{center}
\caption{Graphical results as provided by the \code{plot.AROC} function for an object of class \code{AROC.sp}. Estimate and 95\% pointwise bootstrap confidence interval of the age/gender adjusted ROC curve (AROC) and corresponding AUC.}
\label{AROC_sp_plot}
\end{figure}

\begin{Schunk}
\begin{Sinput}
R> # Kernel-based approach
R> set.seed(123, "L'Ecuyer-CMRG") # for reproducibility
R> AROC_kernel_men <- AROC.kernel(marker = "bmi", covariate = "age",
+    group = "cvd_idf", tag.h = 0, data = subset(endosyn, gender == "Men"),
+    p = seq(0, 1, l = 101), B = 500, ci.level = 0.95, parallel = "snow", ncpus = 2)
		
R> summary(AROC_kernel_men)
\end{Sinput}
\begin{Soutput}
Call: [...]
		
Approach: AROC Kernel-based
----------------------------------------------
Area under the covariate-adjusted ROC curve: 0.668 (0.636, 0.708)*
 * Confidence level:  0.95

Regression function:

                Group H
Bandwidth:     5.767820

Kernel Estimator: Local-Constant
Bandwidth Selection Method: Least Squares Cross-Validation
Continuous Kernel Type: Second-Order Gaussian

Variance function:

                Group H
Bandwidth:     6.489771

Kernel Estimator: Local-Constant
Bandwidth Selection Method: Least Squares Cross-Validation
Continuous Kernel Type: Second-Order Gaussian

Sample sizes:
                           Group H     Group D
Number of observations         899         418
Number of missing data           0           0
\end{Soutput}
\end{Schunk}

\begin{Schunk}
\begin{Sinput}
R> set.seed(123, "L'Ecuyer-CMRG") # for reproducibility
R> AROC_kernel_women <- AROC.kernel(marker = "bmi", covariate = "age",
+    group = "cvd_idf", tag.h = 0, data = subset(endosyn, gender == "Women"),
+    p = seq(0, 1, l = 101), B = 500, ci.level = 0.95, parallel = "snow", ncpus = 2)

R> summary(AROC_kernel_women)
\end{Sinput}
\begin{Soutput}
Call: [...]
		
Approach: AROC Kernel-based
----------------------------------------------
Area under the covariate-adjusted ROC curve: 0.673 (0.636, 0.716)*
 * Confidence level:  0.95

Regression function:

                Group H
Bandwidth:     3.993242

Kernel Estimator: Local-Constant
Bandwidth Selection Method: Least Squares Cross-Validation
Continuous Kernel Type: Second-Order Gaussian

Variance function:

                 Group H
Bandwidth:     18.250813

Kernel Estimator: Local-Constant
Bandwidth Selection Method: Least Squares Cross-Validation
Continuous Kernel Type: Second-Order Gaussian

Sample sizes:
                           Group H     Group D
Number of observations        1250         273
Number of missing data           0           0
\end{Soutput}
\end{Schunk}
\begin{Schunk}
\begin{Sinput}
R> op <- par(mfcol = c(1,2))
R> plot(AROC_kernel_women, main = "AROC kernel-based \n Women", 
+    cex.main = 1.5, cex.lab = 1.5, cex.axis = 1.5, cex = 1.7)
R> plot(AROC_kernel_men, main = "AROC kernel-based \n Men", 
+    cex.main = 1.5, cex.lab = 1.5, cex.axis = 1.5, cex = 1.7)
R> par(op)
\end{Sinput}
\end{Schunk}
\begin{figure}[h!]
\begin{center}
\includegraphics[width=14cm]{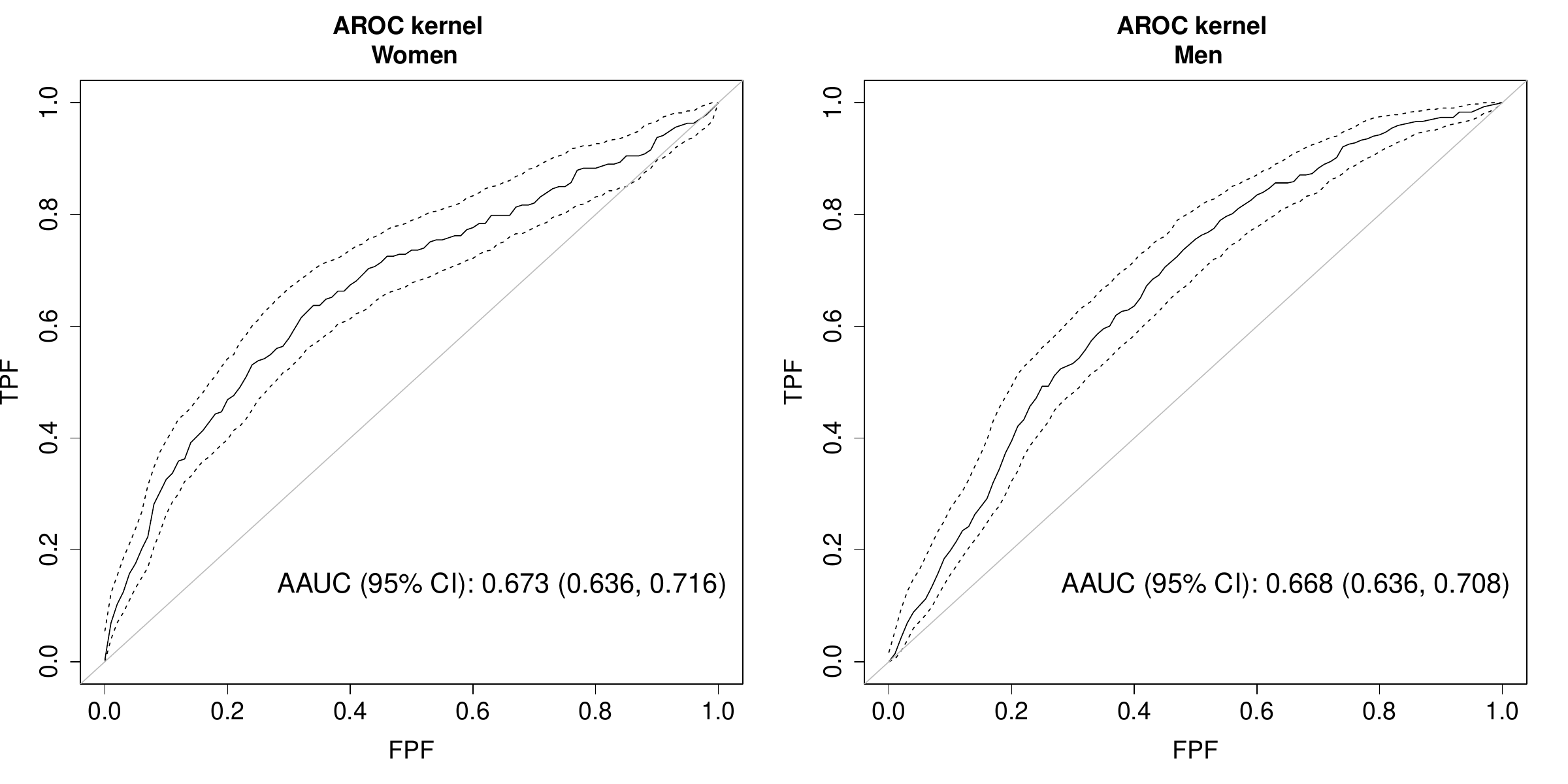}
\end{center}
\caption{Graphical results as provided by the \code{plot.AROC} function for an object of class \code{AROC.kernel}. Estimate and 95\% pointwise bootstrap confidence interval of the age adjusted ROC curve (AROC) and corresponding AUC. Analyses were done separately for women (left plot) and men (right plot).}
\label{AROC_kernel_plot}
\end{figure}
\end{document}